\documentclass[a4paper,11pt]{article}
\pdfoutput=1
\usepackage{jcappub}
\usepackage{bm}
\usepackage{soul}
\usepackage[dvipsnames]{xcolor}
\usepackage{latexsym}
\usepackage{dcolumn}
\usepackage{amsfonts,amssymb,amsmath}
\usepackage{graphicx,epsfig}
\usepackage{psfrag}
\usepackage{braket}
\usepackage{caption}
\usepackage{subcaption}
\usepackage{graphicx}
\usepackage{rotating}
\usepackage{upgreek}
\usepackage{hyperref}
\usepackage{tikz}
\usepackage{stmaryrd} 
\usepackage{comment}
\usepackage{footmisc}
\usepackage{tikz}
\usepackage{booktabs}
\usepackage{siunitx}
\usetikzlibrary{trees, positioning}
\usetikzlibrary{decorations.pathreplacing, calc}
\usepackage[framemethod=default]{mdframed}
\newmdenv[skipabove=7pt,
skipbelow=7pt,
rightline=false,
leftline=false,
topline=false,
bottomline=false,
backgroundcolor=gray!10,
linecolor=gray,
innerleftmargin=5pt,
innerrightmargin=5pt,
innertopmargin=5pt,
innerbottommargin=5pt,
leftmargin=0cm,
rightmargin=0cm,
linewidth=4pt]{eBox}

\hypersetup{
	unicode=false,          
	pdftoolbar=true,        
	pdfmenubar=true,        
	pdffitwindow=false,     
	pdfstartview={FitH},    
	pdftitle={My title},    
	pdfauthor={Author},     
	pdfsubject={Subject},   
	pdfcreator={Creator},   
	pdfproducer={Producer}, 
	pdfkeywords={keyword1} {key2} {key3}, 
	pdfnewwindow=true,      
	colorlinks=true,       
	linkcolor=red,          
	citecolor=cyan,        
	filecolor=magenta,      
	urlcolor=green,           
	linktocpage=true
}

\newcommand{\Mp}{M_{\mathrm{Pl}}} 
\newcommand{\dd}{\mathrm{d}} 
\newcommand{\ii}{\mathrm{i}} 
\newcommand{\ppi}{\uppi}
\newcommand{\etamed}{\eta_{\rm med} }

\makeatletter
\gdef\@fpheader{}
\makeatother

\begin{document}

\title{
Multi-Stage and Multi-Field Inflation in Random Inflationary Landscapes
}

\author[1]{Xingang Chen,}
\author[2]{Lucas Pinol,}
\author[3]{Zhong-Zhi Xianyu,}
\author[3]{Yisong Zhang}

\affiliation[1]{Institute for Theory and Computation, Harvard-Smithsonian Center for Astrophysics, 60 Garden Street, Cambridge, MA 02138, USA}

\affiliation[2]{Laboratoire de Physique de l’École Normale Supérieure, ENS, CNRS, Université PSL, Sorbonne Université, Université Paris Cité, F-75005, Paris, France}

\affiliation[3]{Department of Physics, Tsinghua University, Beijing 100084, China}

\abstract
{
Generic potentials in an inflationary landscape are typically too steep to support a prolonged period of cosmic inflation. Atypical flat regions of the landscape are required to support successful inflation.
The occurrence of such regions leads broadly to two possibilities: a single prolonged period of slow-roll inflation (which we call single-stage inflation), or multiple shorter periods of inflation separated by transient departures from slow roll (which we call multi-stage inflation), giving rise to primordial features and other departures from the behavior of the simplest slow-roll model. The former is the possibility most commonly assumed when deriving the primordial density fluctuations for the Big Bang model. In this paper, using Gaussian random potentials as a simple model of the inflationary landscape, we study the relative occurrence of these two possibilities. We find a substantial fraction of multi-stage inflationary trajectories in landscapes with field-space dimensions one, two, and three, and this fraction increases with dimensionality at fixed typical landscape curvature. In constructing the inflationary landscapes and identifying successful inflationary trajectories, we also define several relevant quantities and classify detailed properties of both multi-stage and multi-field inflationary trajectories.
}

\maketitle
		
\newpage
\setcounter{page}{1}		

\section{Introduction}
\label{sec1}

Cosmic inflation, a sustained period of accelerated expansion of the Universe, is the leading paradigm for our primordial universe history~\cite{Starobinsky:1979ty, Mukhanov:1981xt, Guth1981inflation, Linde:1982uu, Albrecht:1982mp, Hawking:1982cz,Starobinsky:1982ee,Guth1982inflation-perturbations, Vilenkin:1982wt, Linde:1983gd, Mukhanov:1985rz, Sasaki:1986hm}.
Yet, its microphysical details remain to be determined.
So far, the simplest model, the so-called single-field slow-roll inflation model, is phenomenologically good enough to explain all astronomical observations (see, e.g., Planck's legacy~\cite{Planck-legacy-2020}).
However, physicists believe that this apparently simple dynamics must emerge from a richer and more fundamental theory.

The type of flat potential needed to sustain a long period of inflation should have a curvature, $V''(\phi)$, much smaller in magnitude than the squared Hubble parameter, $H^2$. It is found that such potential shapes are not generic in a consistent gravitationally coupled quantum field theory. Loop corrections, couplings to spacetime curvature, or Planck-suppressed interactions, all of which are related to the backreaction from the inflationary background, can introduce contributions to $V''(\phi)$ of order $H^2$, making the potential too steep to support a sustained period of inflation. Equivalent conclusions can also be reached when constructing inflation models from a more UV-complete fundamental theory. This problem has been called the ``$\eta$-problem"~\cite{Copeland:1994vg, Chen:2008hz} because one of the slow-roll parameters, $\eta$, characterizes the curvature of the inflationary potential. On the other hand, this does not mean that viable inflation models are impossible to construct, but rather that the required flat potential shapes are not generic --- symmetries, tunings, or accidental cancellations among different contributions are needed to realize a potential flat enough to support a successful period of inflation. Also, observationally, $\eta$ is constrained to be small only at most of the CMB scales, but does not have to remain so on smaller scales.

This consideration leads broadly to two categories of inflation models.
One possibility is that accidental cancellations result in a potential sufficiently flat to continuously support one stage of slow-roll inflation lasting for at least $\sim 60$ e-folds.\footnote{The precise minimum number of required e-folds depends on the details of reheating, see Appendix \ref{app_efolds}.} In this paper, we refer to such models as ``single-stage inflation models".
Another possibility is that the required $\sim 60$ e-folds of inflation are achieved through several stages of inflation, with the potential during each stage not sufficiently flat to support the full required number of e-folds. We refer to such models as ``multi-stage inflation models".

Although the first possibility is the one most commonly assumed, there are, at least qualitatively, arguments in favor of either possibility. For example, since tuning is required to obtain sufficiently flat potentials, less flat potentials might be expected to occur more commonly. On the other hand, once a stage of inflation ends, inflation may be more likely to terminate altogether, since another accidental condition may be required for the inflaton to encounter a subsequent inflationary stage.
As the inflationary landscape most likely involves more than one field, the situation becomes even more complicated.
Thus, despite the conventional focus on the first possibility, it is actually difficult to determine, without a more detailed analysis, the relative occurrence rates of the two scenarios. Investigating this relative frequency is one of the main goals of this paper.

To model the randomness of the inflationary potential described above, we use a modified Gaussian random potential as a toy model of the inflationary landscape, which we also call a Gaussian random landscape.\footnote{
Inspiration for the construction of Gaussian random landscapes mainly comes from Ref.~\cite{Masoumi:2016eag} (itself using techniques presented in the single-field case in~\cite{Tegmark:2004qd}). Other statistical studies of multifield inflation include~\cite{Frazer:2011tg,Frazer:2011br, McAllister:2012am, Bjorkmo:2017nzd}. Different from the previous works, the focus of our work is statistics of {\em multi-stage} inflation trajectories. More discussions on the differences can be found in Sec.~\ref{sec-conclusion}.
}

A Gaussian random potential is a scalar potential whose values in field space form a zero-mean Gaussian random field, with its statistical properties specified by 
its power spectrum. We further uplift each realization such that its global minimum, identified as the true vacuum, has zero potential energy. The typical curvature of the potential is controlled by the power spectrum and is chosen to be sufficiently large so that the typical value of $\eta$ is of order one and generic regions of the landscape do not support slow-roll inflation. We randomly choose the initial position of the inflaton in field space. The tuning associated with the $\eta$-problem then amounts to the inflaton encountering rare, sufficiently flat regions of the potential landscape.
Using numerical simulations, we study various statistical properties of inflationary trajectories in these landscapes, including the relative occurrence of single- and multi-stage inflation, adopting a flat field-space measure as the starting point for our statistics.
We also study the relative occurrence of multi-field inflation models when the field dimension is more than one.\footnote{In this work, we use the number of e-folds required to solve the horizon problem as the main criteria for a successful inflationary trajectory
and, although we perform a first comparison of predicted primordial power spectra to Planck~\cite{Planck:2018jri}, we leave detailed
statistics after observational constraints for a future work.}

As part of the motivation for this study, we would like to emphasize that multi-stage inflation models may have important phenomenological consequences. The most commonly assumed initial conditions for the Big Bang model, in terms of primordial density perturbations, are those predicted by the simplest single-field slow-roll inflation models: the perturbations are nearly scale-invariant and approximately Gaussian over all scales. Multi-stage inflation necessarily departs from this assumption by introducing time-dependent features during transitions between adjacent inflationary stages, leading to scale-dependent features in the primordial density perturbations. Although the simplest predictions agree well with observations on the scales probed to date by the cosmic microwave background and large-scale structure, such scale-dependent features may become apparent with more precise measurements on these scales or may appear on scales that remain unexplored by current observations. These primordial features may enhance or reduce structure formation, seed primordial black holes, excite primordial standard clocks, generate gravitational waves, or increase the energy scale of the cosmological collider. 

\vskip 4pt
The paper is organized as follows. We start in Sec.~\ref{sec-landscapes} by explaining how to construct random inflationary landscapes and how to characterize the typical value for the potential curvature, for any field-space dimension.
In Sec.~\ref{sec-methods}, we present our numerical procedures and algorithms to systematically explore the physics of inflation on such landscapes.
We start our investigation with the simple single-field case in Sec.~\ref{sec1field}.
Then, in Sec.~\ref{sec2field}, we thoroughly characterize the two-field case.
Finally, in Sec.~\ref{sec3field}, we show some statistical trends when going to the case of three scalar fields.
We finish with conclusions and discussion in Sec.~\ref{sec-conclusion}, and the Appendices provide several technical details.

\section{Random inflationary landscapes}
\label{sec-landscapes}

In this Section, we explain how to construct Gaussian random landscapes with physical properties of interest.
What we require is that the typical curvature of the multi-dimensional scalar potential be of order squared Hubble parameter.
To do so, we will specify the potential through its power spectrum, meaning that it will be decomposed into Fourier modes with random amplitudes and phases that verify certain probability distributions.
By changing the parameters of these distributions, we can vary the statistical properties of the landscapes, such as their typical curvatures.
For a given set of parameters, we generate statistically equivalent but realisation-dependent landscapes on which inflation may happen, thus mimicking the randomness of picking our universe among a very large number of other universes with similar overall properties.  
By doing so, we will answer whether it is \textit{likely} that our universe has emerged from such highly curved primordial landscapes.
We start in Sec.~\ref{subsec:eta-problem} by briefly reviewing the $\eta$-problem as a motivation to construct specific inflationary landscapes, and in Sec.~\ref{subsec2.1} by defining Gaussian random landscapes that we will use throughout our paper, focusing on their general properties.
In Sec.~\ref{subsec2.2} we define more precisely the concrete setup to be used in our statistical exploration as well as the fiducial set of parameters that constitute the spine of our analysis.
Then, in Sec.~\ref{subsec2.3}, we explain how to characterize the typical curvature of a given landscape, and we check whether the landscapes generated with our methodology verify the physical properties that we wish.

\subsection{The $\eta$-problem}
\label{subsec:eta-problem}

Let us start by reviewing the problem which one encounters in constructing inflation models, namely the $\eta$-problem. This problem is an important motivation for the subject of study in this paper and the methodology that will be used in constructing inflationary landscapes.

First, it ought to be noted that the Standard Model of particle physics, minimally coupled to gravity as described by General Relativity, does not provide us with a successful microphysical description of inflation.
Going beyond it opens the Pandora box of questioning the nature of physics at energies so high that they will never be probed in terrestrial experiments.
Therefore, to proceed, physicists would be better off following a few guiding principles.
These may be divided into two classes, commonly dubbed top-down and bottom-up arguments, respectively.
Top-down arguments proceed from our attempts at describing the quantum nature of gravity, chief amongst which are string theories. 
Bottom-up arguments are more generic and simply build upon our understanding of physics at lower energies.

\vskip 4pt
How the inflationary dynamics may be realised in string theory is the topic of several dedicated reviews~\cite{Quevedo:2002xw,Linde:2005dd,McAllister:2007bg,Baumann:2014nda,Cicoli:2023opf}; we here make no attempt to thoroughly review the vast and complex literature on this topic, and we instead focus solely on the conclusions that are relevant for this work while referring the reader to the references for further details.
Although a fully consistent description of gravity in a quantum framework remains elusive so far, recent progress in various fields of string theory hints at the plausible physics of inflation.
A common feature of these theories is the necessity to reduce the number of spacetime dimensions to only four, a popular mechanism being dimensional compactification, wherein extra dimensions are small and compact.
After the process, the properties of these extra dimensions, as well as the objects living in the fundamental theory, are effectively described as fields in the four-dimensional spacetime.
Although the nature of these fields and their interactions depends on the specific choice of a string theory and a compactification scheme, a common trait is the existence of multiple scalars.
The simplest interactions amongst these  scalars are encapsulated within the scalar potential, also called ``landscape'' for it provides us with a visual description of the potential energy in the multi-dimensional field space, but other interactions involving kinetic terms might also appear.
Now, an important observation is that it is very challenging to achieve a stable four-dimensional configuration while maintaining the remaining scalars light, i.e.~with a small curvature of the potential.\footnote{
See also Refs.~\cite{Achucarro:2018vey,Bravo:2020wdr} for related discussions on how turning trajectories in the landscape might respect the criteria enunciated as the string swampland conjectures~\cite{Obied:2018sgi,Klaewer:2016kiy,Grimm:2018ohb}.}

\vskip 4pt
There also exist arguments against flat scalar potentials that are independent of any assumptions about the genuine nature of quantum gravity.
These effective field theory arguments, based on simple power counting and dimensional analysis, are powerful tools to guide us towards a more realistic theory of inflation.\footnote{See~\cite{Weinberg:2008hq} for EFT formalisms applied to the covariant theory of full fields and~\cite{Cheung:2007st,Senatore:2010wk,Noumi:2012vr,Pinol:2024arz} for the ones straight applied to the theory of fluctuations. Note that, because the background evolution plays a crucial role in addressing the questions considered in this work, we need to start from an EFT of inflation formulated in terms of the full fields, rather than working directly with perturbations.}
Indeed, the so-called UV-sensitivity of inflation is the most severe, with even Planck-suppressed operators playing an important role.
The easiest way to realize that is to revisit the $\eta$-problem with fewer assumptions about the UV.
Dimension-six operators, even when suppressed by two powers of the Planck mass in the inflationary Lagrangian, bring corrections to the curvature of the potential of order one in Planck mass units, thus generically ruining flat potentials. 
This flaw is common to any inflationary scenario based on flat potentials, but it is even more dramatic for so-called large-field models, wherein the distances spanned in field space are larger than a Planck mass, with those radiative corrections becoming uncontrollably large for an infinite number of higher-dimensional operators.

To conclude, from the perspective of a consistent gravitationally coupled quantum field theory or a UV-complete fundamental theory, generic inflaton potentials are too steep to support a prolonged period of cosmic inflation. Atypical flat regions in the inflationary landscape, arising from symmetries, tunings, or accidental cancellations, are therefore required to support successful inflation. 
This now called ``$\eta$-problem"~\cite{Copeland:1994vg, Chen:2008hz} ---with reference to $\eta_V(\phi) = \Mp^2 V''(\phi)/V(\phi)$ in single-field inflation---motivates the study of multifield inflation in highly curved potentials.
Below, we model both this generic expectation and the required tuning using a simple Gaussian random landscape.

\subsection{Generalities}
\label{subsec2.1}

The landscape construction methodology is mainly built upon the theoretical framework introduced in \cite{Masoumi:2016eag}, which we promote to a complete numerical implementation. We consider an $N_{\mathrm{f}}$-dimensional field space, with $\boldsymbol{\phi}=\left(\phi^1,\phi^2,\dots,\phi^{N_{\mathrm{f}}}\right)$.
We assume the field space is flat and unbounded, which, in practice, implies that the scalar fields $\phi^a$ with $a \in \{1, \ldots, N_\mathrm{f} \}$ have canonical kinetic terms and span infinite ranges on the real line.\footnote{Multifield models of inflation can always be written in a frame where the kinetic terms of the scalar fields are all canonical, though at the price of generating non-minimal coupling to gravity~\cite{Kaiser:2010ps}. Although we could in principle study these cases, either in the frame where the kinetic terms are non-canonical or in the one where the coupling to gravity is non-minimal, we focus in this work on the simplest models where the scalar potential contains all the information about the theory.}

\vspace{4 pt}
We then consider a random potential landscape $v(\boldsymbol{\phi})$ living on this field space. 
The dimensionless function $v$ is related to the usual scalar potential $V$ via $v(\boldsymbol{\phi})=V(\boldsymbol{\phi})/V_0$ where $V_0^{1/4}$ would be the typical energy scale of inflation.
We assume that $v$, although a random function with field-dependent values, is statistically homogeneous and isotropic.
In practice, this means that there is no preferred location nor direction in field space across landscapes and that, e.g., global or local minima must be looked for case-by-case.
We also assume that $v$ follows a Gaussian distribution with zero mean and a variance set by:
\begin{equation}
\label{eq:corr}
    \braket{v(\boldsymbol{\phi})v(\boldsymbol{\phi}')}=\int\frac{\dd^{N_{\mathrm{f}}}\boldsymbol{\ppi}}{(2\pi)^{N_{\mathrm{f}}}}e^{\ii\boldsymbol{\ppi}\cdot\left(\boldsymbol{\phi}-\boldsymbol{\phi}'\right)}P(\ppi)\,,
\end{equation}
where we defined $\ppi = \sqrt{\boldsymbol{\ppi}\cdot\boldsymbol{\ppi}}$ the modulus of the conjugate momentum $\boldsymbol{\ppi}$ to $\boldsymbol{\phi}$, not to be confused with the number $\pi \simeq  3.14$.
The existence of $P(\ppi)$ as the power spectrum of the landscape is a consequence of statistical homogeneity, while its dependence on the modulus $\ppi$ only is a consequence of isotropy.
The power spectrum is a deterministic quantity which fixes the statistical properties of the landscape, but realizations of the landscape are themselves random.
Indeed, let us construct the landscape step by step.
First, we consider the following Fourier decomposition:
\begin{equation}    
\label{eq:Fourier-decomp}v(\boldsymbol{\phi})=\int\frac{\dd^{N_{\mathrm{f}}}\boldsymbol{\ppi}}{(2\pi)^{N_{\mathrm{f}}}}a_{\boldsymbol{\ppi}}e^{\ii\boldsymbol{\ppi}\cdot\boldsymbol{\phi}},
\end{equation}
where $a_{\boldsymbol{\ppi}}$ is the complex amplitude of the Fourier mode $\boldsymbol{\ppi}$ of the landscape.
Note that 
\begin{equation}
    a_{-\boldsymbol{\ppi}}=a_{\boldsymbol{\ppi}}^*  \,, 
\end{equation}
as imposed by the reality of the scalar potential.
Moreover, in order for $v$ to have zero-mean and to verify Eq.~\eqref{eq:corr}, we require that the Fourier amplitudes verify
\begin{align}
    \braket{a_{\boldsymbol{\ppi}}} &= 0 \,, \\
    \braket{a_{\boldsymbol{\ppi}}a_{\boldsymbol{\ppi}'}^*}&=(2\pi)^{N_{\mathrm{f}}}\delta^{({N_{\mathrm{f}}})}\left(\boldsymbol{\ppi}-\boldsymbol{\ppi}'\right)P(\ppi)\,.
\end{align}
We assume that their higher-order connected correlation functions all vanish, so the $a_{\boldsymbol{\ppi}}$ appearing in Eq.~\eqref{eq:Fourier-decomp} are themselves Gaussian random variables with zero mean and a diagonal covariance set by $P(\ppi)$ only.

\vspace{4pt}
Although well-defined, this setup cannot be used 
in practice.
Indeed, computer simulations of the potential landscape can only manage compact field spaces, meaning that we have to regularize the possible field excursions with a cutoff $\Lambda$, with $\phi^a \in [-\Lambda,\Lambda]$ and periodic boundary conditions $v(\ldots, \Lambda,\ldots)=v(\ldots,-\Lambda,\ldots)$, the total volume being now finite and equal to $(2 \Lambda)^{N_\mathrm{f}}$.
We are led to consider the discrete version of the Fourier decomposition:
\begin{equation}
    \int\frac{\dd^{N_{\mathrm{f}}}\boldsymbol{\ppi}}{(2\pi)^{N_{\mathrm{f}}}}a_{\boldsymbol{\ppi}}e^{\ii\boldsymbol{\ppi}\cdot\boldsymbol{\phi}} \longrightarrow \frac{1}{(2\Lambda)^{N_\mathrm{f}}} \sum_{\substack{\boldsymbol{m} \in \mathbb{Z}^{N_\mathrm{f}}\\ \forall  a,\, |m_a|\leqslant M}} 
   A_{\boldsymbol{m}}\exp\left(\ii\frac{\pi}{\Lambda}\boldsymbol{m}\cdot\boldsymbol{\phi}\right)\,,
\end{equation}
where the Fourier modes $\boldsymbol{\ppi}$ now take discrete values $\ppi_a= m_a\times(\pi/\Lambda)$ in which $m_a\in\mathbb{Z}$ and where $\pi$ in the RHS is $3.14\ldots$
In practice, we have avoided dealing with infinite sums by setting a cutoff in Fourier space as $ |m_a| \leqslant M \in \mathbb{N}$.\footnote{This is a purely technical matter, and for large enough $M$ it is clear that landscape properties become independent of $M$. Indeed, $\Lambda / M$ can be understood as the resolution scale of the landscape in the field space and, for any reasonable power spectrum decreasing with $m$, the structures at smaller scales become more irrelevant and can indeed be truncated.}
The $A_{\boldsymbol{m}}$'s verify statistical properties similar to the ones of the $a_{\boldsymbol{\ppi}}$'s:
\begin{align}
    \braket{A_{\boldsymbol{m}}} &= 0 \,, \\
    \braket{A_{\boldsymbol{m}}A_{\boldsymbol{m}'}^*}&=(2\Lambda)^{N_{\mathrm{f}}}\delta_{\boldsymbol{m},\boldsymbol{m}'}P\left(m \frac{\pi}{\Lambda}\right)\,,
\end{align}
where $m = \sqrt{\boldsymbol{m}\cdot \boldsymbol{m}}$ is the modulus of $\boldsymbol{m}$ (not necessarily an integer) and $P$ is the same power spectrum as before. 
The variance of the landscape is then given by
\begin{equation}
\label{eq:corr-discrete}
    \braket{v(\boldsymbol{\phi})v(\boldsymbol{\phi}')}=\frac{1}{(2 \Lambda)^{N_\mathrm{f}}}  \sum_{\substack{\boldsymbol{m} \in \mathbb{Z}^{N_\mathrm{f}}\\ \forall  a,\, |m_a|\leqslant M}} 
    \exp\left[\ii\frac{\pi}{\Lambda}\boldsymbol{m}\cdot\left(\boldsymbol{\phi}-\boldsymbol{\phi}'\right)\right]
    P\left(m \frac{\pi}{\Lambda}\right)\,.
\end{equation}

\vspace{4 pt}
We can also take advantage of the reality of the landscape to rewrite this sum in terms of explicitly real terms only by grouping non-zero modes into pairs $\left(\boldsymbol{m},-\boldsymbol{m}\right)$ as:
\begin{equation}
    A_{\boldsymbol{m}}\exp\left(\ii\frac{\pi}{\Lambda}\boldsymbol{m}\cdot\boldsymbol{\phi}\right)+A_{\boldsymbol{-m}}\exp\left(-\ii\frac{\pi}{\Lambda}\boldsymbol{m}\cdot\boldsymbol{\phi}\right)=2\rho_{{\boldsymbol{m}}}\cos\left(\frac{\pi}{\Lambda}{\boldsymbol{m}}\cdot\boldsymbol{\phi}+\delta_{\boldsymbol{m}}\right),
\end{equation}
where we wrote the complex amplitude $A_{\boldsymbol{m}}=\rho_{\boldsymbol{m}} e^{i \delta_{\boldsymbol{m}}}$ with real amplitude $\rho_{\boldsymbol{m}}$ and real phase $\delta_{\boldsymbol{m}}$. Note that $A_0$ being already real, it can be written as $A_0=\rho_0$ only, i.e. we always have $\delta_0 = 0$.
Also, the $A_{\boldsymbol{m}}$'s being Gaussian random variables, $\rho_{\boldsymbol{m}}$ with $m>0$ must verify a Rayleigh distribution, while $\delta_{\boldsymbol{m}}$ must be drawn from a flat one:
\begin{align}
\label{eq:pdf-rho}
    \forall m >0\,, \quad p_\rho(\rho_{\boldsymbol{m}}) &= \frac{2 \rho_{\boldsymbol{m}}}{(2\Lambda)^{N_{\mathrm{f}}}P(m \pi/\Lambda)} \exp \left(- \frac{\rho_{\boldsymbol{m}}^2}{(2\Lambda)^{N_{\mathrm{f}}}P(m \pi/\Lambda)} \right)\,, \quad \text{with} \quad \rho_{\boldsymbol{m}} \in [0,\infty[\,, \\
\label{eq:pdf-delta}
    p_\delta(\delta_{\boldsymbol{m}}) &= \frac{1}{2\pi}\,, \quad \text{with} \quad \delta_{\boldsymbol{m}} \in [0,2\pi[ \,,
\end{align}
and $\rho_0$ is drawn from a zero-mean Gaussian with variance $(2\Lambda)^{N_{\mathrm{f}}}P(0)$.
The landscape is then given by:
\begin{equation}
\label{eq:landscape-discrete-box}
    v(\boldsymbol{\phi}) = \frac{1}{(2\Lambda)^{N_\mathrm{f}}} \left[\rho_0 + 2 \sum_{\substack{\boldsymbol{m} \in \mathcal{H}_{N_f}\\ \forall  a,\, |m_a|\leqslant M}} 
    \rho_{\boldsymbol{m}}\cos\left(\frac{\pi}{\Lambda}\boldsymbol{m}\cdot\boldsymbol{\phi}+\delta_{\boldsymbol{m}}\right)
    \right]\,,
\end{equation}
where, $\mathcal{H}_{N_f}$ denotes a half of the integer lattice, chosen such that for every nonzero pair $(\mathbf{m},-\mathbf{m})$, exactly one representative is included.
This is the final form that we use in practice in the following.

\subsection{Choice of power spectrum, fiducial parameters and vertical shift}
\label{subsec2.2}

The only quantity that determines the landscape properties is its power spectrum, $P(\ppi)$.
We choose
\begin{equation}
\label{eq:Gaussian-PS}
    P(\ppi;\xi) = P_0 \exp\left[-\frac{(\xi \ppi)^2}{2}\right] \,,
\end{equation}
which is parameterized by a single parameter $\xi$ and corresponds to a Gaussian with variance $1/\xi^2$.
This choice is interesting because it allows us to tune the typical curvature of the potential with the parameter $\xi$ which corresponds to the absolute correlation length of the landscape.
Moreover, it decays sufficiently fast with $\ppi$ that it quickly becomes irrelevant to add more modes, and therefore cutting the expansion at $|m_a|\leqslant M$ is already justified for not-so-large values of $M$.
Finally, the parameter $P_0$, together with $V_0$, sets the scale of inflation through the Friedman equation, as $H/\Mp \sim (V_0^2 P_0)^{1/4}$.
In App.~\ref{app-differentPS}, we investigate the robustness of our results to changes in the functional form of $P(\ppi)$ that preserve these properties.

For example, our fiducial set of parameters for the two-field case $N_{\rm f}=2$ is as follows:\footnote{The value of $P_0$ does not have an absolute meaning since it degenerates with $V_0$ and can be absorbed by a rescaling of the Hubble parameter $H$ in the equations of motion.}
\begin{equation}
\label{eq:fiducial-params}
    \Lambda = 10\xi\,, \quad M=12 \,, \quad  \xi=\Mp\,, \quad   P_0 = 100 \,.
\end{equation}
In practice, we use dimensionless fields $\tilde{\bm \phi} = {\bm \phi}/\Mp$ with dimensionless landscapes $\tilde{v}(\tilde{\bm \phi})=v(\tilde{\bm {\phi}}\Mp)$ and we need not choose explicitly $V_0$.

Finally, we will be interested in biasing any landscape so that its global minimum is exactly zero. 
By doing so, we ensure that inflation can always proceed around the shifted global minimum and smoothly end when approaching it.
In practice, we perform a vertical shift of the landscape:
\begin{equation}
\label{eq:landscape-rescaling}
    v({\bm \phi}) \longrightarrow u({\bm \phi})=v({\bm \phi})-\underbrace{ \underset{{\bm \phi}}{\rm min} \, v({\bm \phi})}_{v_{\rm min}} \geqslant 0 \,.
\end{equation}
Note that this shift is landscape-dependent and that, strictly speaking, the resulting shifted landscape $u({\bm \phi})$ is not a Gaussian random field any more, since $v_{\rm min}$ is itself a nonlinear and non-local functional of $v({\bm \phi})$.

Realistically, inflation should occur within a local patch of the landscape whose minimum potential energy is zero. This procedure may also be viewed as qualitatively modeling the effect of a long-wavelength mode in the potential landscape that has uplifted such a local patch.

\subsection{Characterizing the typical potential curvature}
\label{subsec2.3}

Now, we define and compute quantities that characterize the shapes of potentials in landscapes.
As usual in the study of multifield inflation, we define the second potential slow-roll parameter as the (relative, dimensionless) Hessian of the potential:
\begin{equation}
    \eta_{IJ} = \Mp^2 \frac{V_{;IJ}}{V} \,,
\end{equation}
where a semicolon represents covariant field-space derivatives.
Note that since we consider a flat field space, those coincide with regular derivatives, and therefore $\eta_{IJ}=\Mp^2 \left(\partial_{I}\partial_{J}v\right)/v$.
But, $\eta_{IJ}$ is a coordinate-dependent $N_{\rm f} \times N_{\rm f}$ matrix, which obscures its physical interpretation. 
We therefore define ``the'' landscape curvature at a given field-space position as:
\begin{equation}
    \eta_V \equiv  \frac{1}{N_{\rm f}} {\rm Tr}\left(\eta_{IJ}\right) = \frac{\Mp^2}{N_{\rm f}} \frac{\nabla^2v  }{v}\,, 
\end{equation}
where we used $\nabla^2 = \delta^{IJ}\partial_{I}\partial_{J} $.
The advantage of using $\eta_V$ is that it is independent of the choice of coordinates; moreover, basing ourselves on statistical isotropy, we can expect its statistics to resemble those of the curvature ``in any direction'' for any field-space dimension $N_{\rm f}$.
Having a correlation length of order Planck mass corresponds to $|\eta_V| \sim 1$ on the landscape.
Let us be more precise.

\paragraph{Mean of $\eta_V$.}
The first obvious quantity to look at is the average of $\eta_V$ over many landscape realisations, at a fixed field-space position.
First, we find that $(v({\bm \phi}),\nabla^2 v({\bm \phi}))$ is a bivariate Gaussian random variable with 
\begin{align}
    \Braket{v({\bm \phi})} &=0 \,, \quad  \quad \quad  \Braket{v({\bm \phi})^2} =  \int \frac{\dd^{N_{\rm f} } {\bm \pi}}{(2 \pi)^{N_{\rm f} }}P(\ppi) \,, \\
    \Braket{\nabla^2 v({\bm \phi})} &= 0 \,, \quad \quad \,\,\, \Braket{\left(\nabla^2 v({\bm \phi})\right)^2} = \int \frac{\dd^{N_{\rm f} } {\bm \pi}}{(2 \pi)^{N_{\rm f} }} \ppi^4 P(\ppi) \,, \nonumber \\
    \Braket{v({\bm \phi}) \times \nabla^2 v({\bm \phi})} &=  - \int \frac{\dd^{N_{\rm f} } {\bm \pi}}{(2 \pi)^{N_{\rm f} }}  \ppi^2 P(\ppi)  \,, \nonumber
\end{align}
which end up all being ${\bm \phi}$-independent (we therefore omit writing them as functions of ${\bm \phi}$ in the following).
These correlations can be calculated explicitly with our choice of landscape power spectrum~\eqref{eq:Gaussian-PS}, giving 
\begin{align}
    \sigma_v^2 \equiv \Braket{v^2}  = P_0(2\pi \xi^2)^{-N_{\rm f}/2} \,, \quad \Braket{\left(\nabla^2 v\right)^2} = \frac{N_{\rm f}(N_{\rm f}+2)}{\xi^4}   \sigma_v^2  \,, \quad \Braket{v \times \nabla^2 v}  = -\frac{N_{\rm f}}{\xi^2} \sigma_v^2 \,.
\end{align}
Equipped with these relations, we can write the joint probability density function for the bivariate variable $\bm{X}=\left(v,\nabla^2 v\right)$ as
\begin{align}
\label{eq:pxX}
    p_{\bm{X}}(\bm{X}) &= \frac{1}{2\pi \sqrt{{\rm det}\Sigma}}\exp\left[-\frac{1}{2}\bm{X} \cdot \Sigma^{-1} \cdot \bm{X}^{\rm T}\right] \,, \,\, \text{with} \\\nonumber \\ 
    \Sigma &= \sigma_v^2 \begin{pmatrix}1 & -{N_{\mathrm{f}}}/\xi^2\\ \\-{N_{\mathrm{f}}}/\xi^2 & \,\, N_{\mathrm{f}}({N_{\mathrm{f}}}+2)/\xi^4\end{pmatrix} \nonumber \,.
\end{align}
Note that ${\rm det} \, \Sigma = 2 \sigma_v^4 N_{\rm f}/\xi^4$ is positive indeed.

We are now ready to compute the average value $\braket{\eta_V}$ across the landscapes.
Using conditional probabilities and $p_v(v)=(2 \pi \sigma_v^2)^{-1/2}\exp\left[-v^2/(2 \sigma_v^2)\right]$, we can write
\begin{align}
    \braket{\eta_V} & \equiv \int \dd {\bm X} \, p_{\bm{X}}(\bm{X}) \, \eta_V(\bm{X}) \\ &=\frac{\Mp^2}{N_{\mathrm{f}}} \int \dd v \, \frac{p_v(v)}{v} \underbrace{\int \dd (\nabla^2 v ) \sqrt{\frac{1}{2\pi \, {\rm det} \Sigma/\sigma_v^2}} \exp\left[-\frac{\left(\nabla^2 v + N_{\rm f} \, v/\xi^2  \right)^2}{2 \, {\rm det} \Sigma /\sigma_v^2}\right] \, \nabla^2 v}_{-N_{\rm f} v/\xi^2}\nonumber \\
    & =-\frac{\Mp^2}{\xi^2} \,. \nonumber
\end{align}
Therefore, in order to have $|\braket{\eta_V}|\sim1$, the field-space correlation length of this Gaussian random landscape needs to be $\xi \sim \Mp$.\footnote{\label{footnote:eta_integral} Note that, strictly speaking, $\eta_V$ seen as a function of $(v,\nabla^2 v)$ is not $L^1$-integrable because of the $1/v$ factor, so $\braket{\eta_V}$ is not absolutely convergent. 
In practice, we were able to perform the integrals by arbitrarily declaring a preferred order of integration: first over $\nabla^2 v$, then over $v$.
Another possibility would be to define $\braket{\eta_V}$ as the principal value of $\int \dd {\bm X} \, p_{\bm{X}}(\bm{X}) \, \eta_V(\bm{X})$, which indeed uniquely gives $- \Mp^2/\xi^2$ again.
Finally, yet another option is to regularise the integral by putting the field space on a box's grid, as we will do in practice for our numerical simulations anyway, in which case the $v=0$ point is never reached and the sum is well defined, yielding $- \Mp^2/\xi^2$ once more.
}

This neat result is however only valid for the Gaussian random landscapes $v({\bm \phi})$ before the vertical lift $v({\bm \phi}) \rightarrow u({\bm \phi})$ that we have described in~\ref{subsec2.2}.
As already discussed, $u({\bm \phi})$ is not a Gaussian random field and discussing its statistical properties is not an easy task.
What we can do, however, is characterizing a proxy for it, defined as 
\begin{equation}
    U({\bm \phi}) = v({\bm \phi}) - \braket{v_{\rm min}} \,,
\end{equation}
where $ \braket{v_{\rm min}}$ should be understood as the average value of the minima across many landscapes.
Then, $U({\bm \phi})$ is a Gaussian random field indeed, as it only differs from $v({\bm \phi})$ by a constant.
For landscape realisations in the discrete box, we find that
\begin{equation}
\label{eq:vmin}
    \braket{v_{\rm min}} \rightarrow -\sigma_v \sqrt{2\log N_{\rm eff}} \left[ 1 + \mathcal{O}\left(\log N_{\rm eff} \right)^{-1} \right] \,, \quad \text{with} \quad N_{\rm eff} = \left(\frac{\Lambda}{\xi}\right)^{N_{\rm f}} \,,
\end{equation}
$N_{{\rm eff}}$ being the effective number of independent variables.
In the infinite box limit, $\Lambda\rightarrow\infty$, we recover $\braket{v_{\rm min} }\rightarrow - \infty$ as expected, but for a finite box it always remains well-defined, although convergence to the asymptotic value quoted above is slow as $\mathcal{O}\left(\log N_{\rm eff} \right)^{-1} \simeq 10\%$ for our fiducial parameter set.
The derivation of this expression is shown in App.~\ref{appB}, where we also compare it to the values found in our numerical simulations.
There, we also prove the following remarkably simple expression for the mean of $\eta_V$ under the $U$ proxy, i.e.
\begin{equation}
\label{eq:mean-etaV-final}
    \braket{\eta_V}_U \rightarrow - \frac{\Mp^2}{\xi^2}\left[ 1 - 2 \sqrt{\log N_{\rm eff}} \, F(\sqrt{\log N_{\rm eff}}) \right]\,,
\end{equation}
where $F(x)=e^{-x^2} \int_0^x \dd t \, e^{t^2} $ is Dawson's $F$ integral and, again, compare it to the numerical values in our simulations.
For our fiducial set of parameters, this corresponds to $ \braket{\eta_V}_U \in\{0.28,\, 0.17,\, 0.10,\, \ldots\} \Mp^2/\xi^2$ for $N_{\rm f} \in \{1,\,2,\,3,\, \ldots \}$, where we remind that we expect a relative error of roughly 10\%.
We conclude that the vertical lift of the landscape effectively decreases its curvature, as measured by $\braket{\eta_V}$, by roughly one order of magnitude on average (and flips its sign).

\paragraph{Median of $|\eta_V|$.}

It is also useful to statistically characterise the landscape's curvature differently, namely with the median of the absolute value of the curvature, that we denote as ${\rm med}\,|\eta_V|$.
The interest is three-fold. 
First, this quantity is by definition strictly positive and does not suffer from cancellations between concave and convex regions of the potential.
Moreover, taking the median is less dependent on rare but extreme fluctuations than taking the mean and, therefore, one can expect the prediction to be more robust.
Finally, it simply gives a different characterization of the landscape's curvature which provides one with an interesting complementary check.
Obtaining an exact analytical formula for the shifted landscape—even for the simpler proxy $U({\bm \phi})$ and the corresponding ${\rm med}\,|\eta_V|_U$—is challenging; therefore, we use the following approximate scheme.
We have seen that $-\braket{v_{\rm min}}/\sigma_v$ equals a few, which means that the spread of $U$ is on average small in units of its centre.
We therefore approximate 
\begin{equation}
    {\rm med}\,|\eta_V|_U \simeq \frac{\Mp^2}{N_{\rm f}} \, \frac{{\rm med} \, \left|\nabla^2 U \right|}{\braket{U}}   \,,
\end{equation}
i.e. we neglect the fluctuations of $U$ compared to the ones of $\nabla^2 U$ and simply evaluate it at its mean $\braket{U}$. Since $\nabla^2 U$ is a Gaussian random variable with same statistics as $\nabla^2 v$, we can simply write the cumulative distribution function of  $\, \left|\eta_V\right|_U$ as
\begin{align}
    F_{\left|\eta_V\right|_U}(z) = \mathbb{P}\left(-z < (\eta_V)_U < z \right)
    \simeq \Phi\left(\frac{\braket{U} N_{\mathrm{f}}}{\Mp^2} \frac{z}{ \sigma_{\nabla^2 U} }\right)  - \Phi\left(-\frac{\braket{U} N_{\mathrm{f}}}{\Mp^2}  \frac{z}{ \sigma_{\nabla^2 U} }\right) 
   \,,
\end{align}
with $\Phi$ the standard cumulative distribution function of a Gaussian (here, centred and with spread $ \sigma_{\nabla^2 U}$).
Using $ \sigma_{\nabla^2 U} =  \sigma_{\nabla^2 v}= (N_{\rm f} \sigma_v / \xi^2) \sqrt{1+2/N_{\rm f}} $ and $\braket{U}=-\braket{v_{\rm min}}\geqslant 0$, turning to the error function, we have
\begin{equation}
     F_{\left|\eta_V\right|_U}(z) \simeq {\rm erf} \left(\frac{\xi^2}{\Mp^2}\frac{-\braket{v_{\rm min}}}{\sqrt{2(1+2/N_{\rm f})} }z \right) \,. 
\end{equation}
It is now immediate to read the median by solving $F_{\left|\eta_V\right|_U}({\rm med} \, \left|\eta_V\right|_U ) = 1/2 $. 
In the discrete box, using Eq.~\eqref{eq:vmin}, we find:
\begin{equation}
     {\rm med}\,|\eta_V|_U \rightarrow 0.67 \sqrt{\frac{N_{\rm f}+2}{2 N_{\rm f} \log N_{\rm eff}}} \frac{\Mp^2}{\xi^2} \,. 
\end{equation}
With our fiducial set of parameters, we predict $ {\rm med}\,|\eta_V|_U  \in \{0.54,\,  0.31 ,\, 0.23,\, \ldots \} \Mp^2/\xi^2$ for $N_{\rm f} \in \{1,\, 2,\, 3,\, \ldots\}$ with a $10\%$ relative precision, roughly twice of $\braket{\eta_V}_U$.
As a comparison, the values that we find in our numerical simulations is $ {\rm med}\,|\eta_V|_U  \in \{0.67,\,  0.38 ,\, 0.28,\, \ldots \} \Mp^2/\xi^2$, confirming the reliability of the above arguments.\footnote{In the rest of the paper, we will use the values of ${\rm med}\,|\eta_V|_U$ from numerical simulations. To fix the value of ${\rm med}\,|\eta_V|_U$ numerically, we generate a sample of realizations with $\xi/\Mp=1$ and numerically obtain ${\rm med}\,|\eta_V|_U$ of this sample, then we use a scaling property of the landscape to rescale the value of $\xi$ to obtain the intended value of ${\rm med}\,|\eta_V|_U$ (see Sec.~\ref{subsec3.1}).}
We again conclude that $\xi \sim \Mp$ is the correct order of magnitude to obtain random landscapes with $\eta\sim 1$ statistically. To simplify the notation, from now on we will use the notation 
\begin{equation}
    \etamed \equiv {\rm med}\,|\eta_V|_U ~.
\end{equation}

\section{Numerical methods}
\label{sec-methods}

In this section, we describe the methods that allow us to draw conclusions about inflation on highly curved landscapes. 
We describe the generation of landscapes and their vertical shifting, how we sample initial conditions in phase space and evolve the system dynamically, finally how we collect statistical data.

\subsection{Generating landscapes}
\label{subsec3.1}

A first landscape is generated by taking our fiducial set of parameters~\eqref{eq:fiducial-params}, drawing the random real numbers corresponding to the amplitudes $\rho_{\bf m}$ and phases $\delta_{\bf m}$ from their distribution functions~\eqref{eq:pdf-rho}--\eqref{eq:pdf-delta} and the power spectrum~\eqref{eq:Gaussian-PS}, and building $v(\bm \phi)$ with~\eqref{eq:landscape-discrete-box}.
One can then repeat this operation many times to generate a large sample of landscapes.

To generate a new sample of landscapes with a different value of $\etamed$, the most straightforward way is to change the value of $\xi$ and generate many new landscapes.
Another possibility consists in rescaling an already existing old landscape, $\phi\to c\phi$ and $\Lambda \to c\Lambda$, which gives a new landscape with an effective correlation length $\xi^\prime = c\xi$\footnote{Note, however, that this new landscape will look just like the old one in rescaled coordinates $\phi/\Lambda$, in which the dimensionless correlation length $\xi/\Lambda$ is also invariant.
This enables us to study landscapes with different physical correlation lengths but otherwise similar features.
} and repeat for many landscapes, a procedure that we expect will give a new $\etamed' \simeq \etamed/c^2$ from the previous section.

\subsection{Finding the true vacuum and setting initial conditions}
\label{subsec3.2}

Let us focus on one landscape realization.
As already mentioned, we want the potential to be everywhere non-negative, so we define a vertically shifted landscape $u({\bm \phi})$ as~\eqref{eq:landscape-rescaling}.
Finding the global minimum in multiple dimensions can be numerically demanding, but since this procedure only needs to be done once per landscape, we prefer robustness over efficiency, and we always perform a thorough search with fine gridding.
After the vertical shifting, the minimum value of $u$ is virtually zero and, therefore, it is expected that inflation dynamically ends as a trajectory approaches it.
We call this global minimum the ``true vacuum'' while we call other local minima with strictly positive potential energy ``false vacua''.
Initial conditions for inflation are chosen as follows.

First, we randomly sample $n$ initial positions on the landscape in a $N_{\rm f}$-ball of radius $\lambda$ and centred on the position of the global minimum.
We should choose $\lambda$ to be a substantial fraction of $\Lambda$ in order not to miss any inflationary trajectory that would end in the true vacuum.
However, for numerical efficiency, we should also choose $\lambda$ to be not too large in order to avoid describing many trajectories falling into false vacua.
Indeed, we do not describe the possible decay of false vacua into the true one via tunnelling and simply dismiss these trajectories as inadmissible.
We leave to future work to include those interesting non-perturbative aspects.
The optimal choices for $n$ and $\lambda$ are found empirically; e.g., for $N_{\mathrm{f}} =2$, we find that $n=1000$ and $\lambda = 3 \Lambda /4$ provide us with a good balance between being conservative and efficient, allowing us to probe both the microstructures and the large-scale correlations of the landscape.

Second, we set the initial velocity vector to be exactly vanishing. 
Indeed, any sizeable initial condition should be quickly washed out by the Hubble friction.\footnote{
Strictly speaking, if the initial velocity is very large, it could very well impede inflation.
Indeed, if the initial equation of state is kinetic-dominated, the inflaton trajectory could ``roll over'' the landscape without seeing its features at all.
To be more precise, we therefore restrict ourselves to the class of initial conditions that result in a potential-dominated equation of state $w < -1/3$ so that inflation always washes out any initial velocity.
It would be interesting to extend our study to randomly chosen initial velocities and characterize the probability for inflation to still happen. We leave this investigation for future work.
}
A small velocity naturally settles after a short period of transition when an attractor trajectory is reached—which we find always happens for admissible trajectories—so that the precise choice of the initial velocity is irrelevant, and zero becomes the most economical one.

In the absence of a universal consensus on the appropriate statistical measure, and since the internal field space studied in this paper is flat, we adopt the flat field-space measure as the starting point for our statistics. E.g.~in the two-field case, this corresponds to $\dd\phi_1\dd\phi_2$, which can be straightforwardly generalized to multi-field cases and with non-canonical field spaces.

\subsection{Equations of motion and time evolution}
\label{subsec3.3}

\paragraph{Background evolution.}

Using the elapsed $e$-folding number $N$ as the time variable, the equations of motion for the fields ${\bm \phi}$ on the landscape $u({\bm \phi})$, with canonical kinetic terms and minimally coupled to gravity are
\begin{equation}
\label{eq:background}
H^2\boldsymbol{\phi}''+(3-\epsilon)H^2\boldsymbol{\phi}'+  V_0\nabla_{\boldsymbol{\phi}} u(\boldsymbol{\phi})=0\,,
\end{equation}
where ${}^\prime$ denotes the derivative with respect to $N$, $\nabla_{\boldsymbol{\phi}}$ is the gradient in the field space, and $\epsilon \equiv -H'/H$.
Moreover, the Hubble scale evolves as $H^\prime = -H\boldsymbol{\phi}'^2 /( 2 \Mp^2)$, but in practice we will simply use the energy constraint
\begin{equation}
    3H^2 \Mp^2=\frac{1}{2}H^2\boldsymbol{\phi}'^2+V_0 u(\boldsymbol{\phi}) \,.
\end{equation}
Note how these background equations of motion are invariant under co-rescaling of $H^2$ and $V_0$.
This explains why the choice of $V_0$ is irrelevant to the dynamics.
Numerically, we use dimensionless variables $\tilde{\bm \phi} = {\bm \phi}/\Mp$, as well as $\tilde{H}=H \Mp/\sqrt{V_0}$, such that the equations read
\begin{equation}\label{eq:eom}
    \left\{\begin{aligned}
        &\tilde{H}^2=\frac{\tilde{u}}{3-\tilde{\boldsymbol{\phi}}'^2/2}\,,\\
        &\tilde{\boldsymbol{\phi}}''+ (3-\epsilon) \tilde{\boldsymbol{\phi}}'+ \frac{\nabla_{\boldsymbol{\tilde{\bm \phi}}} \tilde{u}}{\tilde{H}^2}=0\,,
    \end{aligned}\right.
\end{equation}
with $\tilde{u}=u(\tilde{\boldsymbol{\phi}}\Mp)$.
We define initial conditions at $N=0$ as described in the previous section, and we evolve the system according to the above equations.
There are two possible outcomes for a given trajectory.
Either it will end up in the true vacuum, and the system will start oscillating with increasing frequency and low damping rate.
Or it will end up in a false vacuum, and the system will rapidly damp kinetic energy leading to eternal inflation.
We track the presence of the type of fast oscillations indicating inflation has ended, and we terminate the numerical evaluation of a given trajectory when they appear.
If those increasingly fast oscillations have not appeared yet after a fixed maximal number of elapsed $e$-folds $N_\mathrm{max}$, we terminate the numerical evolution and declare that the trajectory is stuck in a false vacuum.\footnote{This criterion has the limitation that it may be contaminated by trajectories with strong primordial features, which may be mistakenly identified as the end of inflation. However, it can only happen when the oscillatory features are sufficiently strong, which is a rare case in the current setup and do not affect the statistical properties of interest in this work.}
The code of generating landscapes and solving the background dynamics is constructed with \texttt{Mathematica}, and the computation is done on the FASRC cluster with 100 cores.

\paragraph{Dynamics of linear fluctuations.}
Although the vast majority of our results in this paper will concern the background dynamics, we will also be interested in checking the predicted spectra for cosmological fluctuations in a few selected cases.
To do so, we will make use of an independent numerical tool that is freely available online, namely the \texttt{PyTransport} package~\cite{Mulryne:2016mzv}.
This code is based on the transport method for primordial correlation functions that was developed in a series of works~\cite{Seery:2012vj,Mulryne:2013uka, Dias:2016rjq, Butchers:2018hds}.
The main idea is to numerically evolve---in addition to the homogeneous background equations of motion---the set of coupled first-order linear equations for the power spectra of $N_{\rm f}$ fields' fluctuations $Q^a$ in the flat gauge, the $\braket{Q^a Q^b}$'s, and then perform a gauge transformation on super-Hubble scales to infer the power spectrum of the adiabatic curvature perturbation $\zeta$.
Interestingly, \texttt{PyTransport} also allows to compute the tensor power spectrum as well as the fields' fluctuations bispectra in the flat gauge and therefore the primordial bispectrum $\braket{\zeta^3}$. 
We will indeed evolve linear gravitational waves but we will not make use of the bispectrum option in this work.
The transport approach has been built for, and \texttt{PyTransport} has been coded for, the general class of non-linear sigma models of inflation which consist in any number $N_{\rm f}$ of scalar fields with potential and kinetic interactions.
To use it, we perform the following steps:
\begin{itemize}
    \item once a landscape with interesting properties has been identified, we export the corresponding potential into \texttt{PyTransport} and fix the field space to be trivial (i.e., in this work, we do not consider kinetic interactions);
    \item once some trajectories with interesting properties have been identified on this landscape, we export the corresponding initial conditions into \texttt{PyTransport};
    \item we check that the background evolution with \texttt{PyTransport} is consistent with the one we have independently solved (final endpoint of the trajectory, duration of inflation, etc.);
    \item by using the correspondence between wavenumbers and the elapsed time between Hubble crossing and the end of inflation, we select $k$ modes of interest (e.g. CMB scales);
    \item for each of these modes $k$, we evolve the fields' fluctuations and tensor modes power spectra with \texttt{PyTransport} and extract the corresponding values for the tensor and scalar power spectra $\mathcal{P}_\zeta(k)$ and $\mathcal{P}_\gamma(k)$;
\end{itemize}
This task being numerically demanding we will only do so for a small subset of the admissible trajectories, namely the successful ones, see Sec.~\ref{subsec3.4} below.

\subsection{A working example for $N_{\rm f}=2$}\label{working_example}
To gain some insight in our methodology, we present here a working example of a landscape realization for $N_{\rm f}=2$.
For the purpose of illustration, the example is chosen to be a special inflationary trajectory featuring a duration and statistics of linear fluctuations compatible with CMB constraints, taken from the vast samples we have investigated.

\begin{figure}[ht!]
    \centering 
    \raisebox{10pt}{\includegraphics[width=0.45\textwidth]{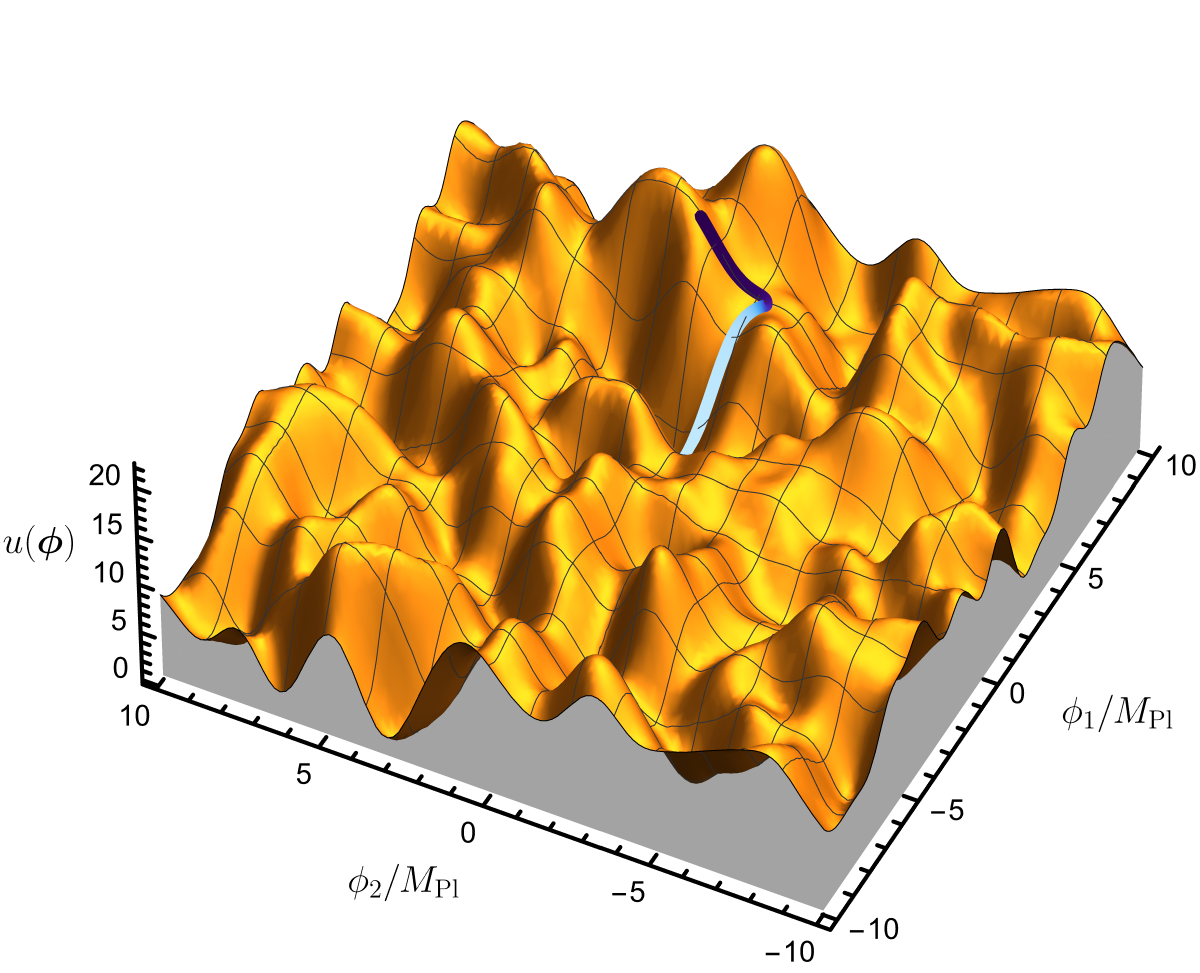}}
    \hspace{23pt}
    \includegraphics[width=0.45\textwidth]{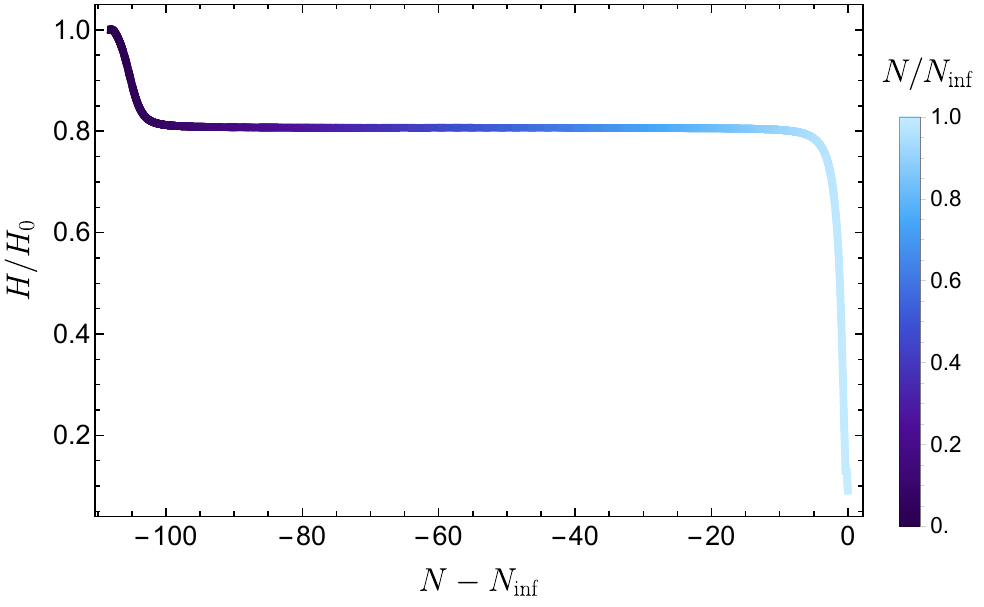} 

    \vspace{10pt}
    
    \includegraphics[width=0.45\textwidth]{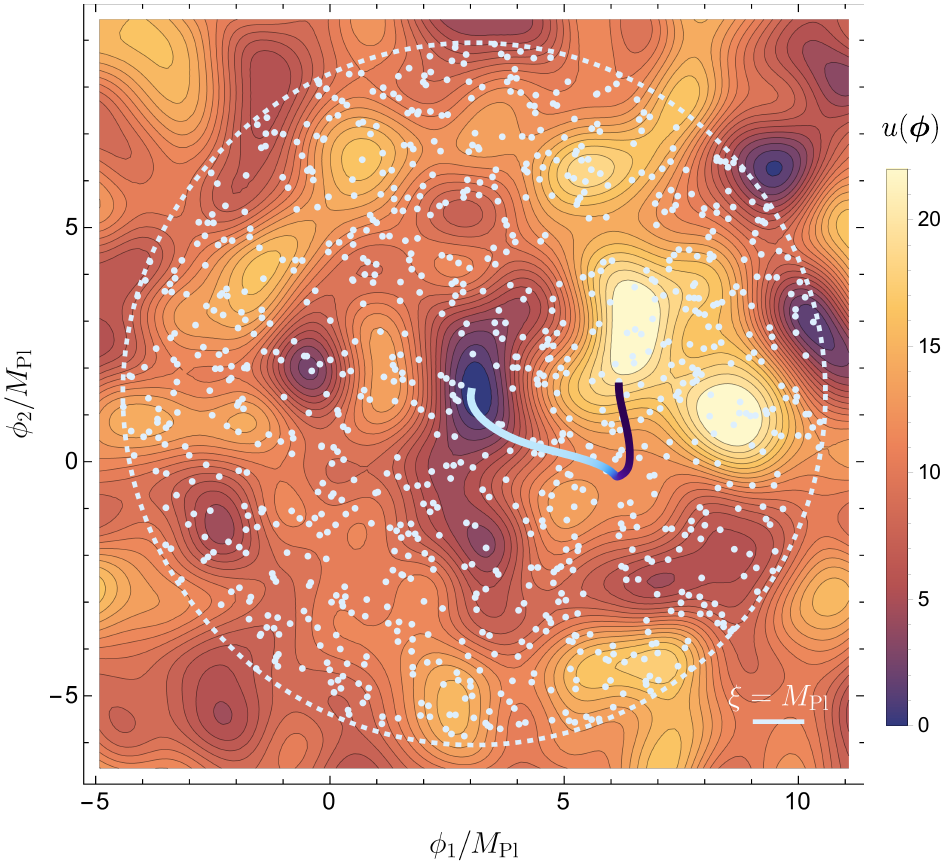} 
    \hspace{6pt}
    \raisebox{15pt}{\includegraphics[width=0.45\textwidth]{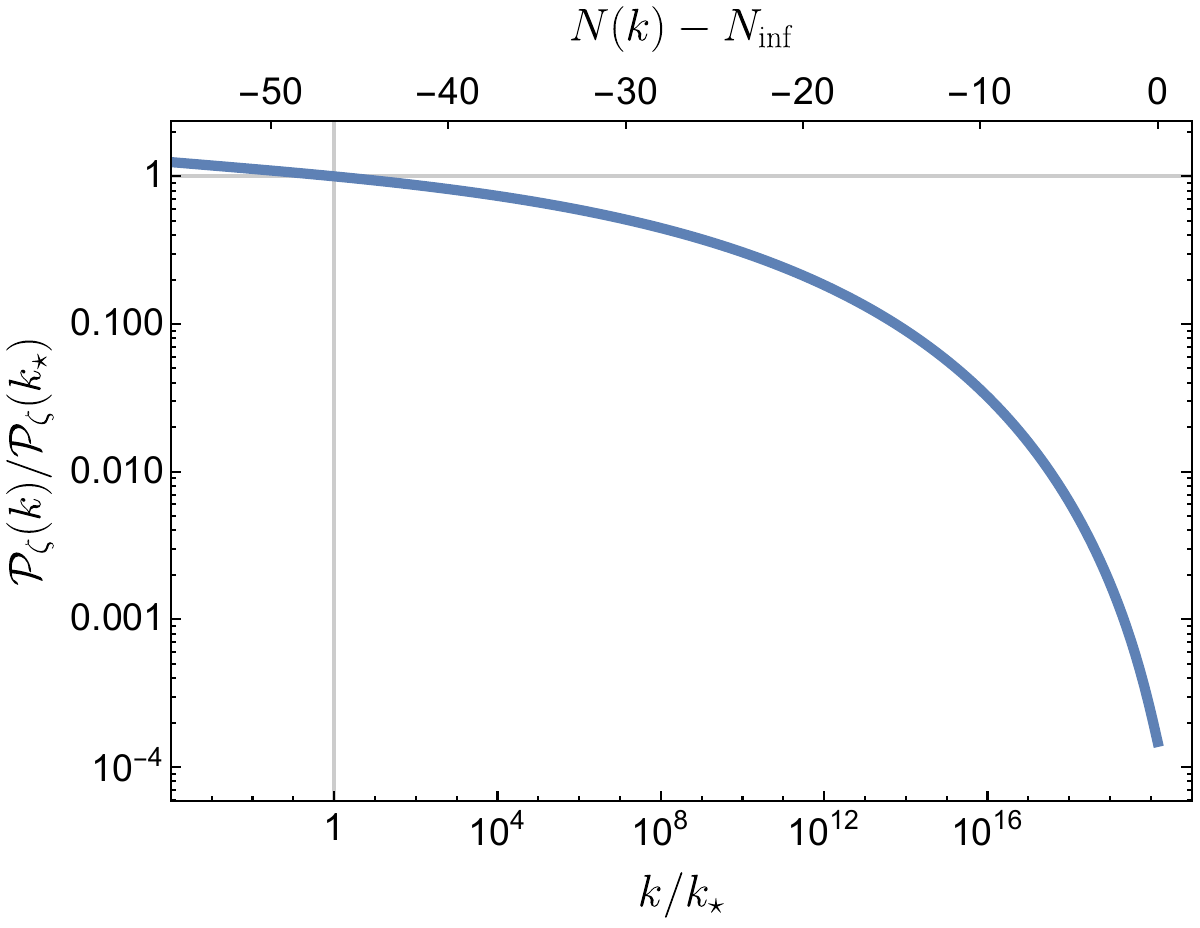}}
    \caption{
    Upper left panel: global view of the shifted landscape $u(\boldsymbol{\phi})$ in which the global minimum is $u_{\rm min}=0$ and of the trajectory on it. 
    Lower left panel: set of 1000 initial conditions in the disk of radius $\lambda=7.5\Mp$ centred on the global minimum, plotted on a 2d contour plot of the landscape, together with the trajectory on it.
    Upper right panel: evolution of the Hubble parameter $H$ as a function of $e$-folding number to the end of inflation, $N-N_{\mathrm{inf}}$, in units of $H_0$ its initial value. In all three previously described panels, the colour on the trajectory denotes the elapsed $e$-folding number normalized by the duration of inflation, $N/N_{\rm inf}$. 
    Lower right panel: power spectrum normalized to its pivot scale value, $\mathcal{P}_\zeta(k)/\mathcal{P}_\zeta(k_\star)$, calculated using \texttt{PyTransport}, where the pivot scale $k_\star$ is chosen so that the power spectrum is CMB-compatible for three decades of $k$ centred on it. The upper horizontal axis represents the time in $e$-folds $N(k)-N_{\mathrm{inf}}$ at which a given comoving wavenumber $k$ exited the Hubble scale with reference to the end of inflation; in this particular example the pivot scale exited the Hubble scale approximately $45$ $e$-folds before the end of inflation.}
  \label{fig:trajectory_exp}
\end{figure}

\paragraph{Landscape construction and initial conditions.}
Using the fiducial parameters (\ref{eq:fiducial-params}), it is straightforward to generate a large number of landscape realizations. Here we only present a particular landscape that contains a trajectory satisfying the \textit{Planck} constraints on inflation. We plot a global three-dimensional view of the shifted potential $u(\boldsymbol{\phi})$ (as defined in (\ref{eq:landscape-rescaling})) in the upper left panel of Figure \ref{fig:trajectory_exp}. Since we have imposed periodic boundary conditions, there are actually no boundaries in the landscape. Next, we sample initial conditions in a 2-ball (disk) of radius $\lambda=3\Lambda/4=7.5\Mp$ centred on the global minimum; the $n=1000$ initial points are shown as white dots on this patch of the landscape (represented as contours) in the lower left panel of Figure \ref{fig:trajectory_exp}.

\paragraph{The inflation trajectory and the power spectrum.}
It is straightforward to solve the background equations of motion (\ref{eq:eom}) numerically for all the initial points above. Amongst all trajectories we obtain, most of them fail to generate realistic inflation, either stuck in false vacua or having too small values of $N_{\rm inf}$. However, in this particular landscape realization, there exists a small set of successful trajectories.
These trajectories happen to pass through a near-flat region on the landscape, which serves as a slow-roll attractor. 
We illustrate several properties of one such inflationary scenario in Figure \ref{fig:trajectory_exp}, including its trajectory through the landscape (left panels), the time evolution of the Hubble parameter $H$ (upper right panel) and the resulting power spectrum (lower right panel).

This trajectory has a total $e$-folding number $N_{\rm inf}= 108$, making it capable of solving the horizon problem. From the appearance of the trajectory on the landscape, we find that most of the $e$-folds are elapsed in the near-flat region halfway up the ``mountain", where slow roll occurs.\footnote{In all panels (except the bottom right one) of Figure \ref{fig:trajectory_exp}, the colour on the trajectory denotes the time variable $N/N_{\rm inf}$. Most of inflation happens at the segment of the trajectory where the colour gradient is large.
This is where the inflaton stays for a long time, on the slow-roll attractor. 
On the other hand, the trajectory has large field excursions within very short times away from the attractor.} 
It turns out that this inflationary scenario is of small-field type, in the sense that the majority of the $e$-folding numbers have elapsed with a field excursion $\Delta\phi\ll\Mp$, despite the inflaton taking a long way to the true vacuum at the end of inflation.
Finally, we can calculate the power spectrum of the trajectory with \texttt{PyTransport} introduced above, and the result is shown in the bottom right panel of Figure \ref{fig:trajectory_exp}. 
We found a pivot scale $k_\star$ verifying the properties enunciated below in Eq.~\eqref{CMB_comp_condition} to declare the trajectory compatible with Planck constraints.

\subsection{Collecting statistical data}\label{subsec3.4}

The techniques introduced above can be applied to generate a large sample of landscape realizations, with a large number of trajectories on each realization, and this practice can be conducted with different values of $N_\mathrm{f}$.
It can be expected that among the vast ensemble of trajectories, only a small portion of them satisfy the requirements of realistic inflation, which are constrained by current observations of CMB and other cosmological probes. 

We define \textit{admissible} trajectories as the ones that naturally end inflation, by which we mean that they terminate in the true vacuum (global minimum of the landscape). 
The trajectories stuck in false vacua are dismissed from the rest of the analysis.
Of course, not all admissible trajectories are realistic inflationary scenarios.

The first criterion that we will discuss is that the duration of inflation in units of $e$-folding number is larger than the minimum amount required to solve the horizon (and related) problem(s).
We call \textit{successful} trajectories the ones that verify this.
The subset of successful trajectories still exhibits some statistical variability in which phenomenological interest may reside. 
In particular, a trajectory can possess multiple inflation stages, which leave characteristic features in the primordial power spectrum.
Additionally, a trajectory on a multi-dimensional field space can have multi-field nature, either due to a sharp turn at the transition between two inflation stages, or being a curved slow-roll trajectory in the multi-field inflation or quasi-single-field inflation framework~\cite{Chen:2009we,Chen:2009zp}. 

Finally, we will be interested in knowing whether trajectories that successfully realise an inflationary background can also fit the detailed CMB constraints on the $(n_s,r)$ plane describing linear fluctuations.
Those are dubbed \textit{CMB-compatible} trajectories.
A statistical study is crucial to determine whether phenomenologically interesting trajectories are frequent in the ensemble of successful ones. 
In order to extract the statistical information from the numerical samples, we now introduce several important quantities for our study.

\subsubsection{Ending in the true vacuum}

As explained in Sec.~\ref{subsec3.3}, we track the appearance of increasingly fast oscillations in the background evolution.
Indeed, those signal that the trajectory has reached the true vacuum and that inflation has already ended.
We then stop the numerical evolution and declare this trajectory as admissible:
\begin{eBox}
    \begin{equation}
    \text{admissible trajectory} \iff \, \text{reaches the true vacuum before $N_{\rm max}$}.
\end{equation}
\end{eBox}
Instead, if the trajectory has not reached the true vacuum after $N_{\rm max}$ $e$-folds of expansion, we consider it is stuck in a false vacuum and we dismiss it for the remainder of the analysis.

\subsubsection{Duration of inflation}

The first and most straightforward quantity characterising an admissible trajectory is the total number of $e$-folds of expansion $N_{\rm inf}$.
We track back from the end of the simulation (when increasingly fast oscillations appear) the $e$-folding time at which $\epsilon = -H^\prime/H = 1$, defining it to be the total duration of inflation, $N_{\rm inf}$.
Note that the choice of $N_{\rm max}$ to declare a trajectory admissible or not must be such that trajectories with $N_{\rm inf}\geqslant N_{\rm max}$ are too rare to affect the statistical properties of the ensemble of trajectories.
Then, by construction, we can only find $N_{\rm inf}< N_{\rm max}$.
In practice, we find for our fiducial set of parameters that $N_{\rm max} = 1000$ is a good conservative choice to reject trajectories classically stuck in false vacua without missing any long-lasting ones that eventually fall into the true vacuum.

For a trajectory to successfully solve the observed horizon problem, it is necessary to have $N_{\rm inf}$ larger than a critical value $N_{\rm horizon}$. 
In Appendix~\ref{app_efolds}, we remind that $N_{\rm horizon}$ critically depends on the scale of reheating, which is not well constrained at all, leaving the large range of possible values $30 \leqslant N_{\mathrm{horizon}} \leqslant 61$.
In this work, we remain conservative by allowing inflation to finish even at a very low scale corresponding to $N_{\mathrm{horizon}}=30$, and therefore we declare:
\begin{eBox}
\begin{equation}
    \text{successful trajectory} \iff N_{\mathrm{inf}} \geqslant 30 \,.
\end{equation}
\end{eBox}
By analogy, we call a \textit{successful landscape} one that features at least one successful trajectory. 

\subsubsection{Multi-stage trajectories} 
\label{subsubsec:mul_stage_def}

We now enter the core of the phenomenological interest of this work, namely the characterisation of the successful inflationary trajectories.
Here, we investigate the possibility of finding multiple inflationary stages along a single trajectory.
Indeed, any successful trajectory should feature at least one extended epoch verifying the usual slow-roll conditions, namely $\epsilon,\eta \ll 1$ with $\eta= \epsilon^\prime/\epsilon$, that we call a \textit{stage}.
To determine the number of such stages given a trajectory, it is convenient to detect the transitions between different slow-roll attractors; the number of stages is then simply the number of transitions plus one. 

\begin{figure}[ht!]
    \centering  
    \includegraphics[width=\textwidth]{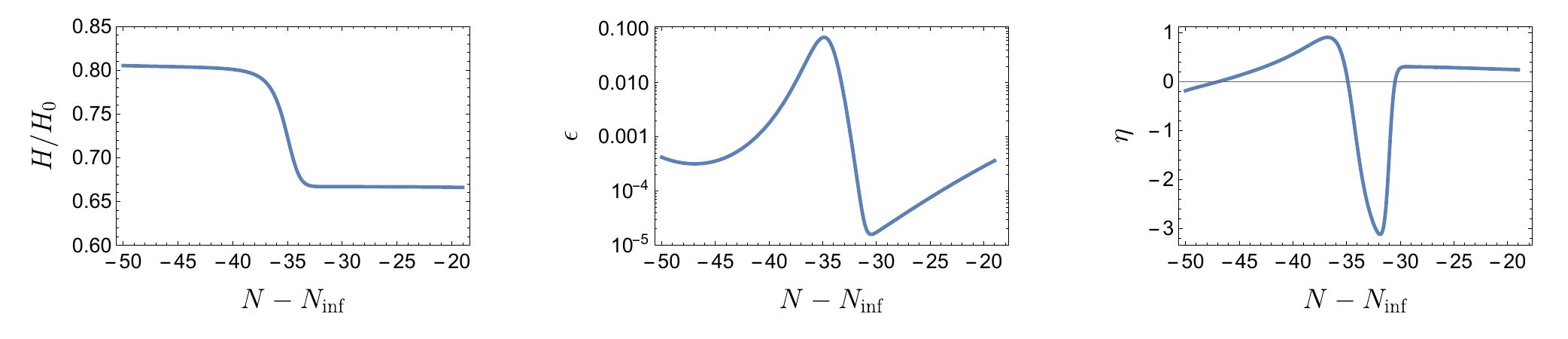}
    \caption{An illustration of a typical transition between two slow-roll stages, in which the evolution of $H$, $\epsilon$ and $\eta$ are shown as a function of the elapsed $e$-folding number $N$, where $H$ is normalized with respect to its value at the initial value $N$ in these figures. Note that $N$ is counted from the start of the trajectory, and therefore not directly related to the number of $e$-folds to the end of inflation.
    }
  \label{fig:Transition_demo}
\end{figure}

Clearly, transitions happen when the inflaton temporarily encounters a non-slow-roll section, such as gaining kinetic energy by reaching a high-slope section of the landscape, between two flat regions where slow roll occurs.
In turn, the slow-roll conditions are temporarily violated during the transition. 
An illustration of a typical transition is shown in Figure \ref{fig:Transition_demo} with the respective behaviours of $H, \epsilon, \eta$.
As expected, $\epsilon$ has a bump with a maximum in the midst of the transition.\footnote{In particular, the existence of the maximum of $\epsilon$ implies the existence of two consecutive slow-roll stages, which distinguishes a transition from the onset or the end of the inflation.} Meanwhile, the value of $\eta$, dictated by the derivative of $\epsilon$, appears to have a more complicated bump, with a sign change from positive to negative at the maximum of $\epsilon$. 

In this work, we apply two technically different methods to automatically determine the number of stages based on the time dependencies of $\epsilon$ and $\eta$, respectively.

\paragraph{Method I: slow-roll violation for $\eta$.} The observation is that during a transition, the value of $\eta$ temporarily reaches $\mathcal{O}(1)$ values, which means $\epsilon$ changes $\mathcal{O}(1)$ times its size within 1 $e$-fold. Based on this observation, we use the event that $|\eta|>\eta_c$, where $\eta_c=\mathcal{O}(1)$ is a critical value, as a criterion of the appearance of a transition.

But we have also seen that $\eta$ changes its sign during a transition, and therefore $|\eta|$ may exceed $\eta_c$ twice during a single transition.
In order not to double-count transitions, we add another criterion: two transitions must be separated by a minimum amount of $e$-folds $\Delta N$.
In turn, the quantity $\Delta N$ can be thought of as the minimal duration of a stage in our setup.
Empirically, we find that the choices $\eta_c =0.5$ and $\Delta N = 4$ are good enough for our purpose.
We denote the number of stages detected by this method as $n_{\rm stage, I}$.
We acknowledge a slight arbitrariness, so we now introduce a second method to compare with.

\paragraph{Method II: local maximum for $\epsilon$.} In this method, we simply identify a transition by the event that $\epsilon$ reaches its local maximum, which corresponds to $\eta=0$, as illustrated in the middle and right panels of Figure \ref{fig:Transition_demo}.
As in Method I, we take $\Delta N=4$ as the minimal duration of a stage, and two maxima with an interval smaller than $\Delta N$ will not be considered as two independent transitions, but rather as a single transition with a complicated structure.
We denote the number of stages detected by this method as $n_{\rm stage, II}$.

\vspace{4pt}
The two methods are slightly different.
Method I detects a violation of the slow-roll conditions, while Method II detects a characteristic dynamical feature of transitions.
The two methods provide consistent results in most cases. However, there are certain occasions where they can disagree with each other.
For example, there is a chance that a transition is so smooth that the maximal value of $|\eta|$ never exceeds $\eta_c$ but does go through zero, which Method I does not recognize as a transition while Method II does. 
Another possibility is that, during a transition, the width of the bump of $\epsilon$ is larger than $\Delta N$, which is counted as two transitions with Method I and one transition with Method II.
Despite these rare occasions, as we shall see, the two methods give statistically consistent conclusions, which can be seen as a robustness check of our methodology.
In the following, we make the conservative decision that we refer to the \textit{minimal} value given by Methods I and II as the number of stages assigned to a trajectory:
\begin{equation}
    n_{\rm stage}=\min\{n_{\rm stage, I}, n_{\rm stage, II}\}.
\end{equation}
Naturally, we then declare
\begin{eBox}
    \begin{equation}
    \text{multi-stage trajectory} \iff n_{\rm stage} > 1  \,.
\end{equation}
\end{eBox}

An important remark is that a multi-stage trajectory found by the methods above does not necessarily lead to \textit{observable} consequences. 
First, a transition happening more than 61 $e$-folds before the end of inflation will not be detectable by any cosmological probe.
Second, transitions between $30$ and $61$ $e$-folds before the end of inflation may or may not be detectable depending on the scale of reheating, or could even be already excluded by current CMB observations. 

\subsubsection{Multi-field trajectories}

A multi-dimensional field space does not necessarily guarantee genuine multi-field inflation, since the inflation attractor may only stretch along a specific direction in the field space, effectively leading to single-field inflation.
We propose here a way to statistically determine whether trajectories with genuine multi-field inflation are frequent.
This requires us to introduce a quantity to automatically characterize the multi-field nature of successful trajectories.\footnote{Although in some literature the terminology ``multi-field inflation models'' refers to models with more than one slow-roll direction, here we use the term in a more general sense, referring to any inflation model whose effective theory involves more than one field.}

To do so, we will use the well-known adiabatic-isocurvature decomposition~\cite{GrootNibbelink:2000vx,GrootNibbelink:2001qt} and later generalized in~\cite{Kaiser:2012ak,Achucarro:2018ngj,Pinol:2020kvw}.
First, we introduce the adiabatic vielbein
\begin{equation}
    \boldsymbol{e}_{\sigma}=\frac{\dot{\boldsymbol{\phi}}}{\dot{\sigma}} \,, \quad \text{with} \quad \dot{\sigma} = \sqrt{2\epsilon}H\Mp \,,
\end{equation}
which is the unit vector pointing in the direction of the background trajectory. 
Note that $\dot{\sigma}=\boldsymbol{e}_{\sigma}\cdot  \dot{\boldsymbol{\phi}}$ is the velocity \textit{along} the inflationary trajectory.
Second, we introduce the first isocurvature vielbein $\boldsymbol{e}_{s_{1}}$ as
\begin{equation}\label{def_es1}
    \frac{\dd \boldsymbol{e}_{\sigma}}{\dd N}= \eta_\perp \boldsymbol{e}_{s_{1}} \,,
\end{equation}
which defines the first isocurvature direction and the \textit{dimensionless turn rate} of the trajectory, $\eta_\perp$.
One can define $N_{\rm f}-1$ independent isocurvature directions $\boldsymbol{e}_{s_{\alpha}}$ which correspond exactly to all directions orthogonal to the adiabatic one.
By projecting the background equations of motion~\eqref{eq:background} onto each of these $\boldsymbol{e}_{s_{\alpha}}$ directions, one finds~\cite{Pinol:2020kvw}
\begin{equation}\label{def_etaperp}
   V_{,\alpha} + H \dot{\sigma} \eta_{\perp} \delta_{\alpha,1} =0 \quad \text{with} \quad V_{,\alpha} = \boldsymbol{e}_{s_{\alpha}} \cdot \nabla_{\boldsymbol{\phi}} V \,,
\end{equation}
i.e. $\eta_\perp$ can be written in terms of the projection of the gradient of the potential in the first isocurvature direction.
Note that the other projections vanish, so that, in practice, one may determine the absolute value of the dimensionless turn rate of the trajectory as
\begin{align}
\label{eq:abs-eta-perp}
    \left|\eta_\perp\right|  = \frac{\left[\left(\nabla_{\boldsymbol{\phi}}V\right)^2-V_{,\sigma}^2 \right]^{1/2}}{H \dot{\sigma}}\quad \text{with} \quad V_{,\sigma} = \boldsymbol{e}_{\sigma} \cdot \nabla_{\boldsymbol{\phi}} V \,,
\end{align}
which is enough for most purposes.

With this instantaneous turning rate, we can define the \textit{total absolute curvature} of the trajectory between two moments $N_1$ and $N_2$: 
\begin{equation}\label{theta_abs}
    \theta(N_1,N_2)\equiv\int_{N_1}^{N_2}|\eta_\perp|\dd N.
\end{equation}
This quantity is positive and has a simple geometrical interpretation: it corresponds to the accumulated angle of turning of a trajectory between moments $N_1$ and $N_2$.

With the quantity $\theta(N_1,N_2)$, we can further define a unique number $\Theta$ for each trajectory as
\begin{equation}\label{Theta}
    \Theta=\theta(N_{\rm ini},N_{\rm fin})\,,
\end{equation}
in which $N_{\rm ini}$ and $N_{\rm fin}$ are the initial and final times. It turns out to be more physically relevant to choose $N_{\rm ini}$ and $N_{\rm fin}$ differently from, respectively $0$ and $N_{\rm inf}$.
First, the beginning of the simulation as $N \rightarrow 0$ can result in a highly curved trajectory before an attractor is reached, and is therefore initial-condition-dependent.
Moreover, the end of inflation as $N\rightarrow N_{\rm inf}$ necessarily features strong turns as the inflaton approaches the minimum of the true vacuum, but the precise value depends on the reheating scenario which we do not model here.
Instead, $N_{\rm ini}$ and $N_{\rm fin}$ should be chosen so that the first and final $e$-folds of inflation are excluded. In practice, we choose $N_{\rm ini}=0.2  N_{\rm inf}$ and $N_{\rm fin}=0.95 N_{\rm inf}$, which is an empirically effective choice to discard the unwanted sections of a trajectory while preserving the interesting ones as much as possible.

For each trajectory, this quantity is unique and characteristic, indicating the ``degree" of multi-field nature.
Note that the only strict criterion to declare that a trajectory is \textit{multi-field} would be $\Theta \neq 0$. 
However, in practice, we would find all trajectories to be multi-field as long as $N_{\rm f} > 1$, which is not illuminative.
Instead, for the purpose of our statistical study, we propose the following criterion
\begin{eBox}
\begin{equation}
    \text{multi-field trajectory} \iff \Theta> \frac{\pi}{10} \,,
\end{equation}
\end{eBox}
which corresponds to a total angle of $18 {}^\circ$.
We acknowledge that this criterion is arbitrary and we will often present the distribution of $\Theta$ for all trajectories as the complete result, while the multi-field nature is only proposed as compressed information.

Obviously, the notion of multi-field trajectory does not apply to the case of $N_{\rm f}=1$. When $N_{\rm f}>1$, there are two different sources of the multi-field nature of trajectories.
The first one is related to the presence of multiple stages, as two consecutive slow-roll attractors are unlikely to be aligned in the field space and therefore are likely to introduce a turning during the transition.\footnote{Also, transverse oscillations can be generated at the end of a transition, which is another source of multi-field nature. A detailed discussion can be found in \ref{subsec:feature}.}
The second one is the possibility that the slow-roll attractor itself is curved, as in multi-field slow-roll models or quasi-single-field inflation models. 
While we cannot tell whether a given trajectory is multi-field from one source or the other, or both, based one $\Theta$ alone, it is possible to conduct a joint analysis of multi-stage and multi-field properties of trajectories, which will be presented in the next sections.

Additionally, we note that the $N_{\rm f}=2$ case is special in the sense that the codimension-1 submanifold of the field space is one dimensional. As a result, we can define \textit{net} angle of turning between two moments $N_1$ and $N_2$ as
\begin{equation}\label{tilde_theta}
    \tilde\theta(N_1,N_2)=\int_{N_1}^{N_2}\eta_\perp\dd N,
\end{equation}
and hence we can define a quantity $\tilde{\Theta}$ in the same way as $\Theta$:
\begin{equation}\label{tilde_Theta}
    \tilde\Theta=\tilde{\theta}(N_{\rm ini},N_{\rm fin})\,,
\end{equation}
which is the \textit{net} angle of turning of a trajectory between $N_{\rm ini}$ and $N_{\rm fin}$, \textit{i.e.} the difference between the direction of field-space velocity $\dot{\boldsymbol\phi}$ at $N_{\rm ini}$ and $N_{\rm fin}$.
Since Eq.~\eqref{eq:abs-eta-perp} does not give the sign of $\eta_\perp$, we use for the $N_{\rm f}=2$ case the explicit expressions $\boldsymbol{e}_s=(-\dot{\phi}_2,\dot{\phi}_1)/\dot{\sigma}$, $V_{,s}=\boldsymbol{e}_s\cdot\nabla_{\boldsymbol{\phi}}V$ and $\eta_\perp = - V_s/(H \dot{\sigma})$.

\subsubsection{Primordial power spectrum and $(n_s,r)$ constraints}\label{subsubsec:CMB_comp_intro}

A last feature of successful trajectories that we investigate is whether they can be compatible with the latest CMB constraints.
Note that, the value of the amplitude of the power spectrum $\mathcal{\tilde P}_\zeta(k_\star)=\tilde A_s$ at any pivot scale $k_\star$, computed from our dimensionless equations, is related to the physical value $\mathcal{P}_\zeta(k_\star)= A_s$ by a rescaling, $A_s=\tilde A_s (V_0/M_{\rm pl}^4)$. So, for any value of $\tilde A_s$, we can choose $V_0$ to match the data $A_s\approx 2.1\times 10^{-9}$, as long as $V_0$ and $k_\star$ satisfy the very flexible constraints in App.~\ref{app_efolds}.
Therefore, in our case, the main constraints come from those in the $(n_s,r)$-plane.
First, we remind that we calculate the scalar and tensor power spectra of all successful trajectories using \texttt{PyTransport}, and we define the quantities
\begin{equation}
    n_s(k) = 1+ \frac{\dd \ln \mathcal{P}_\zeta(k)}{\dd \ln k}\,, \quad r(k) = \frac{\mathcal{P}_\gamma(k) }{\mathcal{P}_\zeta(k)}\,.
\end{equation}
Then, we declare a trajectory CMB-compatible if there exists a range of three consecutive decades in $k$ (corresponding to 7 $e$-folds of inflation) that
exited the horizon between $30$ and $61$ $e$-folds before the end of inflation and with $(n_s,r)$ compatible with the latest CMB constraints.
More in detail, we ask that the found $n_s$ and $r$ values are each compatible with their individual Planck-BICEP-Keck-ACT-SPT constraints at $2 \sigma$~\cite{Balkenhol:2025wms}

So, denoting $N_\star(k)$ the time at which the mode $k$ exited the comoving Hubble radius, we declare:
\begin{eBox}
    \begin{align}\label{CMB_comp_condition}
    \text{CMB-compatible trajectory} \iff \exists k_\star\,,\,\, &\forall k_i \in \left\{\frac{k_\star}{e^{3.5}},k_\star,k_\star e^{3.5}\right\}\,, \\
    &N_\star(k_i) - N_{\rm inf} \in \left[-61,\, -30  \right] \,, \,\, \text{as well as} \nonumber\\
        &n_s(k_i) \in \left[0.9618, 0.9746\right] \,\,
    \text{and}  \,\, r(k_i) < 0.034 \nonumber\,.
\end{align}
\end{eBox}

\subsubsection{Summary}

To summarize, we can find the following sets of trajectories with strict inclusion relations:
\begin{eBox}
\begin{equation}
    \text{admissible} \supset \text{successful} \supset   \begin{cases}
    \text{multi-stage}  \\ 
    \text{multi-field}  \\ 
    \text{CMB-compatible} \end{cases} \quad  \,.
\end{equation}
\end{eBox}
Note that the different categories of successful trajectories are not mutually exclusive, and that we will retrieve information about the multi-field nature and the number of stages of all successful trajectories, independently of whether they are CMB-compatible or not.

\section{One-field landscapes}
\label{sec1field}


\begin{figure}[t]
\centering
\providecommand{\Mp}{M_{\rm Pl}}
\providecommand{\etamed}{\eta_{\rm med}}
\definecolor{funnel}{RGB}{58,80,112}   

\resizebox{\textwidth}{!}{%
\begin{tikzpicture}[
  x=1cm, y=1cm,
  bar/.style  ={draw=funnel!55, line width=0.3pt},
  lost/.style ={draw=none, fill=black!6},
  name/.style ={font=\small, anchor=east, inner sep=0pt},
  sub/.style  ={font=\scriptsize, color=black!55, anchor=east, inner sep=0pt},
  num/.style  ={font=\footnotesize, anchor=east, inner sep=0pt},
  tick/.style ={font=\scriptsize, color=black!55, anchor=north, inner sep=2pt},
  grid/.style ={draw=black!10, line width=0.3pt},
]

\def\XT{10.8523} \def\XA{9.8557} \def\XS{7.1667} \def\XC{4.2966}
\def\bh{0.30}

\foreach \d in {0,...,8}{
  \draw[grid] (1.62*\d, 0.45) -- (1.62*\d, -4.05);
  \draw[draw=black!35, line width=0.4pt] (1.62*\d,-4.05) -- (1.62*\d,-4.16);
  \node[tick] at (1.62*\d, -4.16) {$10^{\d}$};}
\draw[draw=black!35, line width=0.4pt] (0,-4.05) -- (12.96,-4.05);
\node[font=\scriptsize, color=black!55, anchor=north] at (6.48,-4.62)
      {number of trajectories};

\def\y{0}
\node[name] at (-0.35,\y+0.10) {Total};
\node[sub]  at (-0.35,\y-0.17) {the whole sample};
\fill[bar, fill=funnel!20] (0,\y-\bh) rectangle (\XT,\y+\bh);
\node[num] at (\XT-0.16,\y) {5\,000\,000};

\def\y{-1.15}
\node[name] at (-0.35,\y+0.10) {Admissible};
\node[sub]  at (-0.35,\y-0.17) {24.25\% of the sample};
\fill[lost] (\XA,\y-\bh) rectangle (\XT,\y+\bh);
\fill[bar, fill=funnel!38] (0,\y-\bh) rectangle (\XA,\y+\bh);
\node[num]  at (\XA-0.16,\y) {1\,212\,705 \textcolor{black!55}{(24.25\%)}};

\def\y{-2.30}
\node[name] at (-0.35,\y+0.10) {Successful};
\node[sub]  at (-0.35,\y-0.17) {0.53\% of the sample};
\fill[lost] (\XS,\y-\bh) rectangle (\XA,\y+\bh);
\fill[bar, fill=funnel!62] (0,\y-\bh) rectangle (\XS,\y+\bh);
\node[num]  at (\XS-0.16,\y) {26\,538 \textcolor{black!55}{(2.19\%)}};

\def\y{-3.45}
\node[name] at (-0.35,\y+0.10) {CMB-compatible};
\node[sub]  at (-0.35,\y-0.17) {0.0090\% of the sample};
\fill[lost] (\XC,\y-\bh) rectangle (\XS,\y+\bh);
\fill[bar, fill=funnel, draw=funnel] (0,\y-\bh) rectangle (\XC,\y+\bh);
\node[num, text=white] at (\XC-0.16,\y) {449 \textcolor{white!78!funnel}{(1.69\%)}};

\draw[draw=black!15, line width=0.4pt] (-2.95,-5.20) -- (12.96,-5.20);
\node[font=\small, anchor=west] at (-2.95,-5.72)
      {\textcolor{black!45}{$\hookrightarrow$}\; Structure of the
       \textbf{successful} trajectories};

\begin{scope}[shift={(3.5,-8.25)}]
  \fill[draw=white, line width=0.8pt, fill=funnel!25] (90.000:0.92) arc[start angle=90.000, end angle=-198.612, radius=0.92] -- (-198.612:1.78) arc[start angle=-198.612, end angle=90.000, radius=1.78] -- cycle;
  \fill[draw=white, line width=0.8pt, fill=funnel!48] (-198.612:0.92) arc[start angle=-198.612, end angle=-270.000, radius=0.92] -- (-270.000:1.78) arc[start angle=-270.000, end angle=-198.612, radius=1.78] -- cycle;
  \draw[funnel!45, line width=1.3pt] (90:0.8) arc[start angle=90, end angle=-270.000, radius=0.8];
  \node[font=\footnotesize, align=center] at (0,0.10) {26\,538};
  \node[font=\scriptsize, color=black!55] at (0,-0.16) {successful};
  \draw[draw=black!50, line width=0.35pt] (-54.306:1.6) -- (-54.306:2.02);
  \node[anchor=west, align=left, inner sep=1.5pt] at (-54.306:2.06) {\small SFSS\\[-1pt]\scriptsize\textcolor{black!60}{80.17\%}};
  \draw[draw=black!50, line width=0.35pt] (-234.306:1.6) -- (-234.306:2.02);
  \node[anchor=east, align=right, inner sep=1.5pt] at (-234.306:2.06) {\small SFMS\\[-1pt]\scriptsize\textcolor{black!60}{19.83\%}};
\end{scope}
\node[font=\scriptsize, color=black!55, anchor=north] at (3.5,-10.75) {single-field \textcolor{black}{100.00\%}\quad multi-field \textcolor{black}{0.00\%}\quad multi-stage (SFMS$+$MFMS) \textcolor{black}{19.83\%}};

\node[draw=funnel, line width=0.5pt, fill=funnel!8, rounded corners=2pt,
      anchor=east, align=left, text width=3.2cm, inner sep=5pt, font=\small]
     at (12.96,-8.25) {\textbf{449} CMB-compatible\\[1pt]
      \scriptsize\textcolor{black!60}{1.69\% of the successful trajectories}};

\end{tikzpicture}}
\caption{{\em $N_{\rm f}=1$ case} $(\xi=1.32\Mp$, $\etamed=0.38)$.
Classification of $5\times10^{6}$ trajectories drawn on $50\,000$ landscape
realisations. \emph{Top:} the successive selection stages; bar lengths are
logarithmic and the figure in parentheses is the fraction kept from the
previous stage, whose extent is shown by the shaded continuation of the bar.
\emph{Bottom:} the successful trajectories resolved by field content and
stage structure. The wedges run clockwise in the order SFSS, SFMS, MFSS,
MFMS, so the two inner arcs group the single-field and the multi-field
categories (the latter being obviously absent in this $N_{\rm f}=1$ case). SFSS: Single-Field-Single-Stage;
SFMS: Single-Field-Multi-Stage; MFSS: Multi-Field-Single-Stage; MFMS:
Multi-Field-Multi-Stage.}
\label{fig:event-hierarchy_1f}
\end{figure}

After the preparations above, in this and subsequent sections, we will present detailed studies on the statistical properties of inflation trajectories on landscapes with increasing field dimensions. The first and simplest case is $N_{\rm f}=1$, in which the potential is a 1-dimensional Gaussian random field $V(\phi)$. This case provides a simple starting point for understanding the trend as $N_{\rm f}$ increases.

\paragraph{Numerical setup.} In the $N_{\rm f}=1$ case, we choose the Fourier space cutoff to be $M=20$, which is larger than that of the $N_{\rm f}=2$ example presented above, since the numerical solution is more efficient for the one-field potential. Since $\etamed$ is a more physical quantity to characterize a realistic landscape than $\xi$, the value of $\xi$ is chosen so that it gives the same $\etamed$ as the fiducial $N_{\rm f}=2$ case (\ref{eq:fiducial-params}), i.e.~$\etamed= 0.38$ from numerical simulation.
Therefore, the statistics of different $N_{\rm f}$ can be compared on the grounds of $\etamed$ being fixed.
The precise value of $\xi$ is found to be $\xi=1.32\Mp$, and we set $\Lambda=10\xi$ and $P_0=1$ as in the fiducial case.
For this simple case, we take the sampling radius $\lambda=\Lambda$ for initial field positions, i.e.~the whole landscape.
We choose to set $n=100$ random initial points in each landscape realization, which is sufficient to probe all sub-Planckian structures of the potential.
The number of realizations is set to be 50 000 in this case, which is a sufficiently large number to acquire good statistical robustness on the properties of interest. As explained in Sec.~\ref{subsec3.3}, we collect all the \textit{admissible} trajectories --- that terminate at the true vacuum --- in each realization. The one field case is not numerically demanding; it takes $\sim20\mathrm{min}$ to generate the 50 000 landscapes, each with 100 initial conditions, and to classify all 5 000 000 trajectories as admissible (1 212 705, i.e.~24.25\%) or not (3 787 295, i.e.~75.75\%).

\subsection{Duration of inflation}
\label{subsec:1fduration}
\begin{figure}[ht!]
    \centering  
    \includegraphics[width=\textwidth]{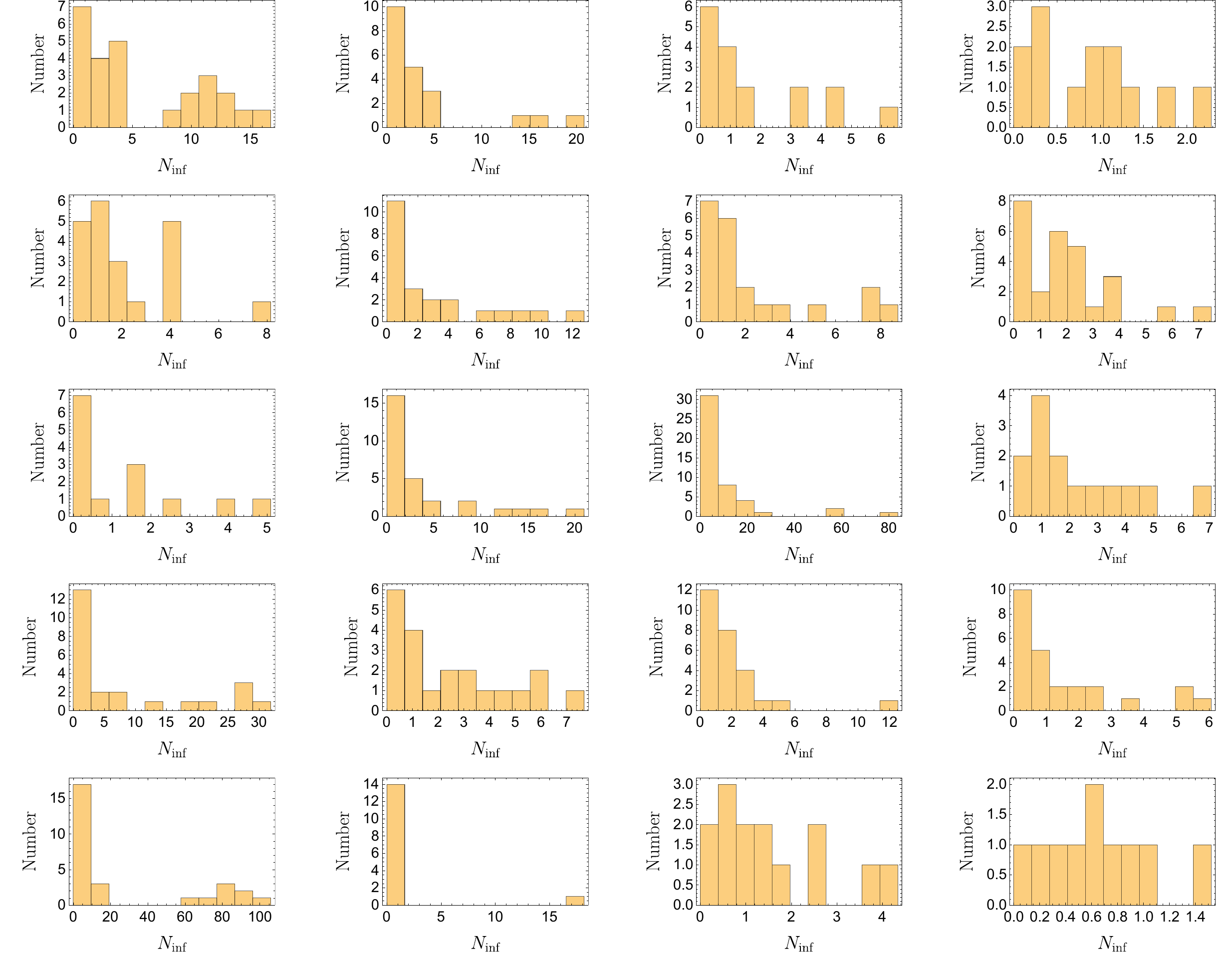}
    \caption{ {\em $N_{\rm f}=1$ case} $(\xi=1.32\Mp$, $\etamed=0.38)$. The distribution of $N_{\rm inf}$ of all admissible trajectories of 20 randomly chosen realizations of the one-field landscape. Amongst all realizations, only a small fraction of them like the bottom left one is capable of supporting successful trajectories.}
  \label{fig:NDist_Individual}
\end{figure}

Due to the stochastic nature of the landscape, the number of admissible trajectories is not the same in each realization, nor is the existence of \textit{successful} trajectories. Indeed, only 5 211 out of the 50 000 realizations have at least one successful trajectory, corresponding to a fraction of $10.4\%$. To exemplify the variance, we plot the distribution of $N_{\rm inf}$ of all admissible trajectories of 20 randomly chosen realizations in Figure \ref{fig:NDist_Individual}. In this figure, it can be clearly seen that most realizations cannot support successful trajectories, except a small fraction of exceptional ones. It is what is expected from the condition $\eta_{\rm med}\sim1$, which means finding a successful trajectory is not a frequent event but requires a bit of fortune. 

Based on the underlying picture that each numerical realization is really a small patch of the ``real" landscape, it is meaningful to combine the trajectories from different realizations together for a joint statistical study. In total, there are 26 538 successful trajectories, making up a fraction of $2.19\%$ of all admissible trajectories, and $0.53 \%$ of all trajectories.

\begin{figure}[ht!]
    \centering  
    \includegraphics[width=\textwidth]{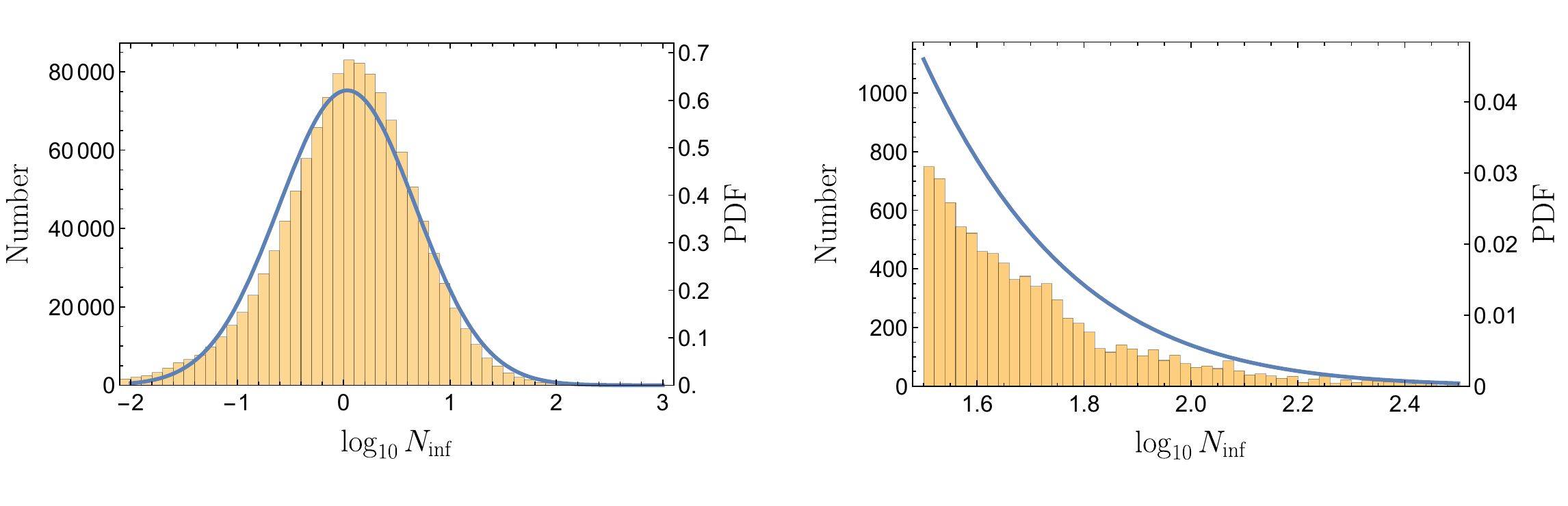}
    \caption{ {\em $N_{\rm f}=1$ case} $(\xi=1.32\Mp$, $\etamed=0.38)$. Left panel: The combined distribution of $N_{\rm inf}$ of all admissible trajectories from all realizations of the one-field landscape. The distribution is displayed in terms of $\log_{10}N_{\rm inf}$, and the profile appears to have a pattern of normal distribution. The blue curve is a fit of the probability density function (PDF) by the normal distribution with mean and standard deviation being 0.20 and 0.67, respectively. Right panel: A zoomed-in view of the rightmost tail of the distribution, corresponding to trajectories with large $N_{\rm inf}$.
    }
  \label{fig:NeDist_xi1_log_1d.pdf}
\end{figure}

The combined distribution of $N_{\rm inf}$ of all admissible trajectories is shown in Figure \ref{fig:NeDist_xi1_log_1d.pdf}, we present the distribution in terms of $\log_{10}N_{\rm inf}$. From the appearance of the histogram, we propose that the profile can be approximated by a normal distribution (\textit{i.e.} a log-normal distribution in term of $N_{\rm inf}$). The mean and standard deviation of $\log_{10}N_{\rm inf}$ are calculated to be 0.20 and 0.67, and the probability density function (PDF) of the normal distribution corresponding to these parameters is shown as the blue curve in the figure. We find that although this PDF captures the peak of the distribution, it fails to describe the tails on both sides. In particular, it overestimates the right tail consisting of the interesting trajectories with large $N_{\rm inf}$, which can be seen clearly from the zoomed-in view in the right panel of Figure \ref{fig:NeDist_xi1_log_1d.pdf}. Despite the imperfection of the approximation, as we will see in the subsequent sections, the log-normal profile is a good benchmark for us to understand certain effects of increasing $N_{\rm inf}$.

\subsection{Multi-stage trajectories}
\label{subsec:1fmultistage}
Heuristically, a multi-stage trajectory arises when the potential $V(\phi)$ has at least two slow-roll attractors connected by an intermediate slope. We would like to find out whether this is a frequent event.
To do so, we study the fraction of multi-stage trajectories among all successful ones.

\begin{figure}[ht!]
    \centering  
    \begin{tabular}{lll}
        \hspace{45pt}One stage & \hspace{43pt}Two stages & \hspace{40pt}Three stages \\
        \hspace{3pt}\includegraphics[width=0.275\textwidth]{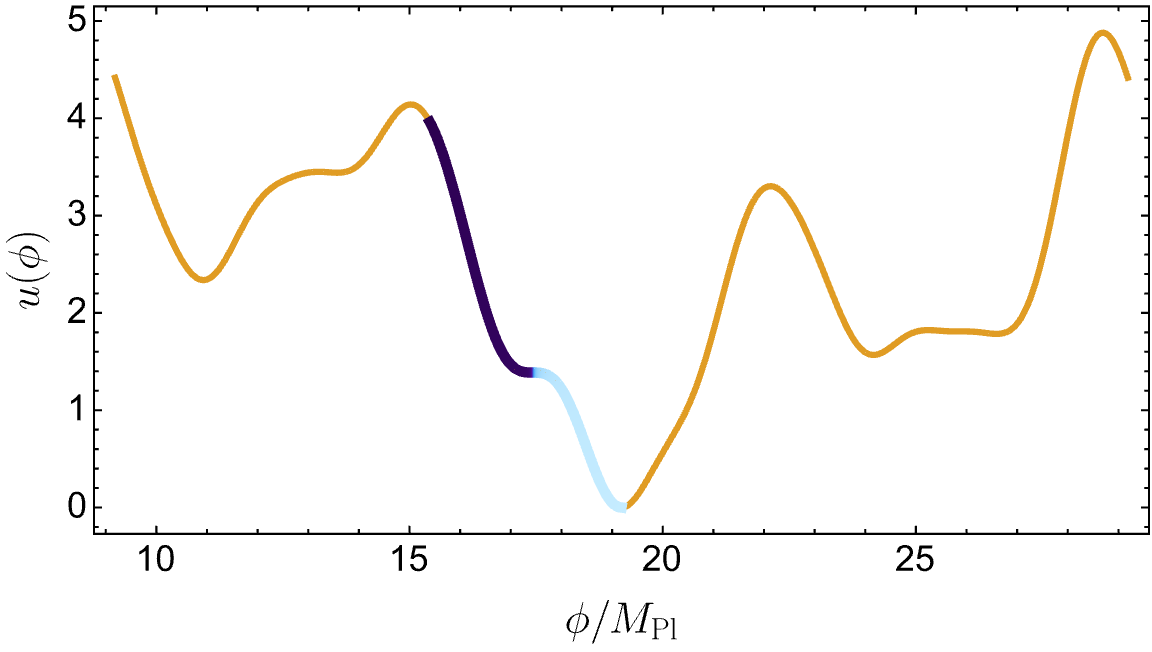} &
        \hspace{3pt}\includegraphics[width=0.275\textwidth]{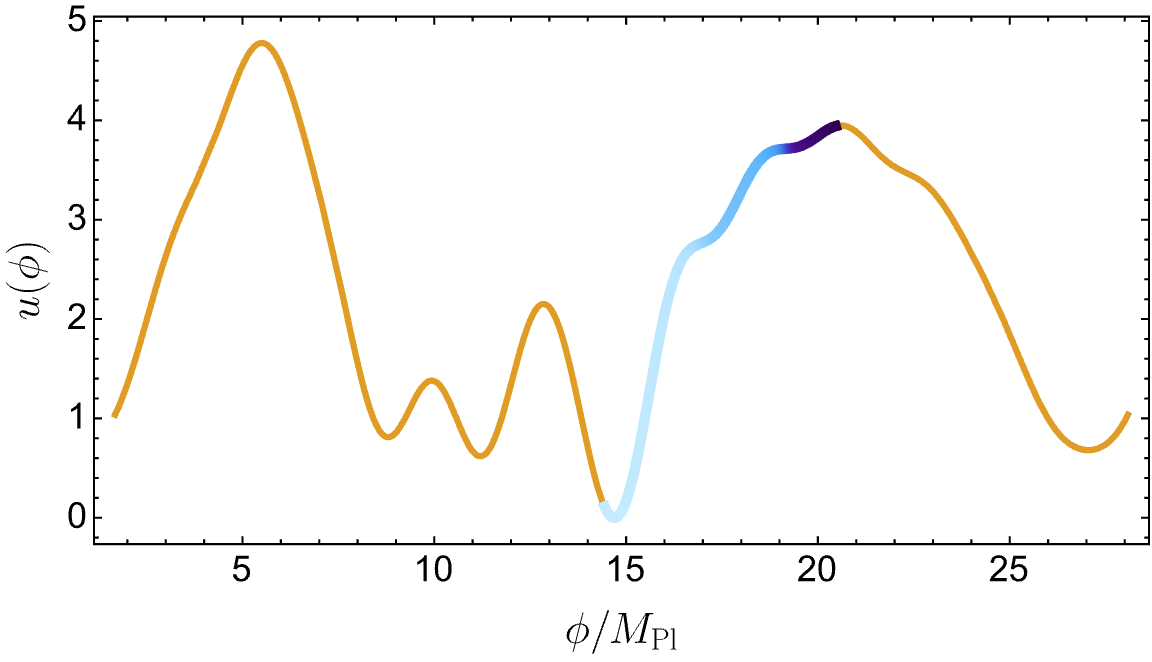} &
        \hspace{3pt}\includegraphics[width=0.275\textwidth]{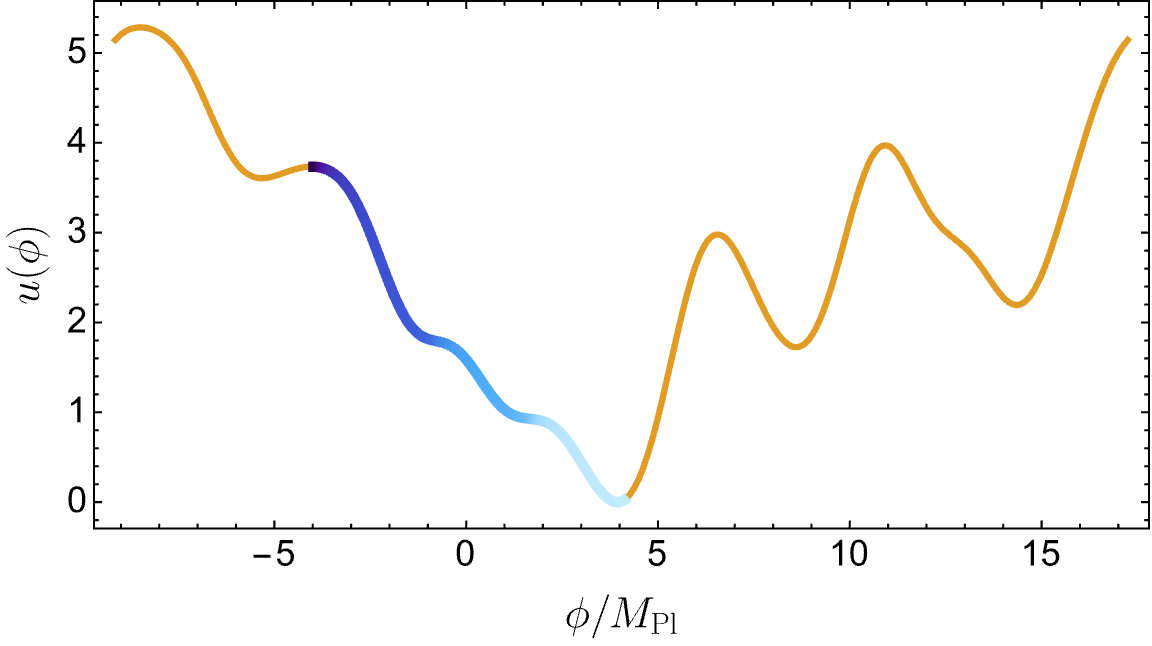}\\
        \includegraphics[width=0.32\textwidth]{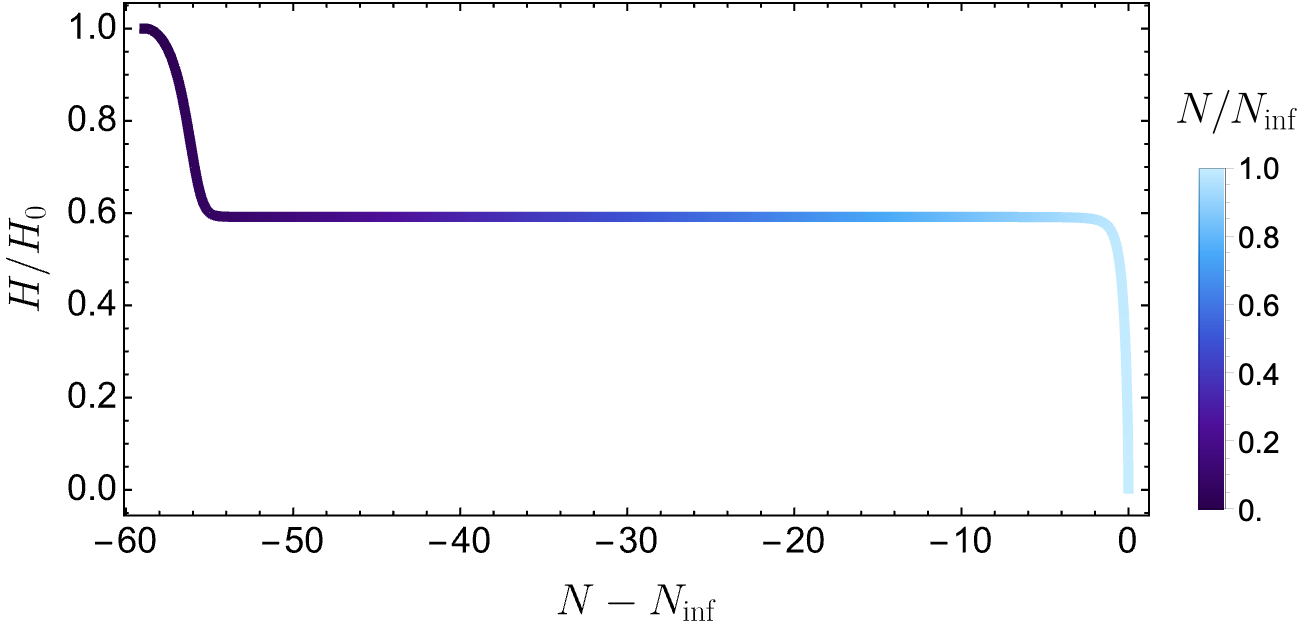} &
        \includegraphics[width=0.32\textwidth]{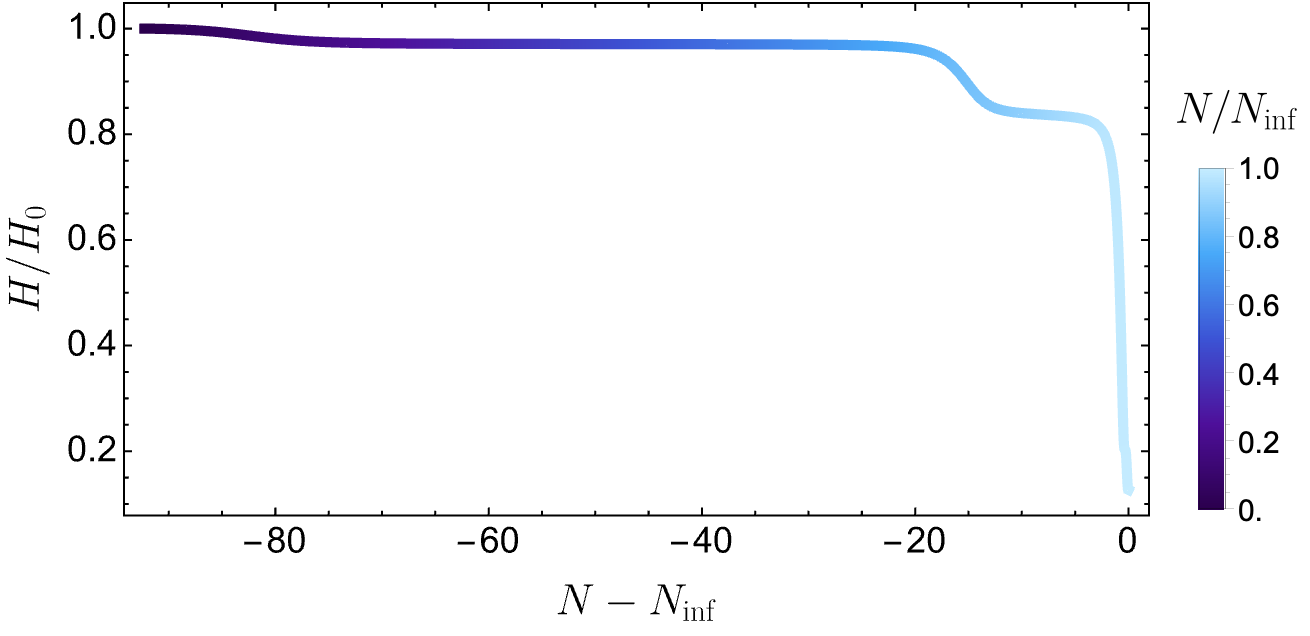} &
        \includegraphics[width=0.32\textwidth]{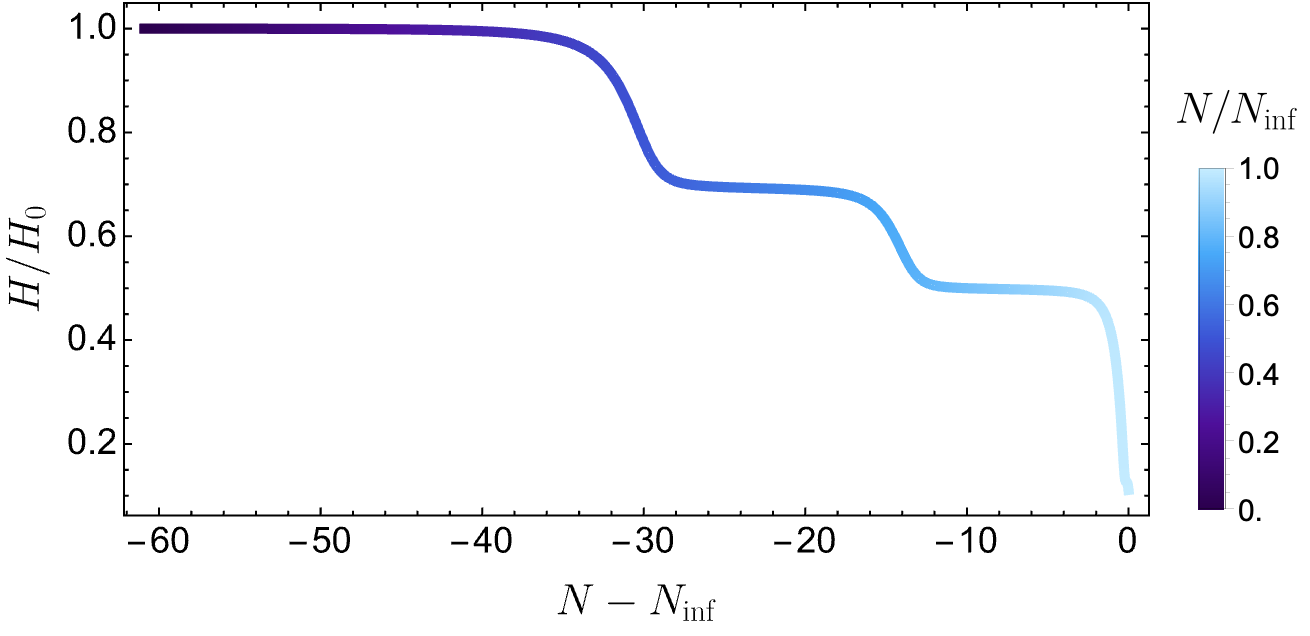}
    \end{tabular}
    \caption{ {\em $N_{\rm f}=1$ case} $(\xi=1.32\Mp$, $\etamed=0.38)$. Three trajectories with different number of stages found on different one-field realizations. The left, middle and right panels are trajectories with 1, 2 and 3 stages, in each of which the upper panel shows the appearance of the trajectory on the landscape and the lower panel shows the evolution of the Hubble parameter $H$. The colour convention of each trajectory follows that in Figure \ref{fig:trajectory_exp}.
    }
  \label{fig:1d_trajectories.pdf}
\end{figure}

Using the methods introduced in Sec.~\ref{subsubsec:mul_stage_def}, we indeed find typical multi-stage trajectories. We present three examples of trajectories with an increasing number of stages in Figure \ref{fig:1d_trajectories.pdf}, each with the shape of the trajectory and the evolution of the Hubble parameter $H$. From these diagrams, we can see clearly how a near-flat region of the potential gives rise to a stage of slow-roll inflation. In particular, the color gradient along the trajectories clearly indicates that most of the $e$-folds are elapsed within those slow-roll stages, whereas transitions are relatively short in $e$-folds. 

\begin{figure}[ht!]
    \centering  
    \includegraphics[width=0.465\textwidth]{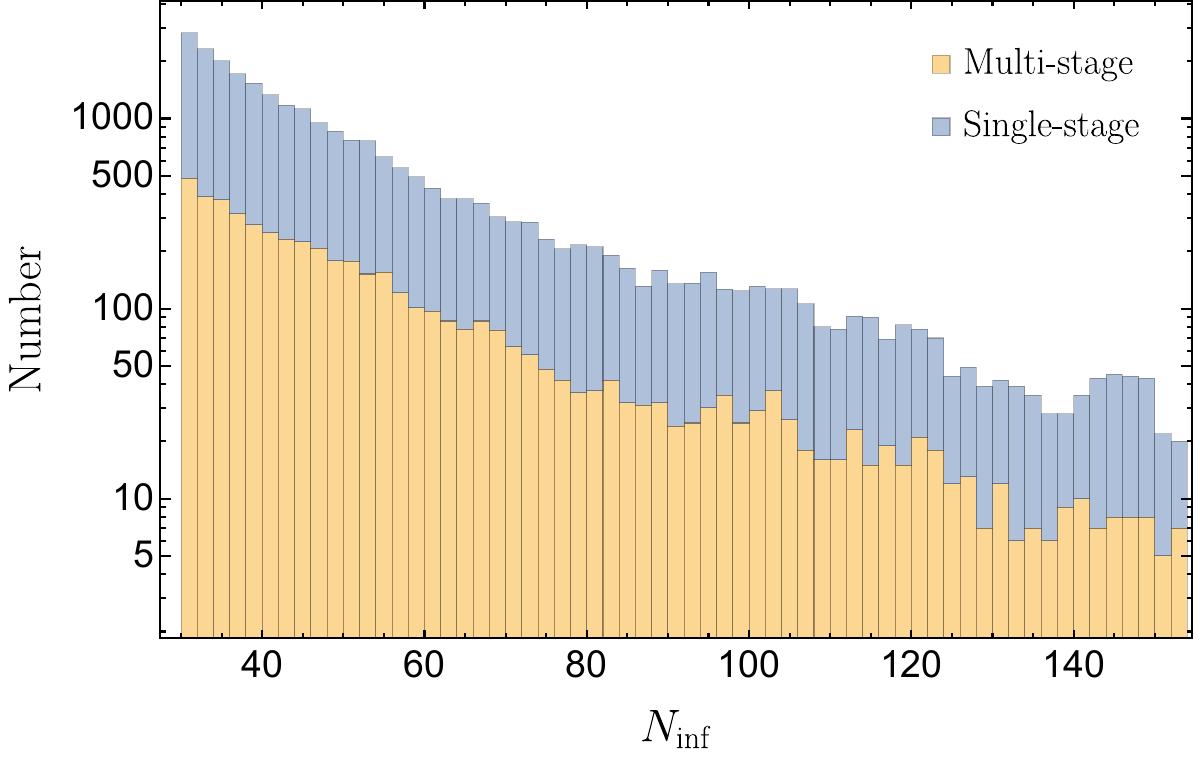}
    \hspace{10pt}
    \includegraphics[width=0.45\textwidth]{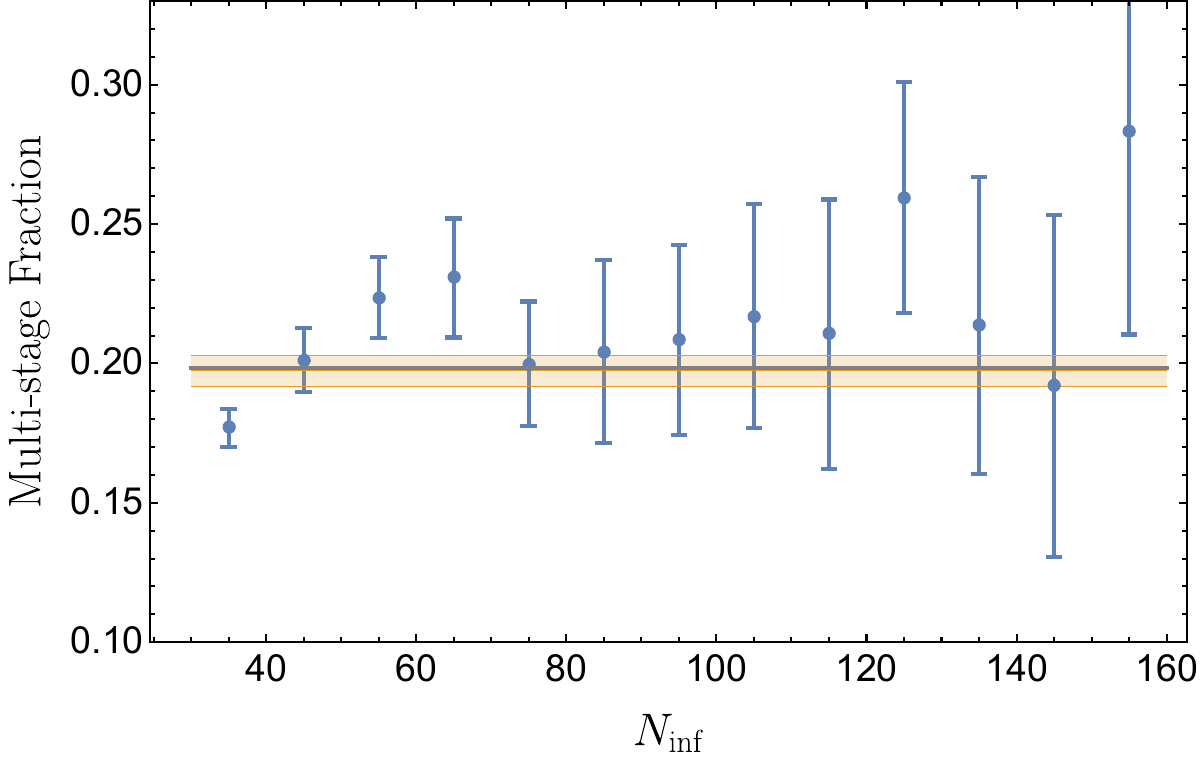}
    \caption{ {\em $N_{\rm f}=1$ case} $(\xi=1.32\Mp$, $\etamed=0.38)$. Left panel: The binned distribution of numbers of single-stage and multi-stage trajectories, in which the bin width is set to be $\Delta N_{\rm inf}=2$. Right panel: The binned fraction of multi-stage trajectories with bin width $\Delta N_{\rm inf}=10$, in which the error bars are estimated by the statistical bootstrap technique. The orange band denotes the overall multi-stage fraction with statistical errors obtained by combining all bins, $(19.7\pm0.5)\%$.
    The gray line is the true overall multi-stage fraction, $19.83\%$.
    }
  \label{fig:Stage_Ne_1d}
\end{figure}

After learning about the particular cases above, we move on to statistical aspects. As in the discussion of durations above, we can merge the multi-stage data of trajectories from all realizations together to obtain larger sets.
We find that 5 263 of all 26 538 successful trajectories are multi-stage, i.e.~a fraction of 19.83\%.
It is a natural question to wonder whether the multi-stage nature is related to the overall duration of inflation.
To test this hypothesis, we divide all successful trajectories into bins of $N_{\rm inf}$ with bin width $\Delta N_{\rm inf}=2$, and we explore the multi-stage nature in each of these bins.
The result is shown in the left panel of Figure \ref{fig:Stage_Ne_1d}. In this histogram, it can be seen that the total number of successful trajectories and the number of multi-stage trajectories decrease simultaneously as $N_{\rm inf}$ increases.
To have a more quantitative answer, we estimate the multi-stage fraction with error bars in multiple $N_{\rm inf}$-bins with the statistical bootstrap technique (summarized in App.~\ref{app_bootstrap}), and the result is shown in the right panel of Figure \ref{fig:Stage_Ne_1d}. In this plot, the bin width is chosen as $\Delta N_{\rm inf}=10$, i.e. five times wider than on the left panel, in order to suppress statistical uncertainty and keep the plot concise. 
It appears that the dependence of multi-stage fraction on $N_{\rm inf}$ is weak, albeit with increasing statistical uncertainties at larger $N_{\rm inf}$.
We also combine these bins and estimate the mean and variance of the multi-stage fraction across all values of $N_{\rm inf}$, giving an estimate of the overall fraction with error bars.
We find an overall multi-stage fraction of $(19.7\pm0.5)\%$ of all successful trajectories, where indeed the central value is perfectly compatible with the exact number we had previously obtained, 19.83\%, up to the error bars.

In summary, we find that there is already a substantial fraction of multi-stage trajectories even in the one-field case. 
More detailed properties of multi-stage inflation will be extensively investigated in the following sections with larger values of $N_{\rm f}$. 
Moreover, since the field content is one-dimensional by construction, there is no notion of multi-field inflation in this case. 

\subsection{CMB compatibility}
\begin{figure}[ht!]
    \centering  
    \includegraphics[width=0.42\textwidth]{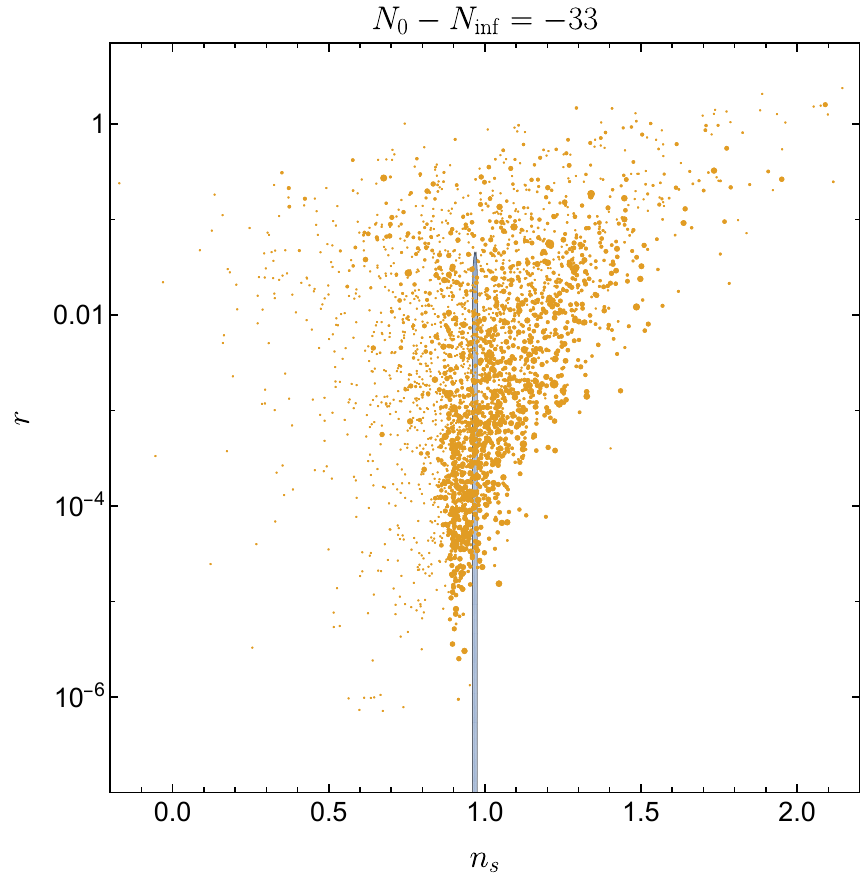}
    \hspace{10pt}
    \includegraphics[width=0.53\textwidth]{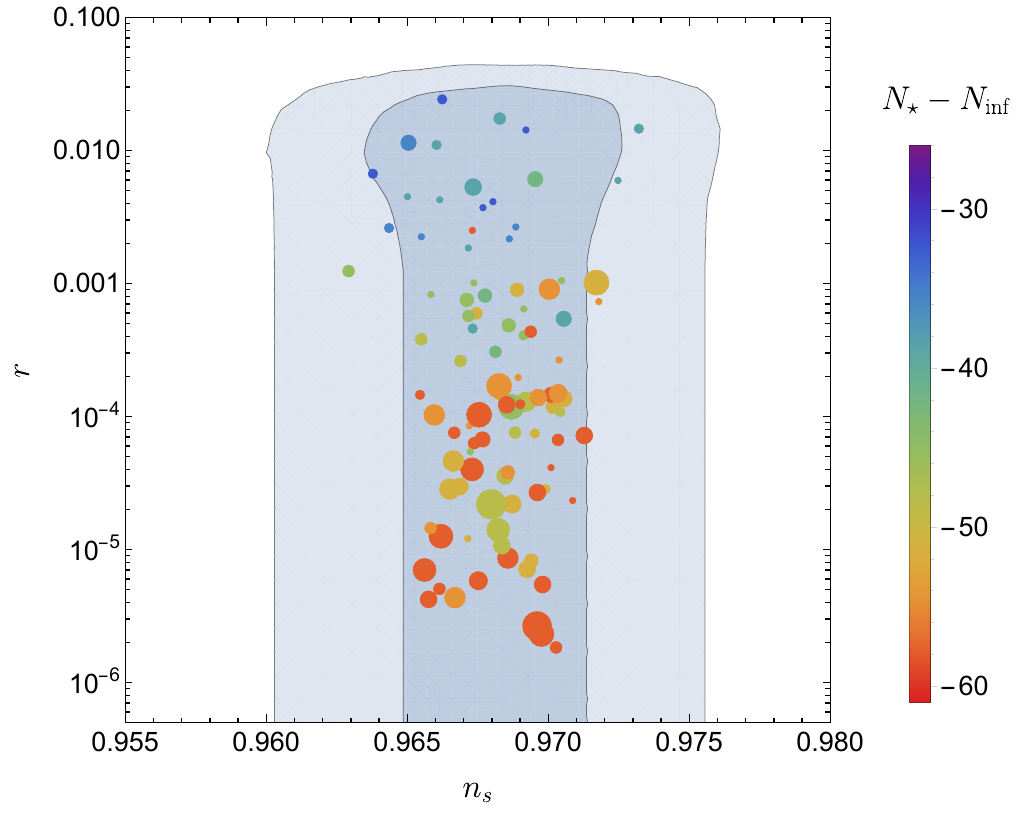}
    \caption{{\em $N_{\rm f}=1$ case} $(\xi=1.32\Mp$, $\etamed=0.38)$. 
    Left Panel: The positions of all trajectories with $N_{\inf}>33$ on the $r$-$n_s$ plane at the fixed scale $N_0-N_{\inf}=-33$. The thin contour is the $95\%$ likelihood of the current CMB constraints. 
    Right Panel: The positions on the $r$-$n_s$ plane of all CMB-compatible trajectories as well as the $68\%$ and $95\%$ CMB constraints. The values of $n_s$ and $r$ of these trajectories are evaluated at the pivot scale $N_*$ at which the mode $k_*$ exits the horizon.
    For both panels,
    we present the position of such a landscape on the $r$-$n_s$ plane by a circle, with its area proportional to the total number of such trajectories in that landscape.
    }
  \label{fig:r_ns_1d}
\end{figure}

With the sample of successful trajectories at hand, we can conduct the CMB compatibility investigation as introduced in \ref{subsubsec:CMB_comp_intro}. As mentioned above, in our sample, there are a total of $9587$ successful trajectories from $1886$ realizations. In order to select from them the CMB-compatible trajectories, we evaluate the scalar power spectrum $P_\zeta(k)$ and the tensor power spectrum $P_\gamma(k)$ of each trajectory on a log-spaced $k$-grid with spacing $\Delta \ln k=3.5$, in the range of $k$ in which the modes exit the horizon between $-61<N-N_{\rm inf}<-26$. Then we calculate the value of $r$ at each point and the average $n_s$ between adjacent points. With these data, we can apply condition (\ref{CMB_comp_condition}) to select the CMB-compatible trajectories. To get a sense of the distribution of $n_s$ and $r$ of the whole sample, we plot a snapshot of ($n_s$,$r$) at a fixed moment $N_0$ that $N_0-N_{\rm inf}=-33$ for trajectories with $N_{\rm inf}>33$ in the left panel of Figure \ref{fig:r_ns_1d}, as well as the contour of $95\%$ likelihood of the latest CMB constraints~\cite{Balkenhol:2025wms}. 
 We note that, in most cases, trajectories with $N_{\rm inf}>33$ in a particular realization share a common section of an attractor trajectory from $N_0$ to $N_{\rm inf}$,\footnote{This is particularly true for the one-field case, in which there are only two directions that a trajectory can reach the true vacuum. The situation can be different for $N_{\rm f}>1$, in which trajectories can reach the true vacuum following very different paths, see Appendix \ref{app_gallery} for related discussions.} 
 so we use a circle to represent each realization, and make the area of this circle proportional to the total number of trajectories with $N_{\rm inf}>33$ in this realization. The snapshot can actually be taken at different values of $N_0-N_{\rm inf}$, and the pattern turns out to be very similar (for example, we can choose $N_0-N_{\rm inf}=-50$, and we will find a similar pattern of the $(n_s,r)$ distribution, despite that there are fewer trajectories with $N_{\rm inf}>50$ than $N_{\rm inf}>33$).

From this plot, it is very interesting to notice that the requirement of being successful trajectories alone would give rise to universes comfortably compatible with the observations of $r$ and $n_s$ --- most of these universes are well within the bound of $r$; and, although the constraint from the precisely measured $n_s$ is very stringent, it represents a fairly generic subset in the population.

The search for CMB-compatible trajectories is straightforward. In a short summary, there are $98$ realizations having at least one CMB-compatible trajectory. In terms of trajectories, there are in total $449$ that are CMB-compatible, comprising $1.7\%$ of all successful trajectories. In almost all cases, different CMB-compatible trajectories on a landscape realization share the common pivot values of $r$ and $n_s$. The positions of these trajectories in the $68\%$ and $95\%$ CMB-compatible contour are plotted in the right panel of Figure \ref{fig:r_ns_1d}, in which each point corresponds to a particular realization, with the area and colour of the points corresponding to the number of CMB-compatible trajectories and the pivot scale $N_\star-N$ at which the $k_\star$-modes exit the horizon, respectively. Since the trajectories are required to be CMB-compatible in the entire range of $7$ $e$-folds, most trajectories reside within the more stringent $68\%$ contour in this graph when $n_s$ and $r$ are taken as the central values evaluated at $k_\star$.

\begin{figure}[ht!]
    \centering  
    \includegraphics[width=0.46\textwidth]{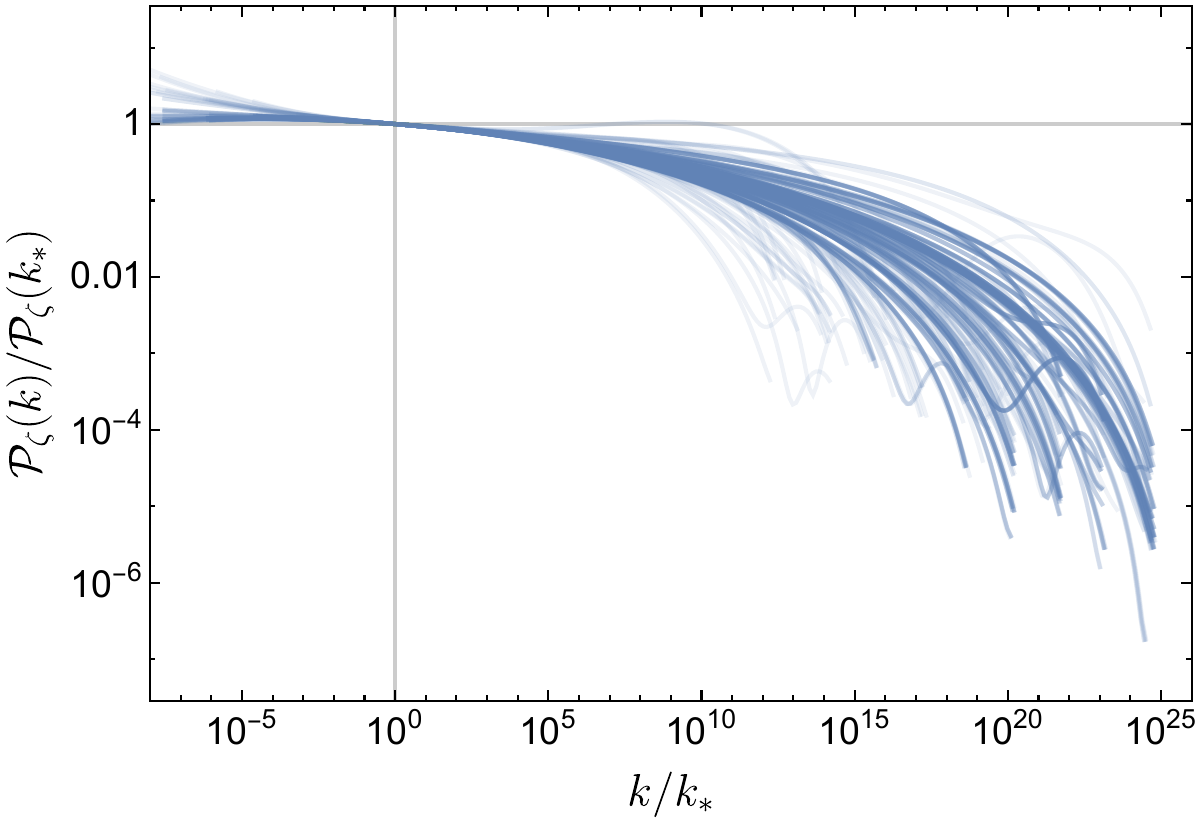}
    \hspace{10pt}
    \includegraphics[width=0.45\textwidth]{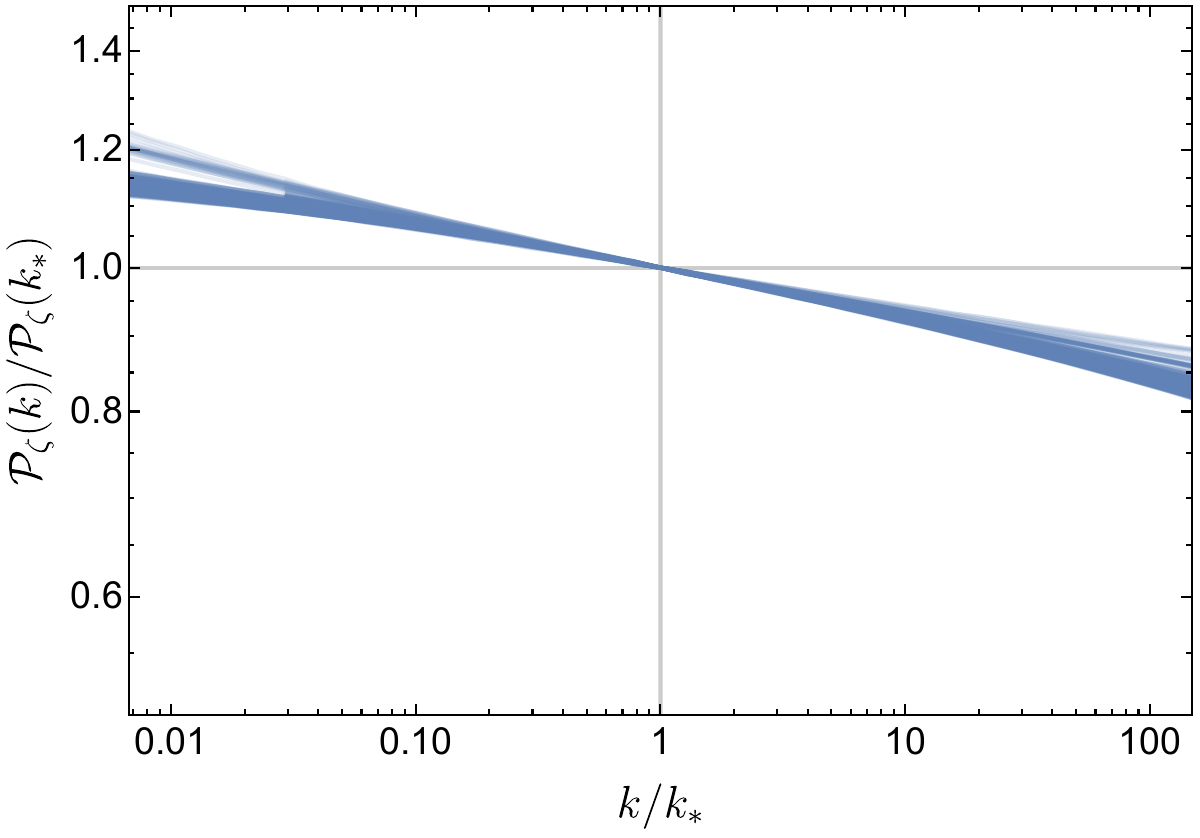}
    \caption{{\em $N_{\rm f}=1$ case} $(\xi=1.32\Mp$, $\etamed=0.38)$. Left panel: The power spectra of all CMB-compatible trajectories. Each trajectory is plotted in a light colour so that the colour is enhanced if there are multiple trajectories found in the same inflation attractor. Right panel: A zoomed-in view of the left panel featuring the vicinity of the pivot scale, where all power spectra satisfy the CMB compatibility constraints.
    }
  \label{fig:PowerSpectra_1d}
\end{figure}

After finding the CMB-compatible trajectories, we can calculate their high resolution power spectra. We plot these power spectra in Figure \ref{fig:PowerSpectra_1d}, in which the values of $P_\zeta(k)$ of each trajectory are normalized at the pivot scale $k_\star$. The full power spectra are shown in the left panel and the CMB-compatible section is highlighted in the right panel. It can be seen that there are a few trajectories showing non-trivial features at small scales, which are related to their multi-stage nature. 
In these landscapes (including the higher-dimensional ones we will study shortly), the most common type of primordial feature generated by multi-stage models is a dip in the power spectrum. This can be understood through the qualitative relation between the curvature perturbation $\zeta$ and the inflaton velocity $\dot\phi$, $\zeta\sim -H\delta\phi/\dot\phi$. During the transition between two adjacent slow-roll stages, the inflaton velocity first increases as it exits the preceding slow-roll stage and rolls down a transient section of steeper potential, and then decreases due to Hubble friction as it enters the subsequent stage.

In this work, however, the number of CMB-compatible trajectories is not sufficiently large to draw meaningful statistical conclusions. 

\section{Two-field landscapes}
\label{sec2field}
In this section, we conduct a comprehensive study on the statistical properties of inflation trajectories on two-field landscapes, where more non-trivial phenomenologies emerge. We will start by the description of basic statistical properties similar to the one-field case, as well as the multi-field property which is new to us. After that, we will take a closer look at the trajectories with non-trivial properties and related phenomenology. We will also discuss the implications of variation of typical potential curvature $|\eta_V|$. At the end of this section, we will study the statistics of the CMB-compatible trajectories.

\paragraph{Numerical setup.} In the two-field case, the samples are constructed in the same manner as the working example of Sec.~\ref{working_example}. To summarize, the realizations are generated according to the set of fiducial parameters (\ref{eq:fiducial-params}), and in each realization we initiate $n=1000$ random points in the 2-ball (disk) of radius $\lambda=3\Lambda/4=7.5\Mp$. The whole sample consists of $10\,000$ realizations and in total $10\,000\,000$ trajectories, among which $1\,214\,786$ (i.e.~$12.15\%$) are admissible. The computation is much more demanding than the one-field case but still quite affordable, requiring $\sim 4\rm h$ to perform the computation using 100 cores on the cluster. 

\subsection{Basic statistical properties}
\label{subsec:2f_basic}
We will study the same types of statistical properties of trajectories, fixing $\etamed=0.38$, as in the one-field case, as well as the new multi-field property. 
These basic properties are summarized in Figure \ref{fig:event-hierarchy_2f}, and we will explain them in more detail in context.


\begin{figure}[t]
\centering
\providecommand{\Mp}{M_{\rm Pl}}
\providecommand{\etamed}{\eta_{\rm med}}
\definecolor{funnel}{RGB}{58,80,112}   

\resizebox{\textwidth}{!}{%
\begin{tikzpicture}[
  x=1cm, y=1cm,
  bar/.style  ={draw=funnel!55, line width=0.3pt},
  lost/.style ={draw=none, fill=black!6},
  name/.style ={font=\small, anchor=east, inner sep=0pt},
  sub/.style  ={font=\scriptsize, color=black!55, anchor=east, inner sep=0pt},
  num/.style  ={font=\footnotesize, anchor=east, inner sep=0pt},
  tick/.style ={font=\scriptsize, color=black!55, anchor=north, inner sep=2pt},
  grid/.style ={draw=black!10, line width=0.3pt},
]

\def\XT{11.34} \def\XA{9.8569} \def\XS{7.1627} \def\XC{3.8351}
\def\bh{0.30}

\foreach \d in {0,...,8}{
  \draw[grid] (1.62*\d, 0.45) -- (1.62*\d, -4.05);
  \draw[draw=black!35, line width=0.4pt] (1.62*\d,-4.05) -- (1.62*\d,-4.16);
  \node[tick] at (1.62*\d, -4.16) {$10^{\d}$};}
\draw[draw=black!35, line width=0.4pt] (0,-4.05) -- (12.96,-4.05);
\node[font=\scriptsize, color=black!55, anchor=north] at (6.48,-4.62)
      {number of trajectories};

\def\y{0}
\node[name] at (-0.35,\y+0.10) {Total};
\node[sub]  at (-0.35,\y-0.17) {the whole sample};
\fill[bar, fill=funnel!20] (0,\y-\bh) rectangle (\XT,\y+\bh);
\node[num] at (\XT-0.16,\y) {10\,000\,000};

\def\y{-1.15}
\node[name] at (-0.35,\y+0.10) {Admissible};
\node[sub]  at (-0.35,\y-0.17) {12.15\% of the sample};
\fill[lost] (\XA,\y-\bh) rectangle (\XT,\y+\bh);
\fill[bar, fill=funnel!38] (0,\y-\bh) rectangle (\XA,\y+\bh);
\node[num]  at (\XA-0.16,\y) {1\,214\,786 \textcolor{black!55}{(12.15\%)}};

\def\y{-2.30}
\node[name] at (-0.35,\y+0.10) {Successful};
\node[sub]  at (-0.35,\y-0.17) {0.26\% of the sample};
\fill[lost] (\XS,\y-\bh) rectangle (\XA,\y+\bh);
\fill[bar, fill=funnel!62] (0,\y-\bh) rectangle (\XS,\y+\bh);
\node[num]  at (\XS-0.16,\y) {26\,389 \textcolor{black!55}{(2.17\%)}};

\def\y{-3.45}
\node[name] at (-0.35,\y+0.10) {CMB-compatible};
\node[sub]  at (-0.35,\y-0.17) {0.0023\% of the sample};
\fill[lost] (\XC,\y-\bh) rectangle (\XS,\y+\bh);
\fill[bar, fill=funnel, draw=funnel] (0,\y-\bh) rectangle (\XC,\y+\bh);
\node[num, text=white] at (\XC-0.16,\y) {233 \textcolor{white!78!funnel}{(0.88\%)}};

\draw[draw=black!15, line width=0.4pt] (-2.95,-5.20) -- (12.96,-5.20);
\node[font=\small, anchor=west] at (-2.95,-5.72)
      {\textcolor{black!45}{$\hookrightarrow$}\; Structure of the
       \textbf{successful} trajectories};

\begin{scope}[shift={(3.5,-8.25)}]
  \fill[draw=white, line width=0.8pt, fill=funnel!25] (90.000:0.92) arc[start angle=90.000, end angle=25.416, radius=0.92] -- (25.416:1.78) arc[start angle=25.416, end angle=90.000, radius=1.78] -- cycle;
  \fill[draw=white, line width=0.8pt, fill=funnel!48] (25.416:0.92) arc[start angle=25.416, end angle=18.288, radius=0.92] -- (18.288:1.78) arc[start angle=18.288, end angle=25.416, radius=1.78] -- cycle;
  \fill[draw=white, line width=0.8pt, fill=funnel!72] (18.288:0.92) arc[start angle=18.288, end angle=-137.808, radius=0.92] -- (-137.808:1.78) arc[start angle=-137.808, end angle=18.288, radius=1.78] -- cycle;
  \fill[draw=white, line width=0.8pt, fill=funnel!100] (-137.808:0.92) arc[start angle=-137.808, end angle=-270.000, radius=0.92] -- (-270.000:1.78) arc[start angle=-270.000, end angle=-137.808, radius=1.78] -- cycle;
  \draw[funnel!45, line width=1.3pt] (90:0.8) arc[start angle=90, end angle=18.288, radius=0.8];
  \draw[funnel!100, line width=2.1pt] (18.288:0.8) arc[start angle=18.288, end angle=-270, radius=0.8];
  \node[font=\footnotesize, align=center] at (0,0.10) {26\,389};
  \node[font=\scriptsize, color=black!55] at (0,-0.16) {successful};
  \draw[draw=black!50, line width=0.35pt] (57.708:1.6) -- (57.708:2.02);
  \node[anchor=west, align=left, inner sep=1.5pt] at (57.708:2.06) {\small SFSS\\[-1pt]\scriptsize\textcolor{black!60}{17.94\%}};
  \draw[draw=black!50, line width=0.35pt] (21.852:1.6) -- (21.852:2.79);
  \node[anchor=west, align=left, inner sep=1.5pt] at (21.852:2.83) {\small SFMS\\[-1pt]\scriptsize\textcolor{black!60}{1.98\%}};
  \draw[draw=black!50, line width=0.35pt] (-59.760:1.6) -- (-59.760:2.02);
  \node[anchor=west, align=left, inner sep=1.5pt] at (-59.760:2.06) {\small MFSS\\[-1pt]\scriptsize\textcolor{black!60}{43.36\%}};
  \draw[draw=black!50, line width=0.35pt] (-203.904:1.6) -- (-203.904:2.02);
  \node[anchor=east, align=right, inner sep=1.5pt] at (-203.904:2.06) {\small MFMS\\[-1pt]\scriptsize\textcolor{black!60}{36.72\%}};
\end{scope}
\node[font=\scriptsize, color=black!55, anchor=north] at (3.5,-10.75) {single-field \textcolor{black}{19.92\%}\quad multi-field \textcolor{black}{80.08\%}\quad multi-stage (SFMS$+$MFMS) \textcolor{black}{38.70\%}};

\node[draw=funnel, line width=0.5pt, fill=funnel!8, rounded corners=2pt,
      anchor=east, align=left, text width=3.2cm, inner sep=5pt, font=\small]
     at (12.96,-8.25) {\textbf{233} CMB-compatible\\[1pt]
      \scriptsize\textcolor{black!60}{0.88\% of the successful trajectories}};

\end{tikzpicture}}
\caption{{\em $N_{\rm f}=2$ case} $(\xi=\Mp$, $\etamed=0.38)$.
Classification of $10^{7}$ trajectories drawn on $10\,000$ landscape
realisations. \emph{Top:} the successive selection stages; bar lengths are
logarithmic and the figure in parentheses is the fraction kept from the
previous stage, whose extent is shown by the shaded continuation of the bar. 
\emph{Bottom:} the successful trajectories resolved by field content and
stage structure. The wedges run clockwise in the order SFSS, SFMS, MFSS,
MFMS, so the two inner arcs group the single-field and the multi-field
categories. SFSS: Single-Field-Single-Stage;
SFMS: Single-Field-Multi-Stage; MFSS: Multi-Field-Single-Stage; MFMS:
Multi-Field-Multi-Stage.}
\label{fig:event-hierarchy_2f}
\end{figure}

\begin{figure}[ht!]
    \centering  
    \includegraphics[width=\textwidth]{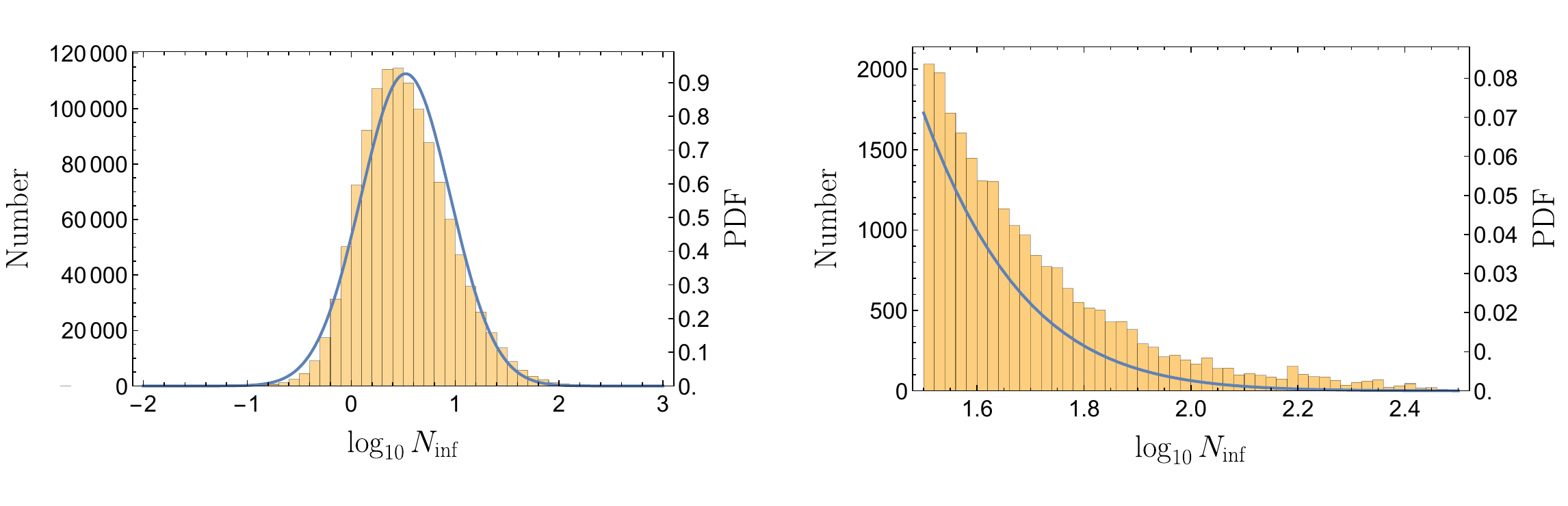}
    \caption{ {\em $N_{\rm f}=2$ case} $(\xi=\Mp$, $\etamed=0.38$).
    Left panel: The combined distribution of $N_{\rm inf}$ of all admissible trajectories from all realizations of the two-field landscape. The distribution is displayed in terms of $\log_{10}N_{\rm inf}$. The blue curve is a fit of the probability density function (PDF) by the normal distribution with mean and standard deviation being $0.52$ and $0.43$, respectively. Right panel: A zoomed-in view of the rightmost tail of the distribution.
    }
  \label{fig:NeDist_xi1_log_2d.pdf}
\end{figure}

\paragraph{Duration of inflation.}
As in the one-field case, the number of admissible trajectories and the distribution of $N_{\rm inf}$ vary a lot between different realizations. Therefore, we study the joint statistics for all admissible trajectories as usual, and the combined distribution of $\log_{10}N_{\rm inf}$ is shown in Figure \ref{fig:NeDist_xi1_log_2d.pdf}. The profile of the PDF can still be approximated by a normal distribution, and the mean and standard deviation take values of $0.52$ and $0.43$, respectively. Compared with the one-field case (which takes values of $0.20$ and $0.67$), we find that the mean value of $N_{\rm inf}$ is larger and the with of the profile is narrower.
We find a similar fraction of successful trajectories among all admissible trajectories. Indeed, there are $2266$ ($22.6\%$) of $10000$ realizations that have at least one successful trajectory, and there are in total $26389$ successful trajectories, comprising $2.17\%$ of all admissible trajectories. We present a gallery of typical successful trajectories in Appendix \ref{app_gallery}, in which one can build an intuition of what these trajectories look like. Moreover, we can see from the right panels of Figure~\ref{fig:NeDist_xi1_log_2d.pdf} that there is an excess of long-lasting trajectories compared to the log-normal prediction.
Therefore, the probability that an admissible trajectory is successful increases as the dimension of the field space increases from one to two, at fixed typical potential curvature.
Heuristically, it seems that higher dimensional landscapes make it easier to achieve successful inflation, and we will return to this topic after we study the three-field case in the next section.

\paragraph{Single and multiple stages.}

\begin{figure}[ht!]
    \centering  
    \includegraphics[width=0.465\textwidth]{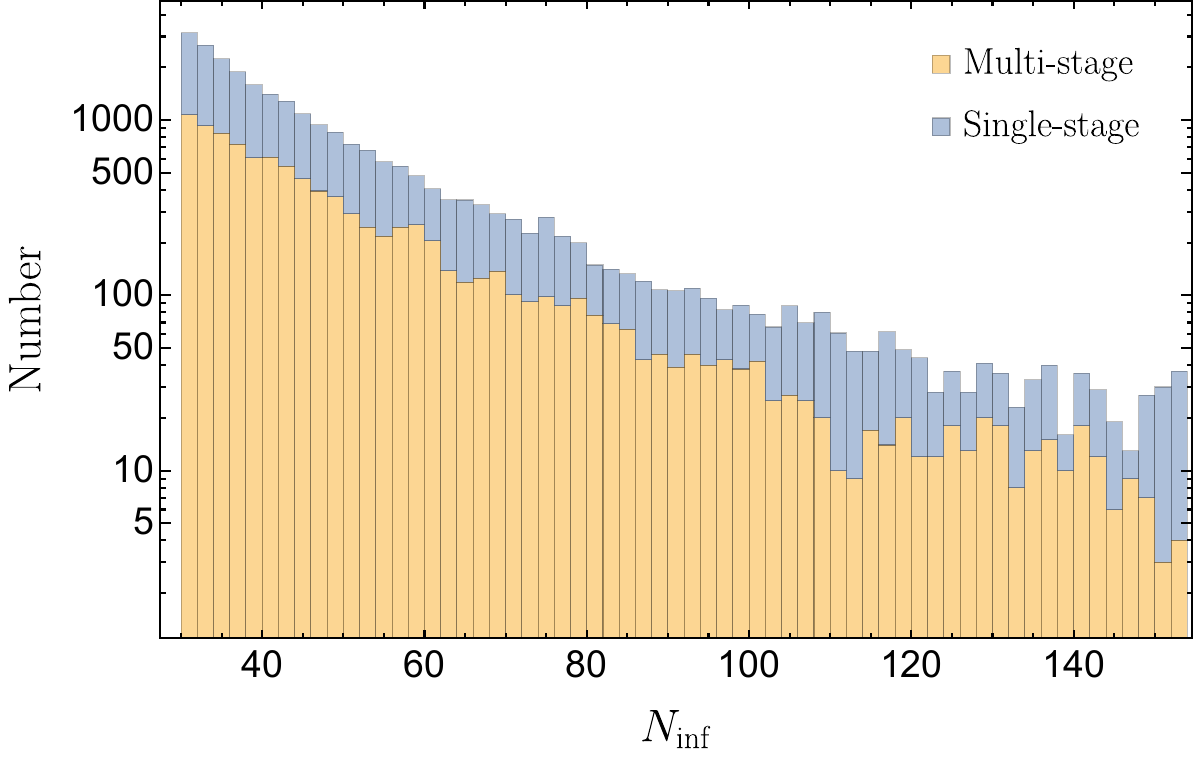}
    \hspace{10pt}
    \includegraphics[width=0.45\textwidth]{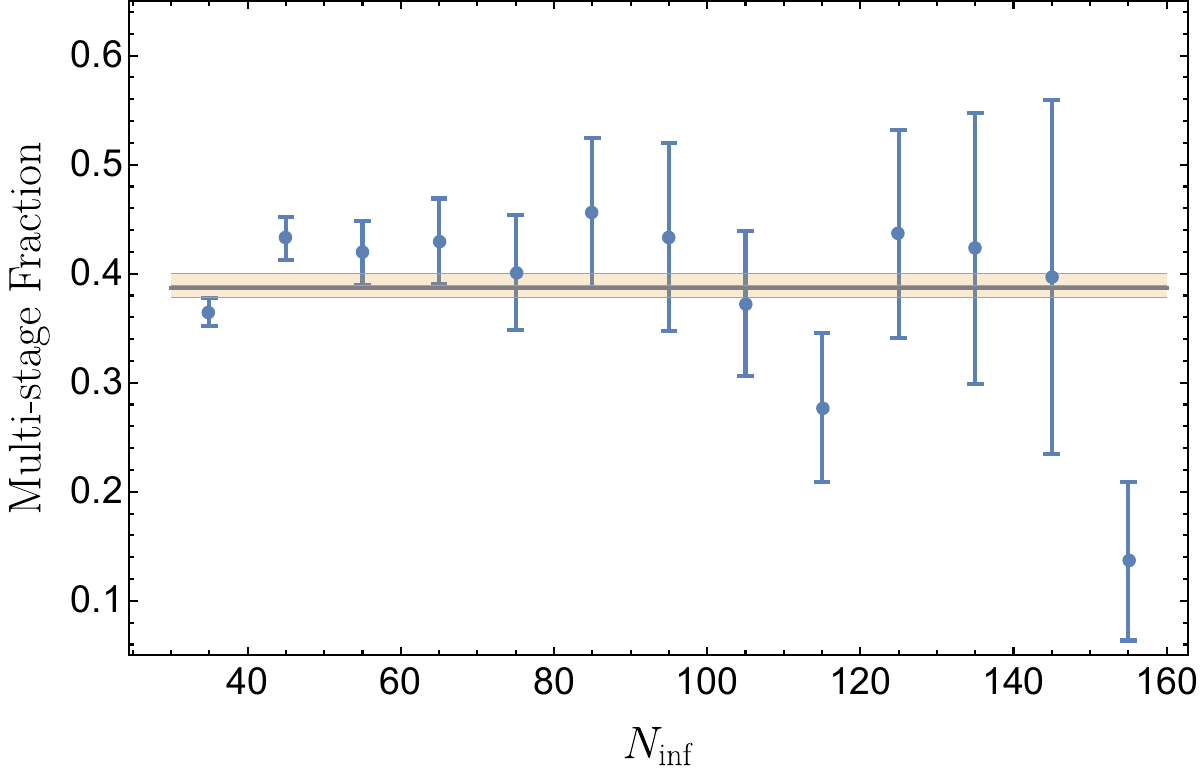}
    \caption{ {\em $N_{\rm f}=2$ case  $(\xi=\Mp$, $\etamed=0.38)$}. Left panel: The binned distribution of numbers of single-stage and multi-stage trajectories, in which the bin width is set to be $\Delta N_{\rm inf}=2$. Right panel: The binned fraction of multi-stage trajectories with bin width $\Delta N_{\rm inf}=10$, in which the error bars are estimated by the statistical bootstrap method. The orange band denotes the overall multi-stage fraction with statistical errors obtained by combining all bins, $(39.0\pm1.1)\%$.
    The gray line is the true overall multi-stage fraction, $38.69\%$.
    }
  \label{fig:Stage_Ne_2d}
\end{figure}

An important and interesting question that we aim to address by moving to higher-dimensional internal field spaces is whether higher dimensionality favours or disfavours multi-stage inflation compared with the lower-dimensional case. As the number of fields increases, there are generically more directions in which the inflaton can fall in a non-slow-roll fashion. This leads to two competing effects. On the one hand, inflation may become easier to end once a given slow-roll stage terminates, reducing the likelihood of multi-stage models. On the other hand, the increased number of non-slow-roll directions may provide more opportunities for generating multi-stage inflation. These effects compete, making it difficult to determine analytically or intuitively which dominates, and by how much. In this work, we use simulations to investigate this highly non-trivial question in Gaussian random landscapes for the first few dimensions, and we speculate possible implications for even higher-dimensional cases.
The statistics of the multi-stage trajectories are shown in Figure \ref{fig:Stage_Ne_2d}.
We find that there is a larger fraction of multi-stage trajectories than in the one-field case. 
In total, there are 10221 multi-stage trajectories, comprising a fraction of $38.7\%$ of all successful trajectories, and after binning of $N_{\rm inf}$, we find that a multi-stage fraction of $\sim40\%$ is nearly invariant for different values of $N_{\rm inf}$.

\begin{figure}[ht!]
    \centering  
    \includegraphics[width=\textwidth]{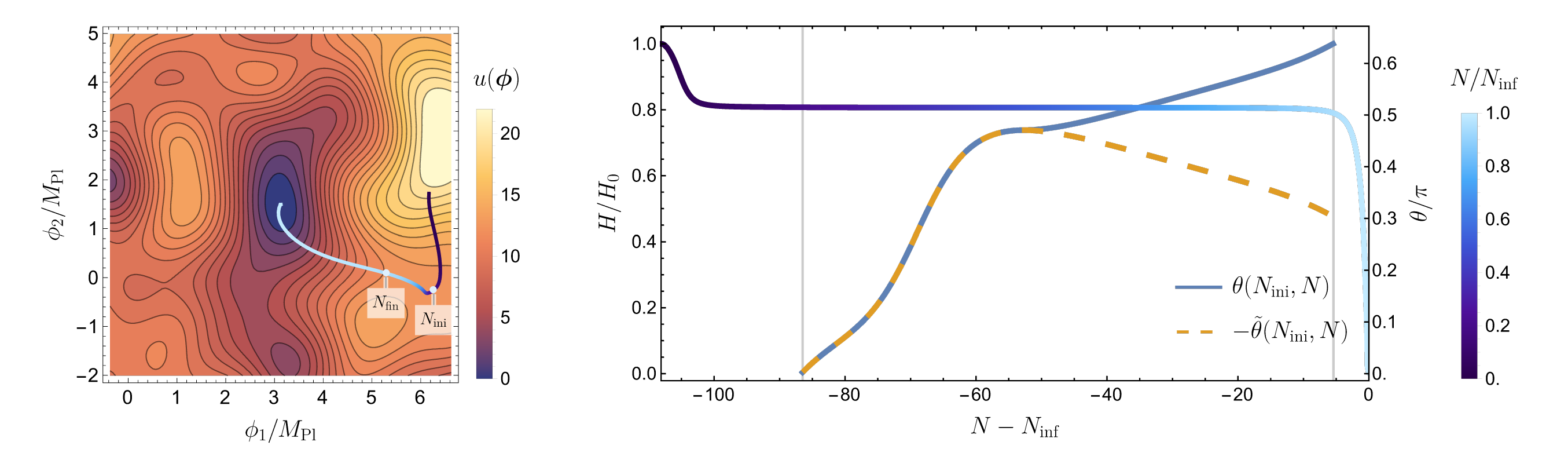}
    \caption{The illustration of geometrical meaning of the angle of turning using the working example of \ref{working_example}. Left panel: The shape of the trajectory on the landscape and two moments $N_{\rm ini}=0.2N_{\rm inf}$ and $N_{\rm fin}=0.95N_{\rm inf}$. Right panel: The evolution of $H(N)$, $\theta(N_{\rm ini}, N)$ and $\tilde{\theta}(N_{\rm ini}, N)$ on the same plot, with the values at the end points of $\theta$ and $\tilde{\theta}$ corresponding to $\Theta=0.64\pi$ and $\tilde{\Theta}=-0.30\pi$. In this particular case, $\tilde{\Theta}$ takes a negative value, so we plot the value of $-\tilde{\theta}$ instead of $\tilde{\theta}$ in this plot.
    }
  \label{fig:Angle_exp}
\end{figure}

\paragraph{Effective single-field and genuine multi-field trajectories.}
The notion of multi-field trajectories starts to play a role for $N_{\rm inf}=2$ and beyond. Each admissible trajectory can be assigned a specific value of the \textit{total} angle of turning $\Theta$ defined as (\ref{Theta}), and the \textit{net} angle of turning $\tilde{\Theta}$ defined as (\ref{tilde_Theta}) which is unique to $N_{\rm f}=2$. In the two dimensional landscape, the geometrical picture of $\Theta$ and $\tilde{\Theta}$ is particularly easy to grab, and we make an illustration using the working example of \ref{working_example} shown in Figure \ref{fig:Angle_exp}. In this example, this trajectory has $\Theta=0.64\pi$ and $\tilde{\Theta}=-0.30\pi$. Intuitively, the trajectory makes two consecutive turns in opposite directions, resulting in the net angle of turning smaller than $\pi/2$ and the total angle of turning larger than $\pi/2$. Since this example satisfies $\Theta>0.1\pi$, it belongs to multi-field trajectories, exhibiting a genuine multi-field nature during inflation.

\begin{figure}[ht!]
    \centering  
    \includegraphics[width=0.45\textwidth]{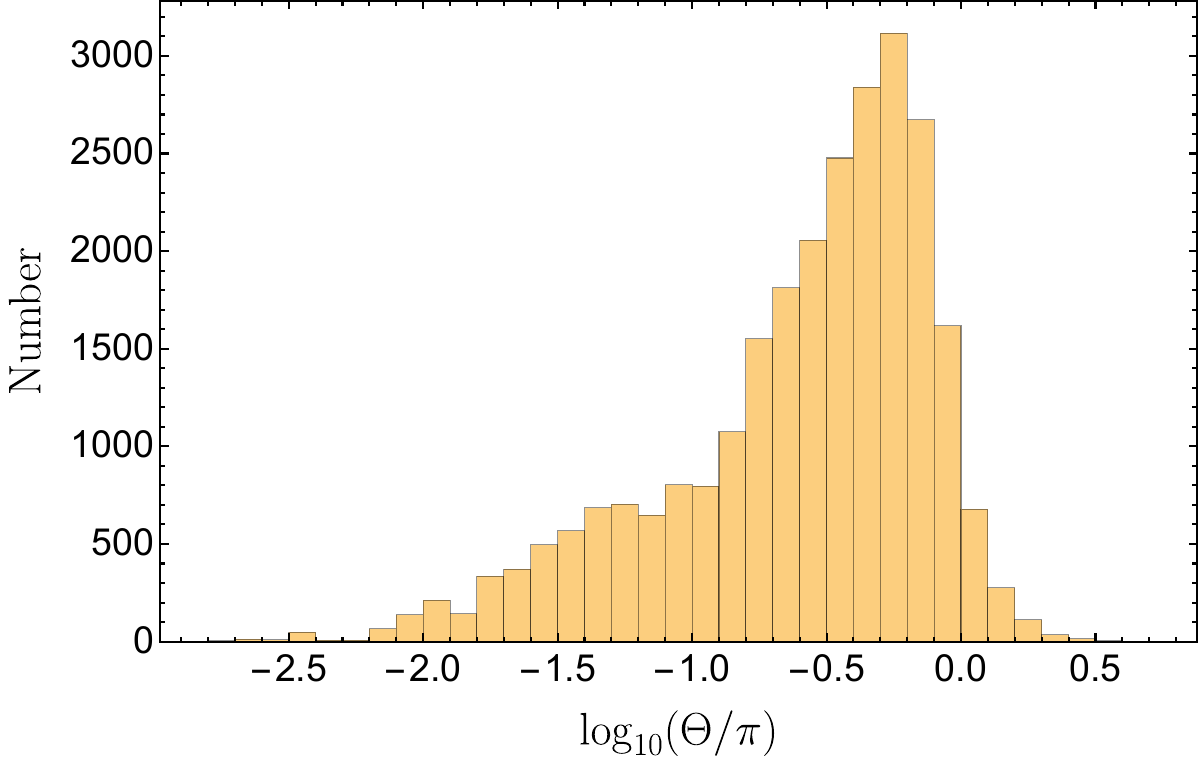}
    \includegraphics[width=0.45\textwidth]{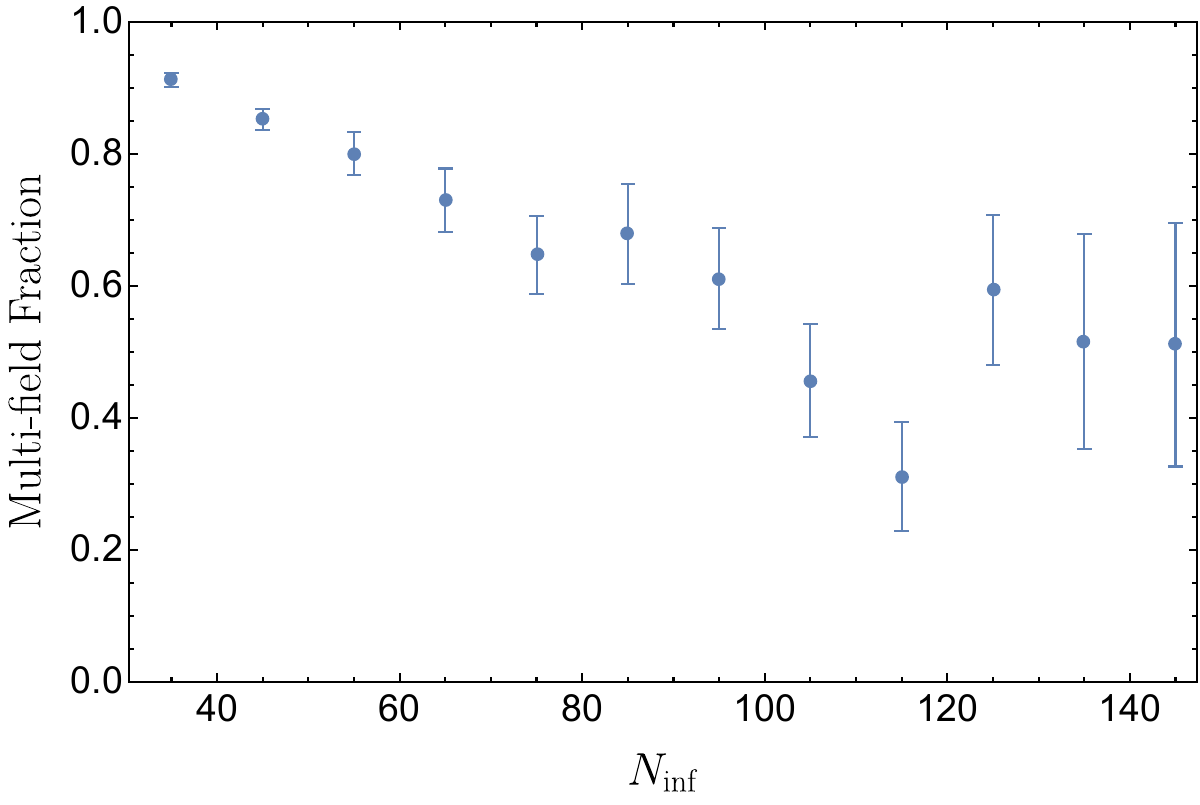}
    \caption{{\em $N_{\rm f}=2$ case $(\xi=1\Mp$, $\etamed=0.38)$}. Left panel: The distribution of $\Theta$ of all trajectories, in which the trajectories with $\log_{10}(\Theta/\pi)>-1$ is categorized as multi-field trajectories. Right panel: The binned fraction of multi-field trajectories with bin width $\Delta N=10$, with error bars estimated by the statistical bootstrap method.
    }
  \label{fig:theta_2d}
\end{figure}

After calculating the value of $\Theta$ of all successful trajectories, we can immediately study the frequency of multi-field trajectories amongst them. We plot the distribution of $\Theta$ in the left panel of Figure \ref{fig:theta_2d}. In total, $21134$ ($80.1\%$) of all successful trajectories have $\Theta>0.1\pi$ and are characterized as multi-field, and the median value of $\Theta$ of is $0.34\pi$. Moreover, we plot the fractions of multi-field trajectories in different bins of $N_{\rm inf}$ in the right panel of Figure \ref{fig:theta_2d}, it indicates a mild decrease of the fraction of multi-field trajectories as $N_{\rm inf}$ increases.

\paragraph{Field-stage cross correlation.}
The discussion on the multi-field and multi-stage properties allows us to ask whether the multi-field and multi-stage properties are correlated. More specifically, all successful trajectories can be classified into four categories: single-field single-stage (SFSS), single-field multi-stage (SFMS), multi-field single-stage (MFSS) and multi-field multi-stage (MFMS). The number and fraction of trajectories in each category is summarized in the rightmost column of Figure \ref{fig:event-hierarchy_2f}. 

In our sample, while SFSS trajectories are not the majority but still comprise a moderate fraction ($17.94\%$), there are relatively few ($1.98\%$) SFMS trajectories, which clearly reflects the fact that a transition between two stages on a multi-dimensional field space is typically associated with a change in the direction of the trajectory. Amongst multi-field trajectories, there are comparable shares of single-stage ($43.36\%$) and multi-stage ($36.72\%$) trajectories. It means that both multi-stage trajectories and single-stage trajectories with curved slow-roll attractors are frequent, both of which have appealing phenomenological implications.

\begin{figure}[ht!]
    \centering  
    \includegraphics[width=0.43\textwidth]{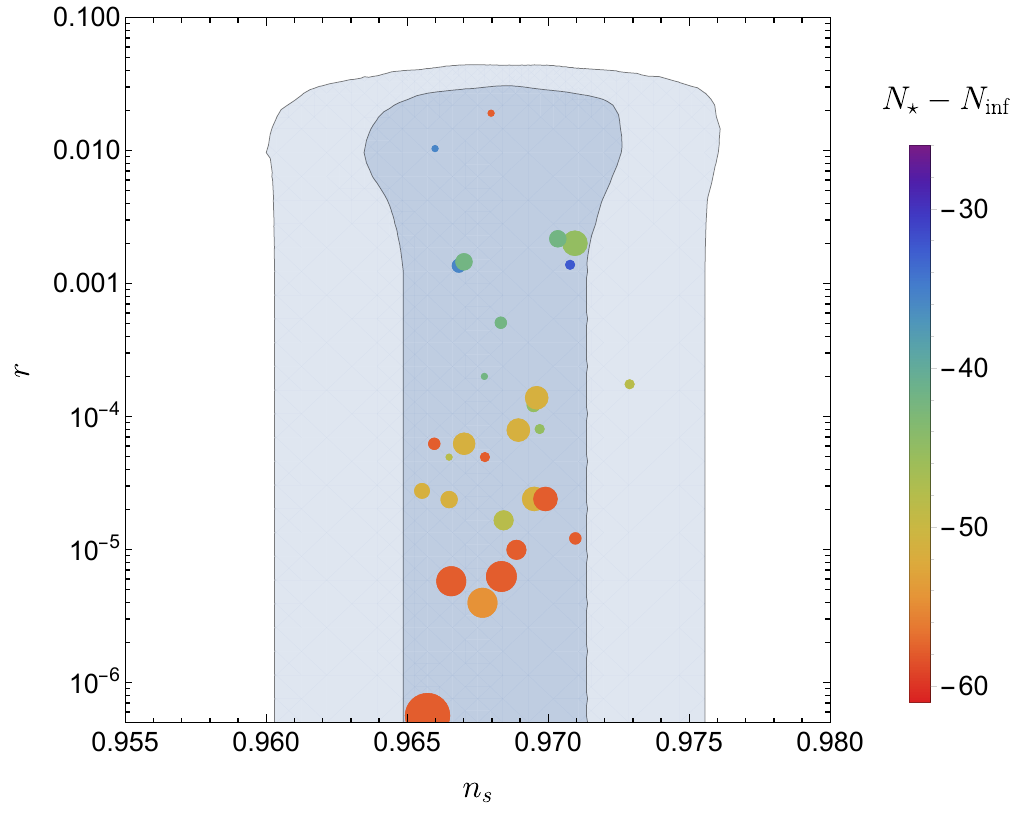}
    \includegraphics[width=0.55\textwidth]{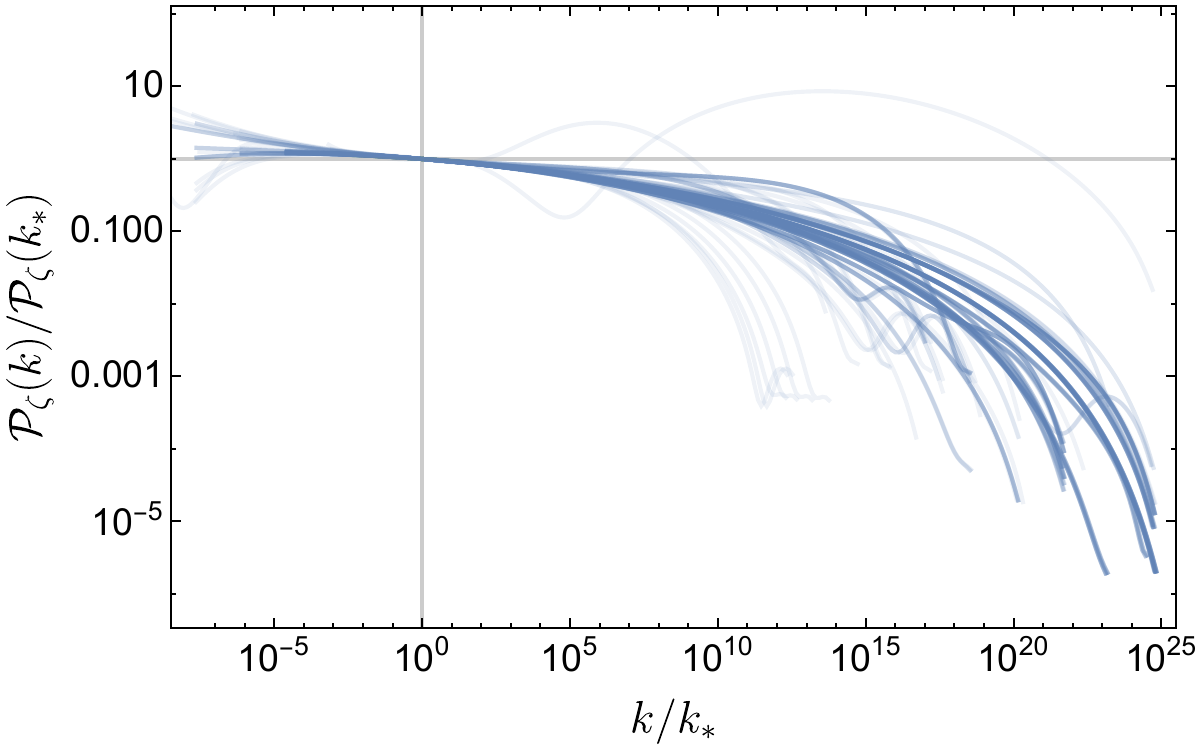}
    \caption{{\em $N_{\rm f}=2$ case} $(\xi=\Mp$, $\etamed=0.38)$. Left panel: The positions on the $r$-$n_s$ plane of all CMB-compatible trajectories as well as the $68\%$ and $95\%$ CMB constraints. Each point denotes a particular realization, with the area proportional to the number of trajectories and the colour corresponding to the pivot scale. Right panel: The power spectra of all CMB-compatible trajectories. Each trajectory is plotted in the thin colour and the colour is enhanced if there are multiple trajectories in the same inflation attractor.
    }
  \label{fig:cmb_compatible_2d}
\end{figure}

\paragraph{CMB compatibility.}
The CMB compatibility search in this case is conducted in the same manner as in the one-field case. After scanning through all successful trajectories, the pattern on the $r$-$n_s$ plane is very much similar to the one-field case (the left panel of Figure \ref{fig:r_ns_1d}), and the condition (\ref{CMB_comp_condition}) selects a total of 233 CMB-compatible trajectories from 29 distinct realizations. The CMB-compatible trajectories comprise a fraction of $0.88\%$, which is very close to the fraction in the one-field case. The distribution of the pivot values of $n_s$ and $r$ of these trajectories are shown in the left panel of Figure \ref{fig:cmb_compatible_2d}, and the corresponding power spectra are shown in the right panel. Among these power spectra, a few exhibit multi-stage features, which have the appearance of a dip and a subsequent slow-roll platform. In particular, there is one trajectory that has a second slow-roll stage whose contribution to the power spectrum is $O(10)$ times larger than that in the CMB scales.

\subsection{Strong slow-roll deviations and primordial features}
\label{subsec:feature}

The multi-dimensional landscape provides a particularly interesting opportunity to investigate the generation of primordial features associated with multi-stage inflation trajectories. During a transition between stages, the scale invariance and slow-roll condition are strongly violated due to sudden changes in the inflaton kinetic energy. Such an event necessarily leaves characteristic imprints on the primordial power spectrum, giving rise to rich phenomenology in cosmological observables.

\begin{figure}[ht!]
    \centering  
    \includegraphics[width=\textwidth]{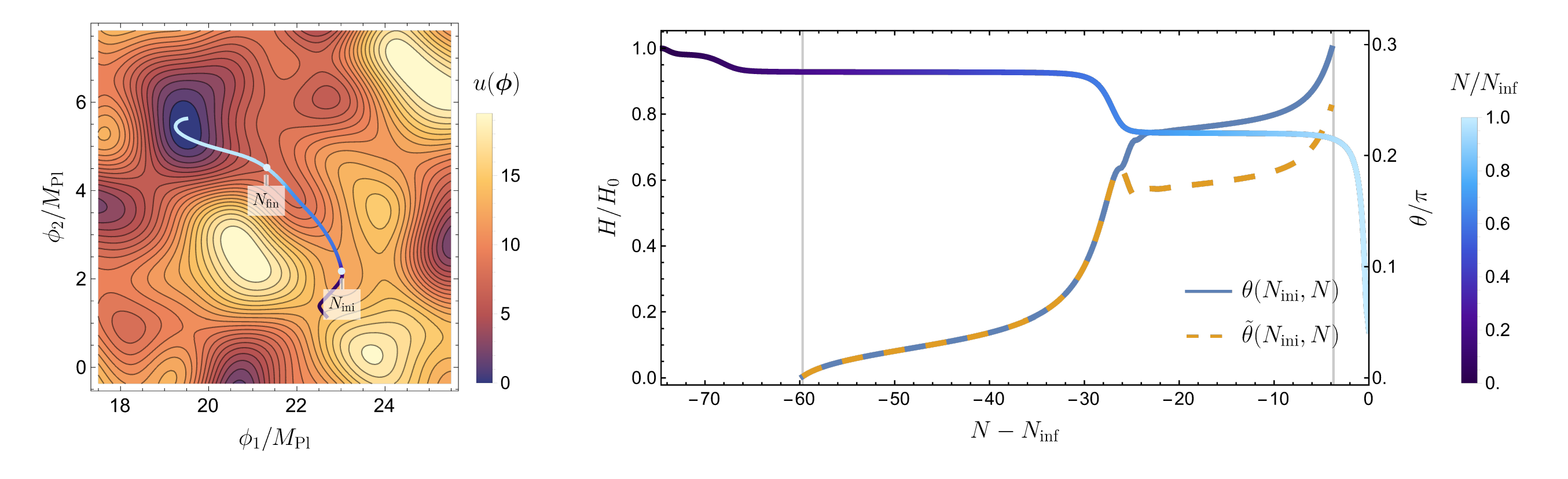}
    \includegraphics[width=0.8\textwidth]{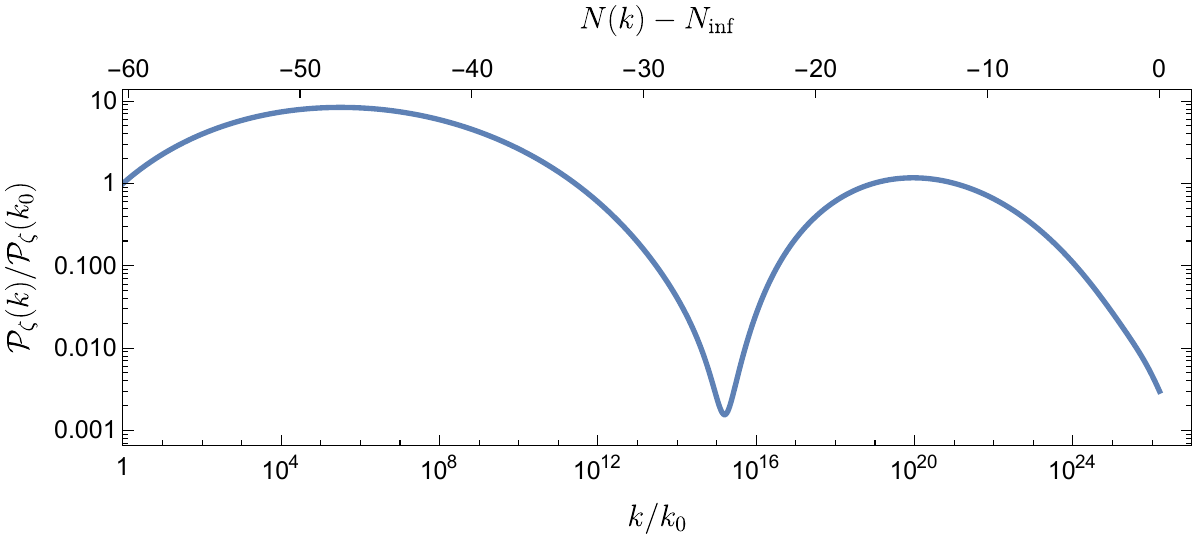}
    \hspace{15pt}
    \caption{An example of $N_{\rm f}=2$ trajectory with two stages. Upper left panel: the shape of the trajectory on the landscape. Upper right panel: The evolution of $H(N)$, $\theta(N_{\rm ini}, N)$ and $\tilde{\theta}(N_{\rm ini},N)$ of this trajectory. Lower panel: The power spectrum of this trajectory. Since this trajectory is not CMB-compatible (because the e-fold span of a potentially compatible region is too short) and does not have the notion of pivot scale $k_\star$, the power spectrum is normalized at the longest mode $k_0$, which is also the convention in Figures \ref{fig:ms2_ps} and \ref{fig:feature_exp}.
    }
  \label{fig:ms1_ps}
\end{figure}

\paragraph{General transitions.} The most common type of primordial features observed in our sample are dips connecting two nearly scale-invariant plateaux, which is the basic feature produced by multi-stage trajectories on the type of landscapes we study in this paper. An example of such trajectories is shown in Figure \ref{fig:ms1_ps}, in which we show the basic information of the trajectory as well as its power spectrum. This trajectory is a typical two-stage trajectory, with a not-too-sharp transition at 30 $e$-folds before the end of inflation. This transition generates a prominent dip in the power spectrum, where the amplitude $\mathcal{P}_\zeta$ drops to $\sim0.1\%$ of the slow-roll level at the minimum. As we also discussed in the $N_{\rm f}=1$ case, such a primordial features is characteristic of multi-stage inflation in our landscapes. Indeed, many power spectra of this type can be found among the CMB-compatible subset of trajectories, as shown in Figures \ref{fig:PowerSpectra_1d}, \ref{fig:cmb_compatible_2d}, and \ref{fig:cmb_compatible_3d}.

\begin{figure}[ht!]
    \centering  
    \includegraphics[width=\textwidth]{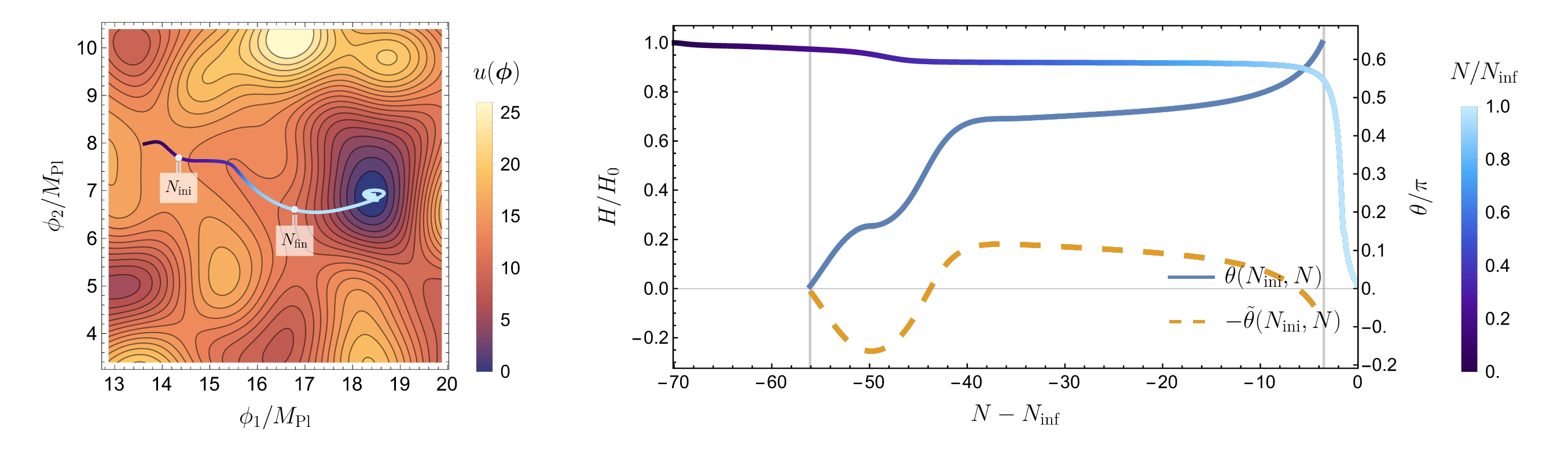}
    \includegraphics[width=0.8\textwidth]{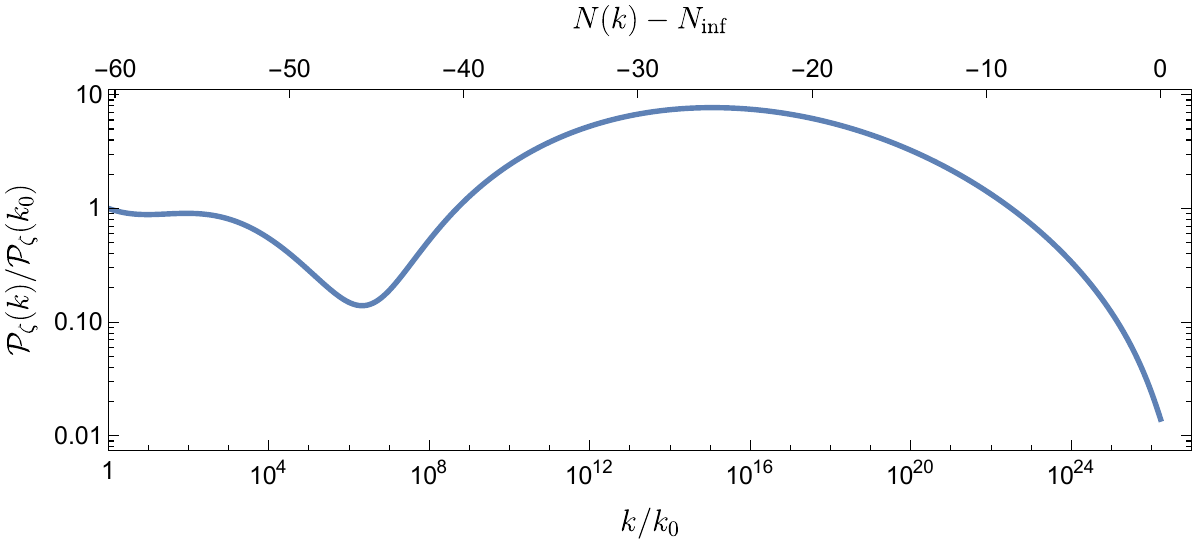}
    \hspace{15pt}
    \caption{An example of $N_{\rm f}=2$ trajectory with enhanced small-scale perturbations. Upper left panel: the shape of the trajectory on the landscape. Upper right panel: The evolution of $H(N)$, $\theta(N_{\rm ini}, N)$ and $\tilde{\theta}(N_{\rm ini},N)$ of this trajectory. Lower panel: The power spectrum of this trajectory, in which the second stage exhibit an $O(10)$ enhancement.
    }
  \label{fig:ms2_ps}
\end{figure}

\paragraph{Small-scale enhancement.} In most cases, the value of $\mathcal{P}_\zeta$ of the subsequent stage is smaller than the preceding one. This may be understood from the schematic estimate $P_\zeta\sim H^2/(8\pi^2\epsilon\Mp^2)$, because the subsequent stage typically has a lower value of $H$ and/or higher value of $\epsilon$ (due to higher inflaton velocity).
However, there are rare cases in which $\mathcal{P}_\zeta$ is enhanced after the transition due to dramatic loss of the inflaton velocity, an example of which is shown in Figure \ref{fig:ms2_ps}. In this example, the value of $\mathcal{P}_\zeta$ of the second stage is enhanced by $\sim10$ times that of the first stage. The enhancement of small-scale density perturbations can potentially enhance gravitational collapse, providing a possible mechanism for the formation of primordial compact objects.

\paragraph{Oscillatory features.} Even though we have not explicitly introduced fields with different mass hierarchies in these potential landscapes, in rare cases we already find inflationary trajectories with \textit{oscillatory} features induced by classical oscillations of heavy fields, also known as primordial standard clock signals~\cite{Chen:2011zf,Chen:2014cwa}. In this scenario, a multi-stage trajectory experiences a sharp transition during which the slow-roll condition is strongly violated and the inflaton acquires a large kinetic energy. At the beginning of the new stage, this motion can excite a massive mode orthogonal to the inflaton direction, causing it to oscillate around its minimum until the oscillations are damped away. These oscillations can in turn generate oscillatory features in the primordial power spectrum.

A useful way to find such trajectories is to use the fact that oscillations introduce rapid growth in $\theta$.
Therefore, we expect that such oscillations are likely to be found among trajectories with $\Theta\gtrsim2\pi$.
However, the fact that $\Theta$ is large alone does not guarantee the existence of such features.
For the purpose of this work, we simply check the properties of all such trajectories by hand to find such cases.

\begin{figure}[ht!]
    \centering  
    \includegraphics[width=\textwidth]{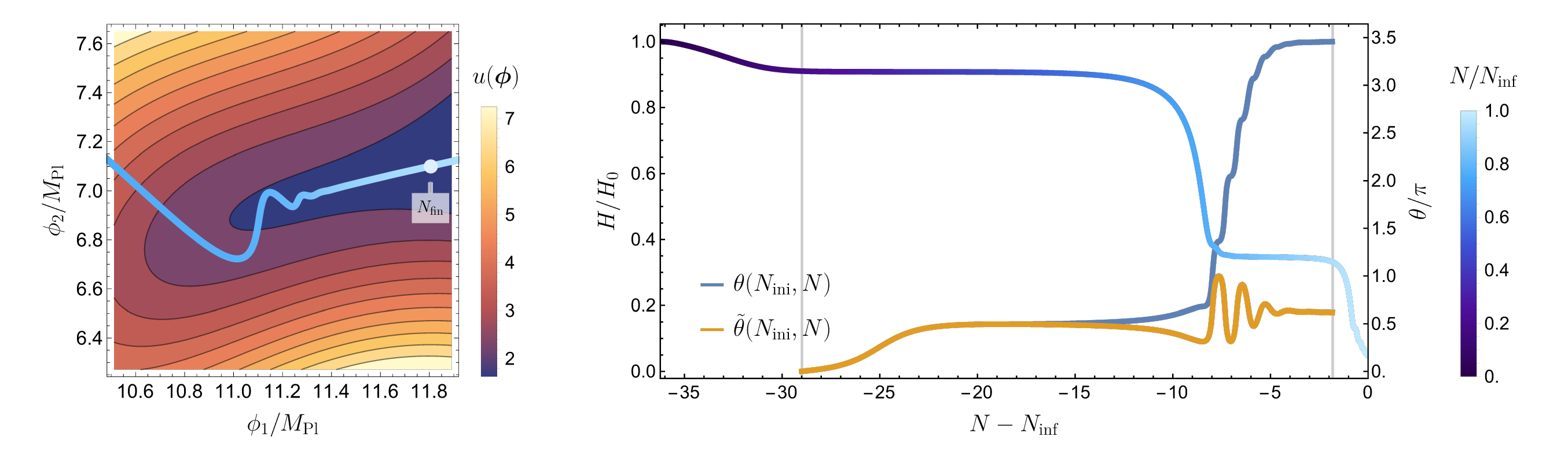}
    \includegraphics[width=0.8\textwidth]{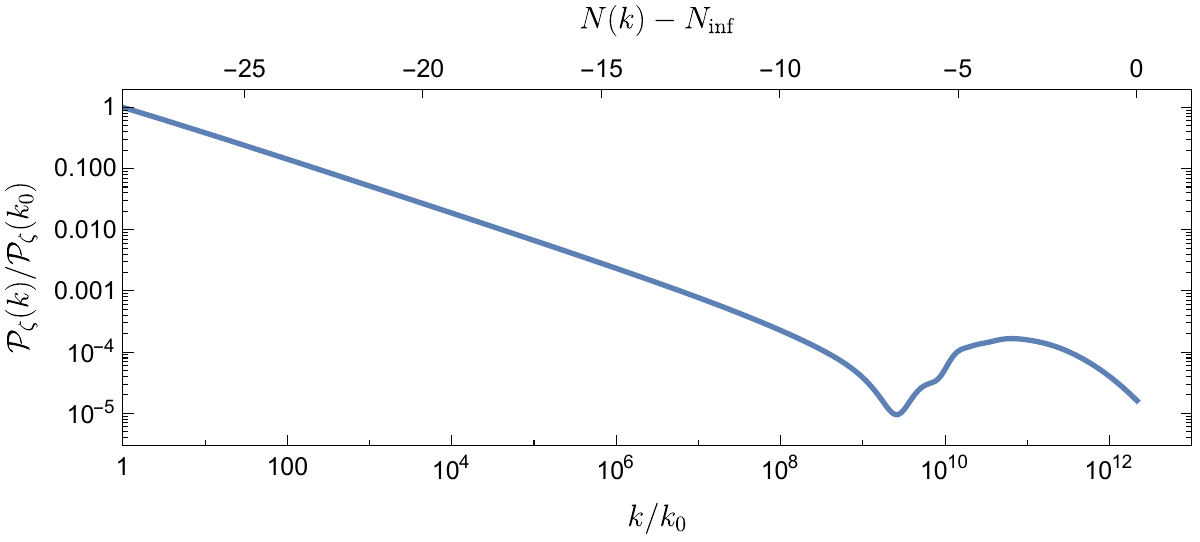}
    \hspace{15pt}
    \caption{An example of $N_{\rm f}=2$ trajectory with oscillatory primordial features. Upper left panel: the shape of the trajectory on the landscape at the location where the oscillation in the massive direction appears. Upper right panel: The evolution of $H(N)$, $\theta(N_{\rm ini}, N)$ and $\tilde{\theta}(N_{\rm ini},N)$ of this trajectory. The oscillation is clearly visible in terms of $\theta$ and $\tilde{\theta}$ between 10 to 5 $e$-folds before the end of inflation. Lower panel: The power spectrum of the trajectory, in which there are oscillatory features on modes that exit the horizon when the background dynamics exhibits oscillations.
    }
  \label{fig:feature_exp}
\end{figure}

An example found in the two-field samples is shown in Figure \ref{fig:feature_exp}. In this example, a transition occurs around $8$ $e$-folds before the end of inflation and induces oscillations that persist for several $e$-folds. The upper-left panel clearly shows that the oscillations occur along the massive direction perpendicular to the slow-roll attractor and gradually damp away due to Hubble friction. In the upper-right panel, one can clearly see a rapid accumulation of $\theta$ as the trajectory oscillates.
As in a typical multi-stage trajectory, the power spectrum of this model exhibits two slow-roll plateaux separated by a dip feature. The oscillatory features begin at the right edge of the dip.

\subsection{Varying the typical potential curvature}
\label{subsec:vary}
In the two-field setup, we now investigate how the statistics change as $\xi$, or equivalently $\etamed$, is varied, since our statistical analyses and comparisons so far have been based on a fiducial choice of $\xi$ (or $\etamed$). The sensitivity of the statistics to variations in $\xi$ also affects the robustness of some of the conclusions drawn from comparisons between cases with different field-space dimensions $N_{\rm f}$.
For this purpose, we conduct two new sets of samples with $\xi=\sqrt{2}\Mp$ and $\xi=\Mp/\sqrt{2}$, keeping all other settings the same as the previous sample of $\xi=\Mp$. In terms of $\etamed$, the three sets of samples with $\xi/\Mp=\{\sqrt{2}, 1, 1/\sqrt{2}\}$ have $\etamed=\{0.19,0.38, 0.75\}$ (obtained by numerical sampling).
We now compare statistical properties of these samples.

\paragraph{Duration of inflation.} If all solutions could be approximated as attractor solutions with negligible kinetic energy, \textit{i.e.} the $\ddot\phi$ terms in the equations of motion are negligible and $\epsilon\ll1$, then landscape statistics for different values of $\etamed$ can be related by a simple rescaling of the landscape, $\phi \to c\phi$, $\Lambda\to c\Lambda$, $\xi \to c\xi$, $\etamed \to \etamed/c^2$, $u\to u$, $H\to H$ and $N\to c^2 N$, which is evident from Eq.~(\ref{eq:eom}). 
This rescaling symmetry is broken if trajectories are not entirely slow-roll attractors. 
On the other hand, if we are only interested in trajectories with a large number of $e$-folds, we expect this scaling symmetry to approximately hold, despite the fact that multi-stage inflation necessarily breaks the attractor approximation. Thus it may be used to understand certain statistical properties of landscapes with different $\etamed$. 

\begin{figure}[ht!]
    \centering  
    \includegraphics[width=0.46\textwidth]{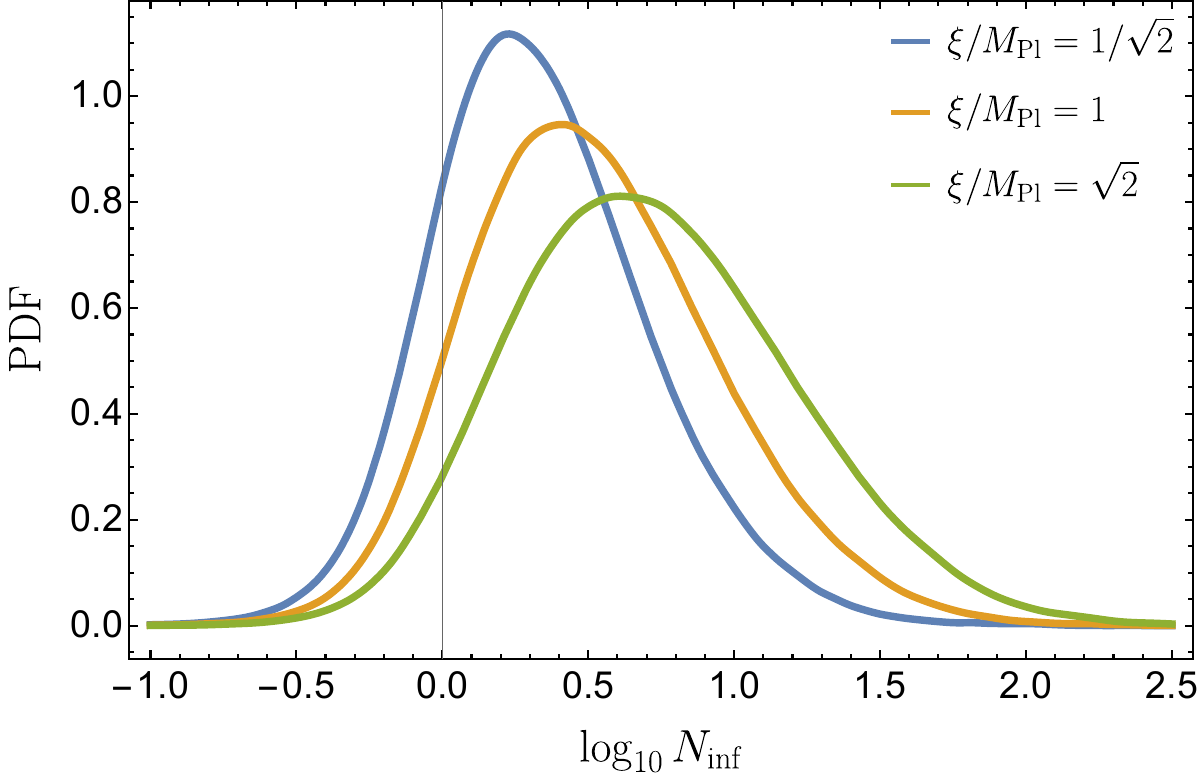}
    \hspace{10pt}
    \includegraphics[width=0.46\textwidth]{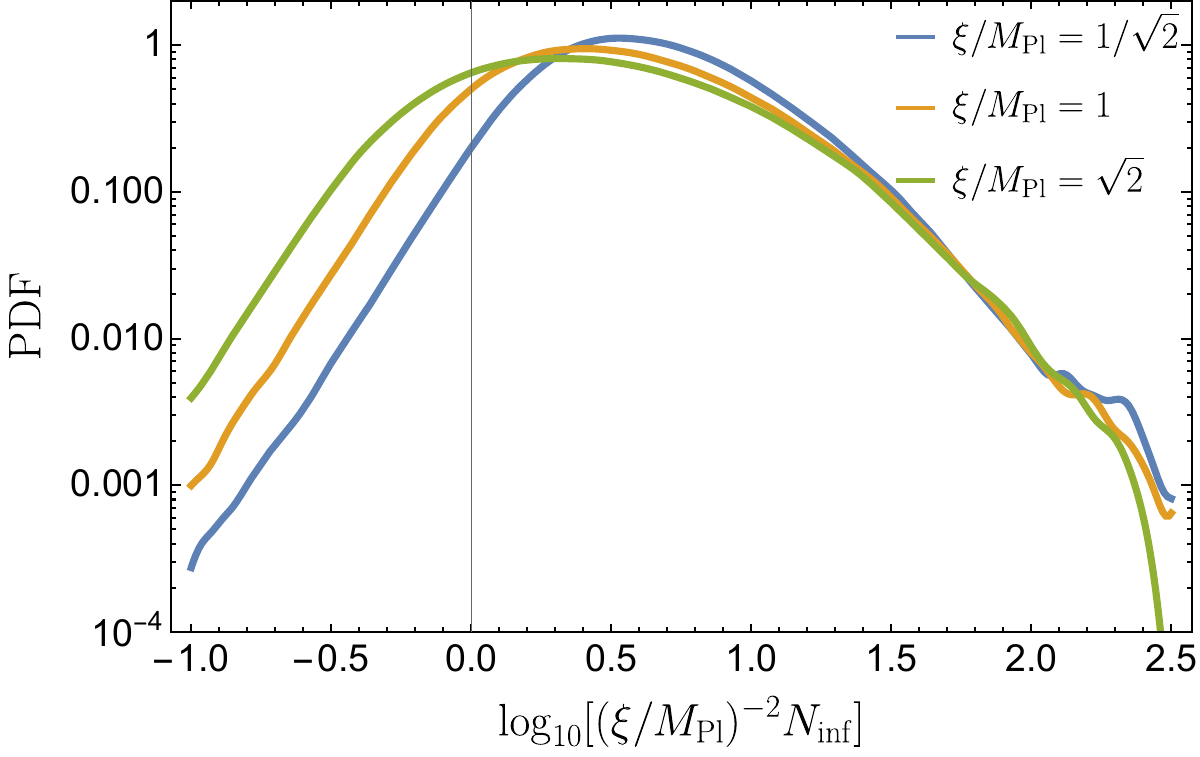}
    \caption{ The comparison of the distribution of $e$-folding numbers $N_{\rm inf}$ of $N_{\rm f}=2$ samples with different values of $\xi$. Left panel: The PDFs of $\log_{10}N_{\rm inf}$ of different samples. Right panel: The PDFs of $\log_{10}\left[(\xi/\Mp)^{-2}N_{\rm inf}\right]$ of different samples plotted in log scale, in which we find that the scaling $N_{\rm inf}\sim\xi^2$ is approximately true for long-lasting trajectories with $(\xi/\Mp)^{-2}N_{\inf}\gtrsim10$.
    }
  \label{fig:NDist_varying}
\end{figure}

A justification and visualization of this scaling behaviour is shown in Figure \ref{fig:NDist_varying}. In the left panel, we plot the PDFs of $\log_{10}N_{\rm inf}$ of all trajectories in each sample, clearly showing that $N_{\rm inf}$ increases with larger $\xi$ (smaller $\etamed$). The numbers of successful ($N_{\rm inf}>30$) trajectories are $6416$ ($0.53\%$) for $\xi/\Mp=1/\sqrt{2}$, $26389$ ($2.2\%$) for $\xi/\Mp=1$ and $84491$ ($6.9\%$) for $\xi/\Mp=\sqrt{2}$. To further demonstrate the approximate scaling property $N_{\rm inf}\propto \xi^2 \propto \etamed^{-1}$, we calculate the PDFs of $\log_{10}\left[(\xi/\Mp)^{-2}N_{\rm inf}\right]$, which are shown in the right panel. We find that the distribution shows different behaviours at different values of $(\xi/\Mp)^{-2}N_{\rm inf}$. 
For the long-lasting trajectories with $(\xi/\Mp)^{-2}N_{\rm inf}>10$, the probability of trajectories with fixed $(\xi/\Mp)^{-2}N_{\rm inf}$ is nearly independent of $\xi$. This is consistent with the scaling symmetry in the slow-roll limit we mentioned above.
On the other hand, as $\xi$ decreases (i.e.~$\etamed$ increases, therefore potentials are steeper), the probability of trajectories as a function of  $(\xi/\Mp)^{-2}N_{\rm inf}$ redistributes, favouring larger values of $N_{\rm inf}$ compared with the scaling behavior in the slow-roll limit, due to contributions from fast-roll models where the kinetic energy becomes an important contributor to $N_{\rm inf}$.

\paragraph{Multi-stage and multi-field properties.}
An interesting question about varying $\xi$ is how it affects the multi-stage and multi-field statistics. 

\begin{figure}[ht!]
    \centering  
    \includegraphics[width=0.7\textwidth]{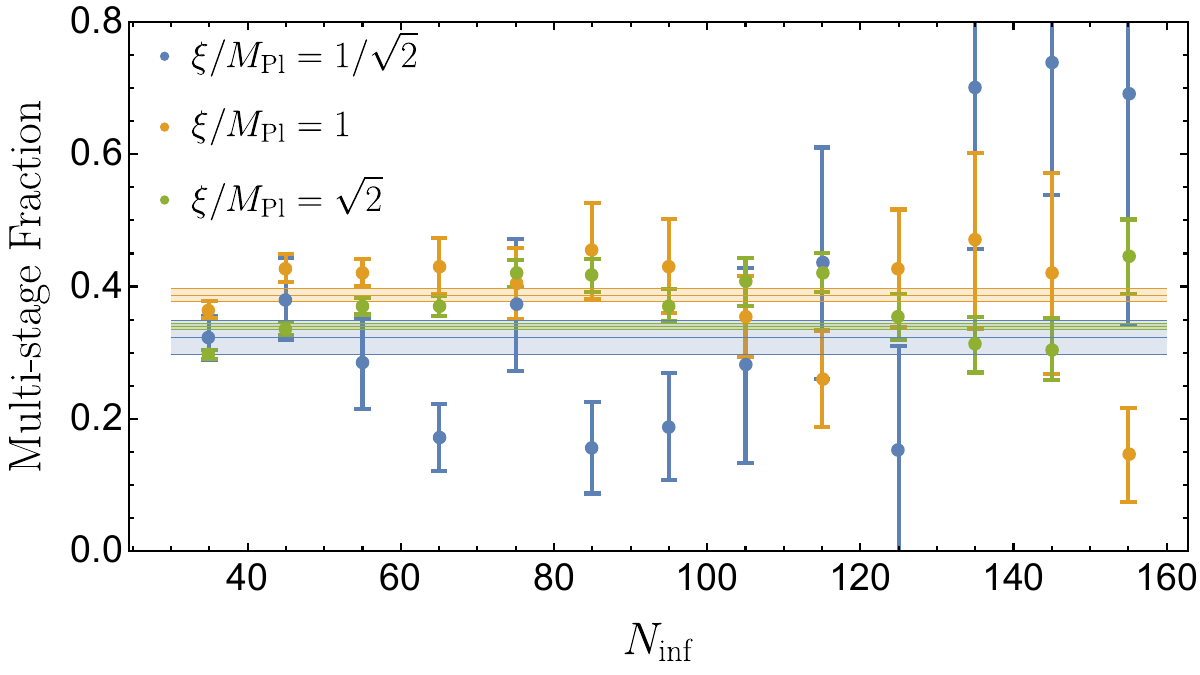}
    \caption{The binned fraction of multi-stage trajectories for $N_{\rm f}=2$ samples with different $\xi$ with bin width $\Delta N_{\rm inf}=10$. The horizontal bands are the total multi-stage fraction and the corresponding error of each sample.
    }
  \label{fig:Stage_Ne_compare}
\end{figure}

Let us first look at the effects on the multi-stage models. 
In Figure \ref{fig:Stage_Ne_compare}, we show the binned multi-stage fraction among successful trajectories for different values of $\xi$, as well as the overall multi-stage fractions which are $32.1\%$ for $\xi/M_{\rm Pl}=1/\sqrt{2}$, $38.7\%$ for $\xi/M_{\rm Pl}=1$, and $33.9\%$ for $\xi/M_{\rm Pl}=\sqrt{2}$.
Although statistical variations increase for smaller $\xi$ due to the smaller total number of successful trajectories, a significant fraction of multi-stage trajectories persists robustly across a range of $N_{\inf}$ and for different choices of $\xi$. 
Also, the observation that the overall fractions vary only weakly with $\xi$ may be understood qualitatively as follows.
Most multi-stage models have at least one prominent slow-roll phase.
Since the rescaling mentioned earlier in this subsection does not change the number of slow-roll attractors and we have observed that the multi-stage fraction varies weakly with $N_{\rm inf}$, it then follows that the multi-stage fraction also varies weakly with $\xi$.

The behavior of multi-field trajectories is somewhat more subtle. On the other hand, as we have seen earlier, the fraction of multi-field trajectories appears to have a mild decrease as $N_{\rm inf}$ gets larger for fixed $\xi$. Therefore, we would expect a statistically smaller $\Theta$ for successful trajectories in samples with smaller $\xi$, due to the scaling $N_{\rm inf}\sim\xi^2$. On the other hand, increasing $\etamed$ has the effect of increasing the kinetic energy of the inflaton and intensifying the transitions between stages due to steeper potentials. This makes it easier to stimulate oscillatory features at the end of transitions, giving rise to statistically larger $\Theta$ for multi-field inflations. 

\begin{figure}[ht!]
    \centering  
    \includegraphics[width=0.46\textwidth]{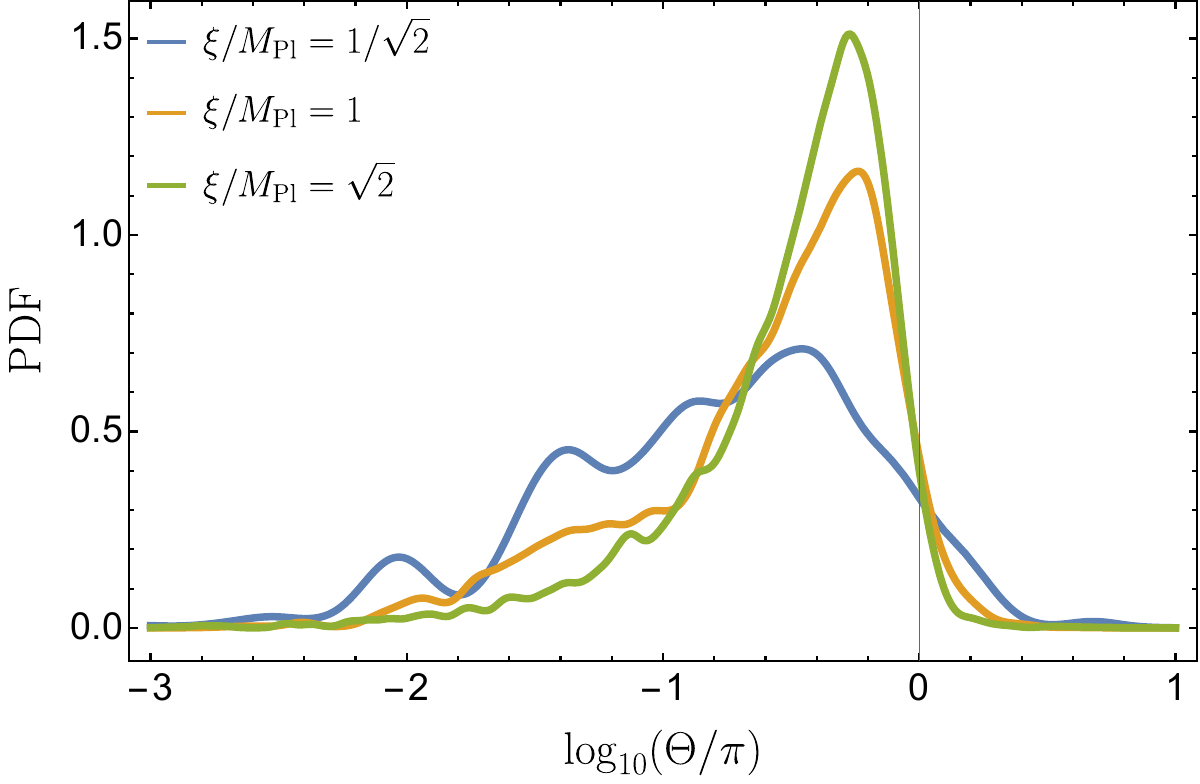}
    \hspace{10pt}
    \includegraphics[width=0.46\textwidth]{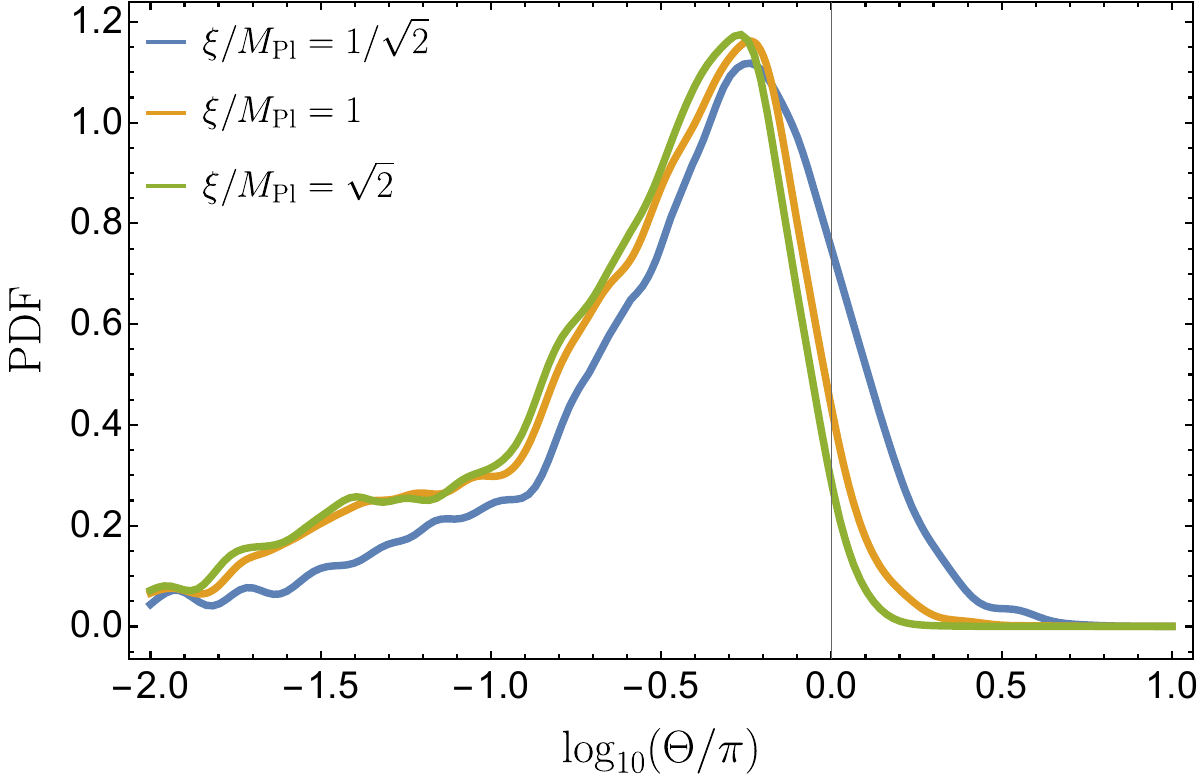}
    \caption{Left panel: The PDFs of $\log_{10}(\Theta/\pi)$ of successful trajectories for all three samples with different values of $\xi$. Right panel: The PDF of $\log_{10}(\Theta/\pi)$ of successful trajectories in the $\xi/\Mp=1$ sample, together with that of the \textit{same} set of trajectories but with $\xi$ rescaled to $\xi/\Mp=1/\sqrt{2}$ and $\xi/\Mp=\sqrt{2}$.
    }
  \label{fig:logTheta_varying}
\end{figure}

Both of the two phenomena are visible in the left panel of Figure \ref{fig:logTheta_varying}, where we show the PDFs of $\log_{10}(\Theta/\pi)$ of successful trajectories for all three samples. We find that as $\xi$ increases, the median values of $\Theta/\pi$ are 0.19, 0.34 and 0.40, respectively, and the peak of the distribution becomes more concentrated in the range $0.1<\Theta/\pi<1$, consistent with the expectation by the scaling behaviour. On the other hand, there is an increase in the population at $\Theta/\pi>1$ for smaller $\xi$, which is a clear evidence that larger $\etamed$ makes highly oscillatory trajectories (which typically contribute to large values of $\Theta$) more frequent. 

To make the second effect more manifest, we can perform a rescaling of $\xi$ to the previous $\xi/\Mp=1$ sample and calculate $\Theta$ of the trajectories obtained by rescaling the initial conditions of the \textit{successful} trajectories in the original sample and recomputing their trajectories in the rescaled landscape.
To explain this procedure in more details, we rescale the landscape realizations by $\boldsymbol{\phi}\to c\boldsymbol{\phi}$ so that $\xi$ is rescaled by $\xi=c \xi$, and rescale the initial conditions of all successful trajectories by $\boldsymbol{\phi}_{\rm{ini}}\to c\boldsymbol{\phi}_{\rm ini}$ at the same time. Then we solve the equations of motion in the rescaled landscape with rescaled initial conditions, and collect the values of $\Theta$. 
In the limit of vanishing kinetic energy, the shape of a trajectory will be invariant under the rescaling, and the value of $\Theta$ should also be invariant. On the other hand, if the kinetic energy cannot be neglected, the rescaled trajectory can have a slightly different value of $\Theta$. More specifically, if $\xi$ becomes smaller ($\etamed$ becomes larger), the inflaton will get a larger kinetic energy, triggering more turning trajectories and lead to a larger $\Theta$. Therefore, the distribution of $\Theta$ is expected to have a shift toward larger $\Theta$ for smaller $\xi$ due to the role of the kinetic energy.

This phenomenon is verified in the right panel of Figure \ref{fig:logTheta_varying}, where we show the distribution of $\log_{10}(\Theta/\pi)$ of successful trajectories in the $\xi/\Mp=1$ sample together with two different rescalings, in which it is evident that smaller $\xi$ features a statistical increase in $\Theta$.

\begin{figure}[t]
\centering
\providecommand{\Mp}{M_{\rm Pl}}
\providecommand{\etamed}{\eta_{\rm med}}
\definecolor{funnel}{RGB}{58,80,112}   

\resizebox{\textwidth}{!}{%
\begin{tikzpicture}[
  x=1cm, y=1cm,
  grid/.style   ={draw=black!10, line width=0.3pt},
  axline/.style ={draw=black!35, line width=0.4pt},
  tick/.style   ={font=\scriptsize, color=black!55, anchor=north, inner sep=2pt},
  axtitle/.style={font=\scriptsize, color=black!55, anchor=north},
  ptitle/.style ={font=\small, anchor=west},
  rname/.style  ={font=\small, anchor=east, inner sep=0pt},
  rsub/.style   ={font=\scriptsize, color=black!55, anchor=east, inner sep=0pt},
  seg/.style    ={font=\scriptsize, anchor=center, inner sep=0pt},
  callout/.style={font=\scriptsize, color=black!65, anchor=south, inner sep=1pt,
                  fill=white},
  lead/.style   ={draw=black!45, line width=0.3pt},
  guide/.style  ={draw=black!45, line width=0.4pt, dotted},
  bnote/.style  ={font=\scriptsize, color=black!50, anchor=north, inner sep=1pt},
  mhead/.style  ={font=\scriptsize, color=black!55, anchor=west, inner sep=0pt},
  mval/.style   ={font=\scriptsize, color=black!75, anchor=west, inner sep=0pt},
  lgroup/.style ={font=\scriptsize, color=black!55, anchor=west, inner sep=0pt},
  lname/.style  ={font=\scriptsize, anchor=west, inner sep=0pt},
]
\node[lgroup] at (0,-0.55) {single-field:};
\fill[draw=funnel!55, line width=0.3pt, fill=funnel!25] (1.72,-0.67) rectangle (2.02,-0.43);
\node[lname] at (2.10,-0.55) {SFSS};
\fill[draw=funnel!55, line width=0.3pt, fill=funnel!48] (3.05,-0.67) rectangle (3.35,-0.43);
\node[lname] at (3.43,-0.55) {SFMS};
\node[lgroup] at (4.70,-0.55) {multi-field:};
\fill[draw=funnel!55, line width=0.3pt, fill=funnel!72] (6.22,-0.67) rectangle (6.52,-0.43);
\node[lname] at (6.60,-0.55) {MFSS};
\fill[draw=funnel!55, line width=0.3pt, fill=funnel!100] (7.55,-0.67) rectangle (7.85,-0.43);
\node[lname] at (7.93,-0.55) {MFMS};
\draw[grid] (0.000,-0.97) -- (0.000,-4.45);
\draw[axline] (0.000,-4.45) -- (0.000,-4.56);
\node[tick] at (0.000,-4.56) {0};
\draw[grid] (2.200,-0.97) -- (2.200,-4.45);
\draw[axline] (2.200,-4.45) -- (2.200,-4.56);
\node[tick] at (2.200,-4.56) {20};
\draw[grid] (4.400,-0.97) -- (4.400,-4.45);
\draw[axline] (4.400,-4.45) -- (4.400,-4.56);
\node[tick] at (4.400,-4.56) {40};
\draw[grid] (6.600,-0.97) -- (6.600,-4.45);
\draw[axline] (6.600,-4.45) -- (6.600,-4.56);
\node[tick] at (6.600,-4.56) {60};
\draw[grid] (8.800,-0.97) -- (8.800,-4.45);
\draw[axline] (8.800,-4.45) -- (8.800,-4.56);
\node[tick] at (8.800,-4.56) {80};
\draw[grid] (11.000,-0.97) -- (11.000,-4.45);
\draw[axline] (11.000,-4.45) -- (11.000,-4.56);
\node[tick] at (11.000,-4.56) {100};
\draw[axline] (0,-4.45) -- (11.00,-4.45);
\node[axtitle] at (5.50,-5.01) {share of the successful trajectories (\%)};
\node[rname] at (-0.35,-1.41) {$\xi=\Mp/\sqrt{2}$};
\node[rsub]  at (-0.35,-1.77) {$\etamed=0.75$};
\fill[draw=none, fill=funnel!25] (0.000,-1.85) rectangle (3.539,-1.25);
\node[seg, text=black] at (1.769,-1.55) {32.17\%};
\draw[white, line width=0.7pt] (3.539,-1.85) -- (3.539,-1.25);
\fill[draw=none, fill=funnel!48] (3.539,-1.85) rectangle (3.791,-1.25);
\draw[lead] (3.665,-1.25) -- (3.665,-1.09);
\node[callout] at (3.665,-1.08) {SFMS\;2.29\%};
\draw[white, line width=1.1pt] (3.791,-1.85) -- (3.791,-1.25);
\fill[draw=none, fill=funnel!72] (3.791,-1.85) rectangle (7.725,-1.25);
\node[seg, text=black] at (5.758,-1.55) {35.77\%};
\draw[white, line width=0.7pt] (7.725,-1.85) -- (7.725,-1.25);
\fill[draw=none, fill=funnel!100] (7.725,-1.85) rectangle (11.000,-1.25);
\node[seg, text=white] at (9.363,-1.55) {29.77\%};
\draw[draw=funnel!55, line width=0.3pt] (0,-1.85) rectangle (11.000,-1.25);
\node[rname] at (-0.35,-2.46) {$\xi=\Mp$};
\node[rsub]  at (-0.35,-2.82) {$\etamed=0.38$};
\fill[draw=none, fill=funnel!25] (0.000,-2.90) rectangle (1.973,-2.30);
\node[seg, text=black] at (0.987,-2.60) {17.94\%};
\draw[white, line width=0.7pt] (1.973,-2.90) -- (1.973,-2.30);
\fill[draw=none, fill=funnel!48] (1.973,-2.90) rectangle (2.191,-2.30);
\draw[lead] (2.082,-2.30) -- (2.082,-2.14);
\node[callout] at (2.082,-2.13) {SFMS\;1.98\%};
\draw[white, line width=1.1pt] (2.191,-2.90) -- (2.191,-2.30);
\fill[draw=none, fill=funnel!72] (2.191,-2.90) rectangle (6.961,-2.30);
\node[seg, text=black] at (4.576,-2.60) {43.36\%};
\draw[white, line width=0.7pt] (6.961,-2.90) -- (6.961,-2.30);
\fill[draw=none, fill=funnel!100] (6.961,-2.90) rectangle (11.000,-2.30);
\node[seg, text=white] at (8.980,-2.60) {36.72\%};
\draw[draw=funnel!55, line width=0.3pt] (0,-2.90) rectangle (11.000,-2.30);
\node[rname] at (-0.35,-3.51) {$\xi=\sqrt{2}\,\Mp$};
\node[rsub]  at (-0.35,-3.87) {$\etamed=0.19$};
\fill[draw=none, fill=funnel!25] (0.000,-3.95) rectangle (1.082,-3.35);
\node[seg, text=black] at (0.541,-3.65) {9.84\%};
\draw[white, line width=0.7pt] (1.082,-3.95) -- (1.082,-3.35);
\fill[draw=none, fill=funnel!48] (1.082,-3.95) rectangle (1.258,-3.35);
\draw[lead] (1.170,-3.35) -- (1.170,-3.19);
\node[callout] at (1.170,-3.18) {SFMS\;1.60\%};
\draw[white, line width=1.1pt] (1.258,-3.95) -- (1.258,-3.35);
\fill[draw=none, fill=funnel!72] (1.258,-3.95) rectangle (7.438,-3.35);
\node[seg, text=black] at (4.348,-3.65) {56.18\%};
\draw[white, line width=0.7pt] (7.438,-3.95) -- (7.438,-3.35);
\fill[draw=none, fill=funnel!100] (7.438,-3.95) rectangle (10.990,-3.35);
\node[seg, text=white] at (9.214,-3.65) {32.29\%};
\draw[draw=funnel!55, line width=0.3pt] (0,-3.95) rectangle (10.990,-3.35);
\node[mhead] at (11.34,-0.71) {multi-stage};
\node[mhead] at (11.34,-0.97) {share};
\node[mval] at (11.34,-1.55) {32.06\%};
\node[mval] at (11.34,-2.60) {38.70\%};
\node[mval] at (11.34,-3.65) {33.89\%};
\end{tikzpicture}}
\caption{Composition of the successful trajectories as the coupling $\xi$ is varied, at fixed $N_{\rm f}=2$; $\etamed$ follows from $\xi$. Bars are stacked in the order SFSS, SFMS, MFSS, MFMS, so the heavier white rule separates the single-field from the multi-field categories. The right-hand column gives the multi-stage share, SFMS$+$MFMS. SFSS: Single-Field-Single-Stage; SFMS: Single-Field-Multi-Stage; MFSS: Multi-Field-Single-Stage; MFMS: Multi-Field-Multi-Stage.}
\label{fig:comp_var_xi}
\end{figure}
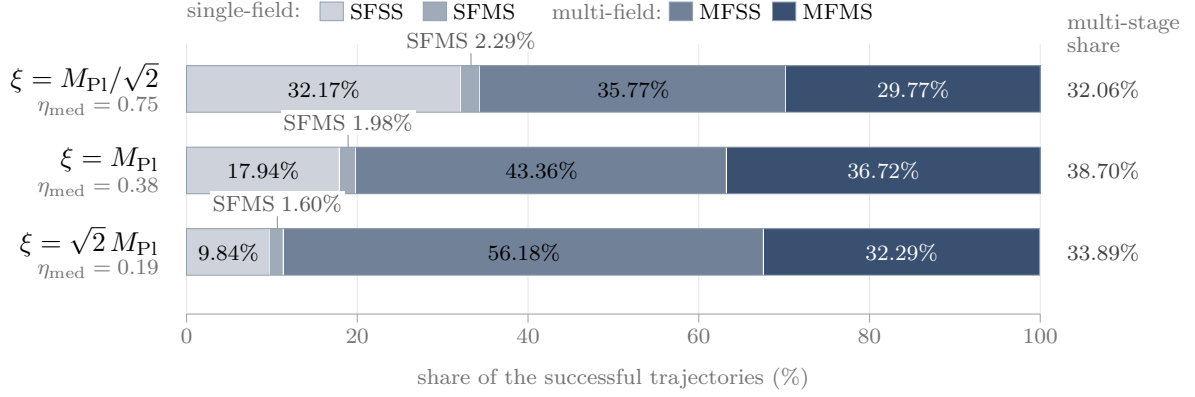

Taking a step further, we can examine the percentages of different categories of successful trajectories with different $\xi$'s, and the result is summarized in Fig.~\ref{fig:comp_var_xi}. 
We notice the following properties: 
(i) the percentage of multi-stage trajectories (or single-stage trajectories) does not change much with $\etamed$; 
(ii) Among all the multi-stage trajectories, the single-field versus multi-field contributions also do not change much with $\etamed$; 
(iii) however, among all the single-stage trajectories, there is a significant increase from the single-field contribution; in other words, the attractors prefer to be ``straight" rather than ``curled" as $\etamed$ increases ($\xi$ decreases), which is the main source of the statistical dependence of $\Theta$ on $\xi$ observed in the left panel of Figure \ref{fig:logTheta_varying}. 
Interestingly, this shows that, of the two possible sources of ``multi-field-ness", multiple stages connected by turns are more robust to changes in the landscape properties than curved attractors.


\section{Three-field landscapes and trends with increasing $N_{\rm f}$}
\label{sec3field}
In this section, we study the statistics of inflation trajectories on the three-field landscapes. The methodology is much the same as in the one-field and two-field cases, albeit with much demanding computational power. 
In organizing the results and conclusions, we will focus more on the trends of the statistical properties as $N_{\rm f}$ increases, in order to build intuition for how these properties may behave in the more realistic regime $N_{\rm f}\gg1$ suggested by string compactifications.

\paragraph{Numerical setup.} In the three-field case, we choose $\xi$ to have the same $\etamed$ as in the two-field case, which is found by numerical sampling to be $\xi=0.85\Mp$. The other parameters are set to be $\Lambda=10\xi$, $M=10$, $P_0=1$, and we take $n=5000$ initial points in a ball of radius $\lambda=0.7\Lambda$ in each realization. Even though we have chosen a smaller value of $M$ to control the numerical load, the computational cost is still much more expensive than the previous case. As a result, we limit the number of realizations to $4000$ in our sample, which still costs several days to perform on the cluster. 

\subsection{Basic statistical properties.}
The basic statistical properties are summarized as usual in Figure \ref{fig:event-hierarchy_3f}, and we will describe these results in more detail in the following context. We will keep the description rather brief, and make a more comprehensive study on the trend of varying $N_{\rm f}$ in the next subsection.


\begin{figure}[t]
\centering
\providecommand{\Mp}{M_{\rm Pl}}
\providecommand{\etamed}{\eta_{\rm med}}
\definecolor{funnel}{RGB}{58,80,112}   

\resizebox{\textwidth}{!}{%
\begin{tikzpicture}[
  x=1cm, y=1cm,
  bar/.style  ={draw=funnel!55, line width=0.3pt},
  lost/.style ={draw=none, fill=black!6},
  name/.style ={font=\small, anchor=east, inner sep=0pt},
  sub/.style  ={font=\scriptsize, color=black!55, anchor=east, inner sep=0pt},
  num/.style  ={font=\footnotesize, anchor=east, inner sep=0pt},
  tick/.style ={font=\scriptsize, color=black!55, anchor=north, inner sep=2pt},
  grid/.style ={draw=black!10, line width=0.3pt},
]

\def\XT{11.8277} \def\XA{9.8416} \def\XS{7.1925} \def\XC{4.3494}
\def\bh{0.30}

\foreach \d in {0,...,8}{
  \draw[grid] (1.62*\d, 0.45) -- (1.62*\d, -4.05);
  \draw[draw=black!35, line width=0.4pt] (1.62*\d,-4.05) -- (1.62*\d,-4.16);
  \node[tick] at (1.62*\d, -4.16) {$10^{\d}$};}
\draw[draw=black!35, line width=0.4pt] (0,-4.05) -- (12.96,-4.05);
\node[font=\scriptsize, color=black!55, anchor=north] at (6.48,-4.62)
      {number of trajectories};

\def\y{0}
\node[name] at (-0.35,\y+0.10) {Total};
\node[sub]  at (-0.35,\y-0.17) {the whole sample};
\fill[bar, fill=funnel!20] (0,\y-\bh) rectangle (\XT,\y+\bh);
\node[num] at (\XT-0.16,\y) {20\,000\,000};

\def\y{-1.15}
\node[name] at (-0.35,\y+0.10) {Admissible};
\node[sub]  at (-0.35,\y-0.17) {5.94\% of the sample};
\fill[lost] (\XA,\y-\bh) rectangle (\XT,\y+\bh);
\fill[bar, fill=funnel!38] (0,\y-\bh) rectangle (\XA,\y+\bh);
\node[num]  at (\XA-0.16,\y) {1\,188\,643 \textcolor{black!55}{(5.94\%)}};

\def\y{-2.30}
\node[name] at (-0.35,\y+0.10) {Successful};
\node[sub]  at (-0.35,\y-0.17) {0.14\% of the sample};
\fill[lost] (\XS,\y-\bh) rectangle (\XA,\y+\bh);
\fill[bar, fill=funnel!62] (0,\y-\bh) rectangle (\XS,\y+\bh);
\node[num]  at (\XS-0.16,\y) {27\,529 \textcolor{black!55}{(2.32\%)}};

\def\y{-3.45}
\node[name] at (-0.35,\y+0.10) {CMB-compatible};
\node[sub]  at (-0.35,\y-0.17) {0.0024\% of the sample};
\fill[lost] (\XC,\y-\bh) rectangle (\XS,\y+\bh);
\fill[bar, fill=funnel, draw=funnel] (0,\y-\bh) rectangle (\XC,\y+\bh);
\node[num, text=white] at (\XC-0.16,\y) {484 \textcolor{white!78!funnel}{(1.76\%)}};

\draw[draw=black!15, line width=0.4pt] (-2.95,-5.20) -- (12.96,-5.20);
\node[font=\small, anchor=west] at (-2.95,-5.72)
      {\textcolor{black!45}{$\hookrightarrow$}\; Structure of the
       \textbf{successful} trajectories};

\begin{scope}[shift={(3.5,-8.25)}]
  \fill[draw=white, line width=0.8pt, fill=funnel!25] (90.000:0.92) arc[start angle=90.000, end angle=55.044, radius=0.92] -- (55.044:1.78) arc[start angle=55.044, end angle=90.000, radius=1.78] -- cycle;
  \fill[draw=white, line width=0.8pt, fill=funnel!48] (55.044:0.92) arc[start angle=55.044, end angle=49.608, radius=0.92] -- (49.608:1.78) arc[start angle=49.608, end angle=55.044, radius=1.78] -- cycle;
  \fill[draw=white, line width=0.8pt, fill=funnel!72] (49.608:0.92) arc[start angle=49.608, end angle=-112.068, radius=0.92] -- (-112.068:1.78) arc[start angle=-112.068, end angle=49.608, radius=1.78] -- cycle;
  \fill[draw=white, line width=0.8pt, fill=funnel!100] (-112.068:0.92) arc[start angle=-112.068, end angle=-270.000, radius=0.92] -- (-270.000:1.78) arc[start angle=-270.000, end angle=-112.068, radius=1.78] -- cycle;
  \draw[funnel!45, line width=1.3pt] (90:0.8) arc[start angle=90, end angle=49.608, radius=0.8];
  \draw[funnel!100, line width=2.1pt] (49.608:0.8) arc[start angle=49.608, end angle=-270, radius=0.8];
  \node[font=\footnotesize, align=center] at (0,0.10) {27\,529};
  \node[font=\scriptsize, color=black!55] at (0,-0.16) {successful};
  \draw[draw=black!50, line width=0.35pt] (72.522:1.6) -- (72.522:2.02);
  \node[anchor=west, align=left, inner sep=1.5pt] at (72.522:2.06) {\small SFSS\\[-1pt]\scriptsize\textcolor{black!60}{9.71\%}};
  \draw[draw=black!50, line width=0.35pt] (52.326:1.6) -- (52.326:2.79);
  \node[anchor=west, align=left, inner sep=1.5pt] at (52.326:2.83) {\small SFMS\\[-1pt]\scriptsize\textcolor{black!60}{1.51\%}};
  \draw[draw=black!50, line width=0.35pt] (-31.230:1.6) -- (-31.230:2.02);
  \node[anchor=west, align=left, inner sep=1.5pt] at (-31.230:2.06) {\small MFSS\\[-1pt]\scriptsize\textcolor{black!60}{44.91\%}};
  \draw[draw=black!50, line width=0.35pt] (-191.034:1.6) -- (-191.034:2.02);
  \node[anchor=east, align=right, inner sep=1.5pt] at (-191.034:2.06) {\small MFMS\\[-1pt]\scriptsize\textcolor{black!60}{43.87\%}};
\end{scope}
\node[font=\scriptsize, color=black!55, anchor=north] at (3.5,-10.75) {single-field \textcolor{black}{11.22\%}\quad multi-field \textcolor{black}{88.78\%}\quad multi-stage (SFMS$+$MFMS) \textcolor{black}{45.38\%}};

\node[draw=funnel, line width=0.5pt, fill=funnel!8, rounded corners=2pt,
      anchor=east, align=left, text width=3.2cm, inner sep=5pt, font=\small]
     at (12.96,-8.25) {\textbf{484} CMB-compatible\\[1pt]
      \scriptsize\textcolor{black!60}{1.76\% of the successful trajectories}};

\end{tikzpicture}}
\caption{{\em $N_{\rm f}=3$ case} $(\xi=0.85\Mp$, $\etamed=0.38)$.
Classification of $2\times10^{7}$ trajectories drawn on $4\,000$ landscape
realisations. \emph{Top:} the successive selection stages; bar lengths are
logarithmic and the figure in parentheses is the fraction kept from the
previous stage, whose extent is shown by the shaded continuation of the bar.
\emph{Bottom:} the successful trajectories resolved by field content and
stage structure. The wedges run clockwise in the order SFSS, SFMS, MFSS,
MFMS, so the two inner arcs group the single-field and the multi-field
categories. SFSS: Single-Field-Single-Stage;
SFMS: Single-Field-Multi-Stage; MFSS: Multi-Field-Single-Stage; MFMS:
Multi-Field-Multi-Stage.}
\label{fig:event-hierarchy_3f}
\end{figure}

\begin{figure}[ht!]
    \centering  
    \includegraphics[width=\textwidth]{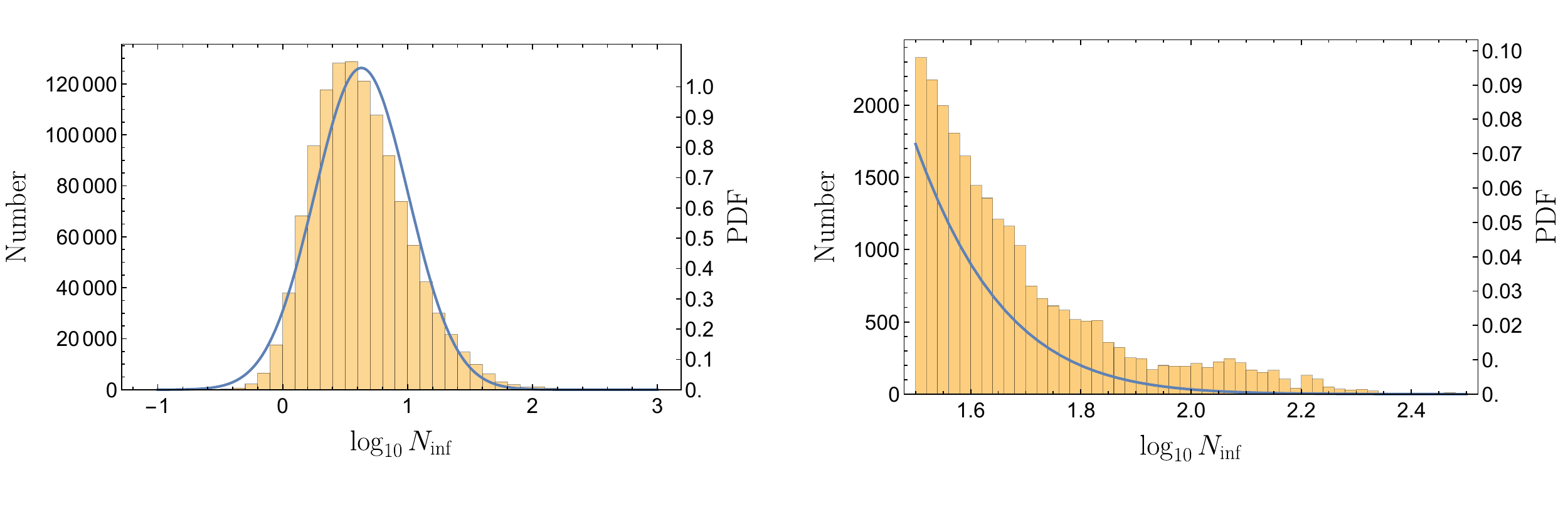}
    \caption{ {\em $N_{\rm f}=3$ case $(\xi=0.85\Mp$, $\etamed=0.38)$}. Left panel: The combined distribution of $N_{\rm inf}$ of all admissible trajectories from all realizations of the two-field landscape. The distribution is displayed in terms of $\log_{10}N_{\rm inf}$. The blue curve is a fit of the probability density function (PDF) by the normal distribution with mean and standard deviation being $0.63$ and $0.38$, respectively. Right panel: A zoomed-in view of the rightmost tail of the distribution.
    }
  \label{fig:NeDist_log_3d.pdf}
\end{figure}

\paragraph{Duration of inflation.} The distribution of $N_{\rm inf}$ of all admissible trajectories in our three-field sample is shown in Figure \ref{fig:NeDist_log_3d.pdf}, together with a log-normal fit as usual. Although the overall distribution of $\log_{10}N_{\rm inf}$ can still be approximated by a normal distribution, there is an even larger excess of number at the rightmost (long-lasting) tail than in the two-field case, as can be seen in the right panel of Figure \ref{fig:NeDist_log_3d.pdf}. We find the median of $\log_{10}N_{\rm inf}$ being $0.63$ with a standard deviation of $0.38$. There are 1493 ($37.3\%$) realizations having at least one successful trajectory, and the total number of successful trajectories is $27 529$, making up of $2.32\%$ of admissible trajectories. 

\begin{figure}[ht!]
    \centering  
    \includegraphics[width=0.465\textwidth]{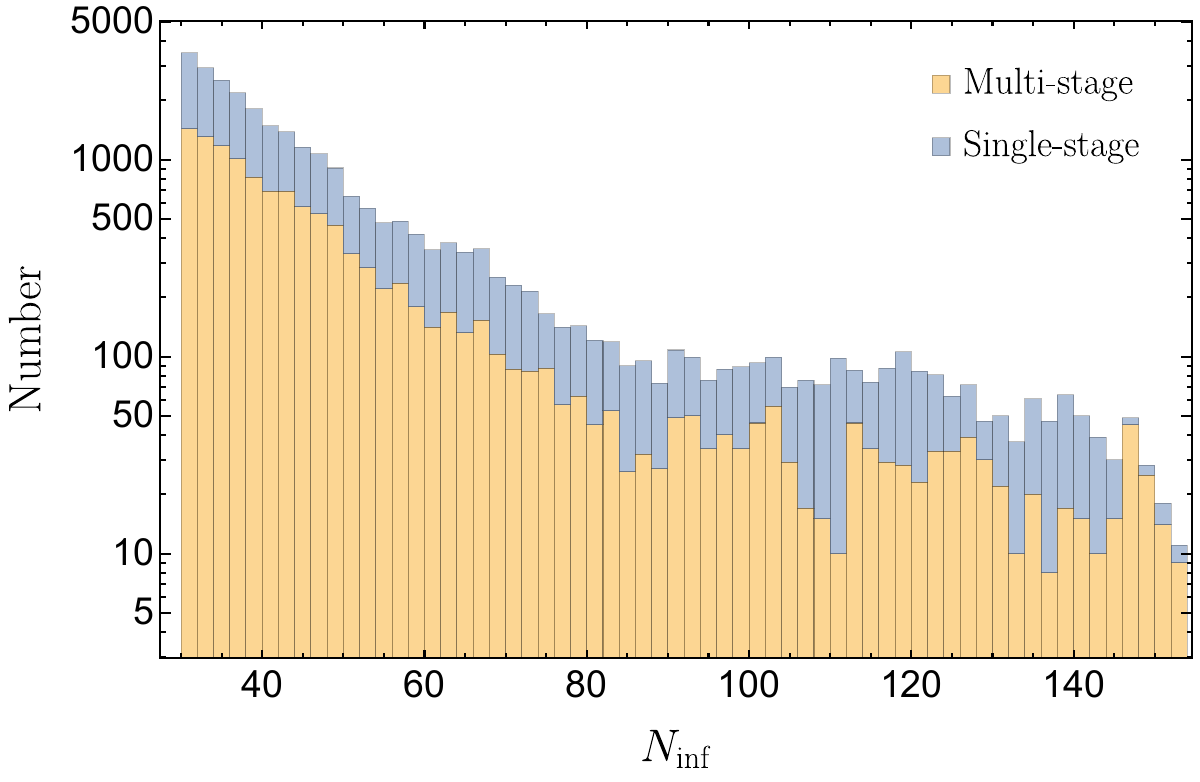}
    \hspace{10pt}
    \includegraphics[width=0.45\textwidth]{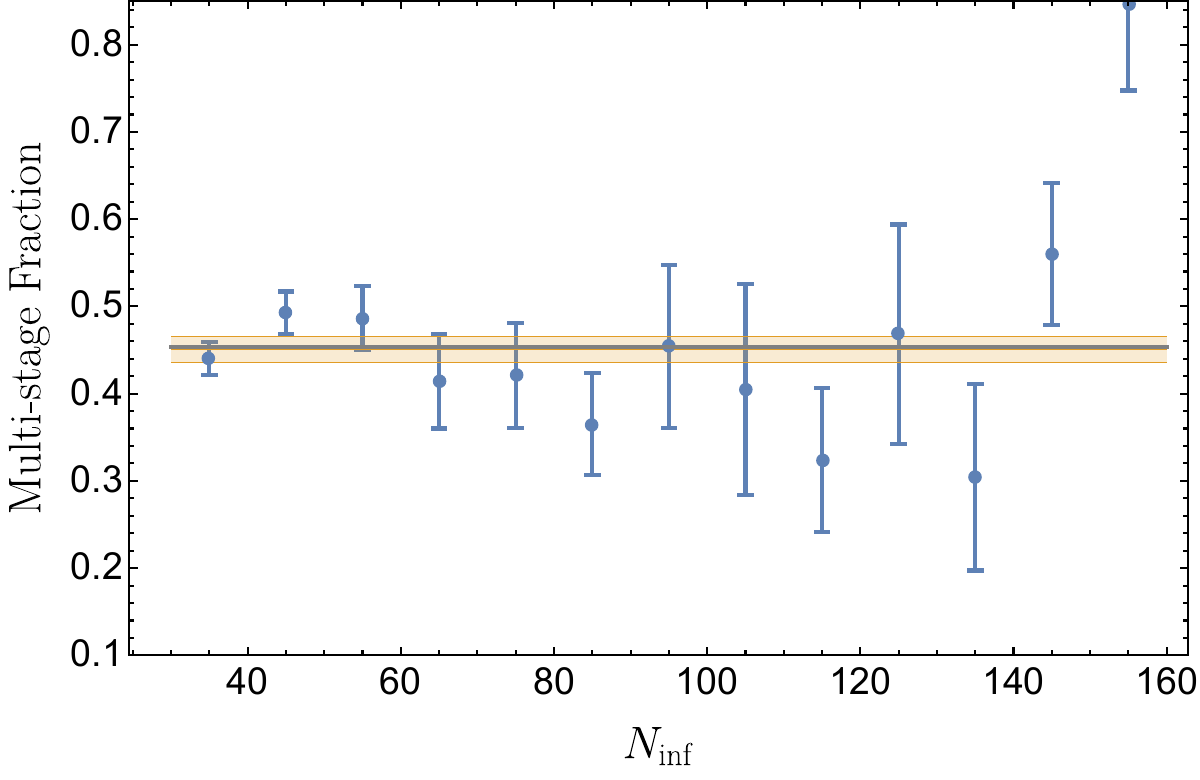}
    \caption{ {\em $N_{\rm f}=3$ case $(\xi=0.85\Mp$, $\etamed=0.38)$}. Left panel: The binned distribution of numbers of single-stage and multi-stage trajectories, in which the bin width is set to be $\Delta N_{\rm inf}=2$. Right panel: The binned fraction of multi-stage trajectories with bin width $\Delta N_{\rm inf}=10$, in which the error bars are estimated by the statistical bootstrap method. 
    The gray line is the overall multi-stage fraction, $45.38\%$;
    the orange band denotes the overall multi-stage fraction with statistical errors obtained by combining all bins, $(45.1\pm1.5)\%$.
    }
  \label{fig:Stage_Ne_3d}
\end{figure}

\paragraph{Single and multiple stages.} The statistics of the single- and multi-stage trajectories are shown in Figure \ref{fig:Stage_Ne_3d}, in the same format as the previous cases. In total, we find $12\,492$ multi-stage trajectories, comprising $45.38\%$ of all successful trajectories, which represents a substantial increase compared with the two-field case. In terms of binned analysis, the statement that the multi-stage fraction is not sensitive to $N_{\rm inf}$ still approximately hold, despite the large variations in the data due to the limited scale of our sample, as seen in the right panel of Figure \ref{fig:Stage_Ne_3d}.

\begin{figure}[ht!]
    \centering  
    \includegraphics[width=0.45\textwidth]{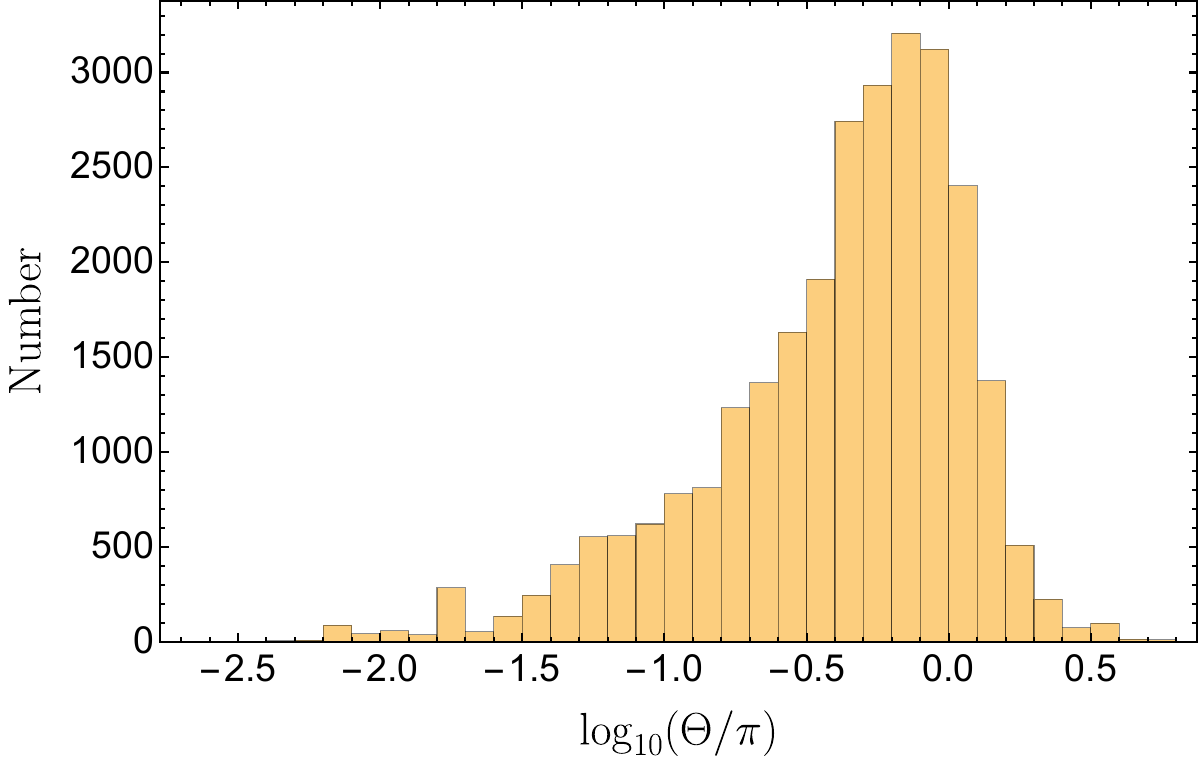}
    \includegraphics[width=0.45\textwidth]{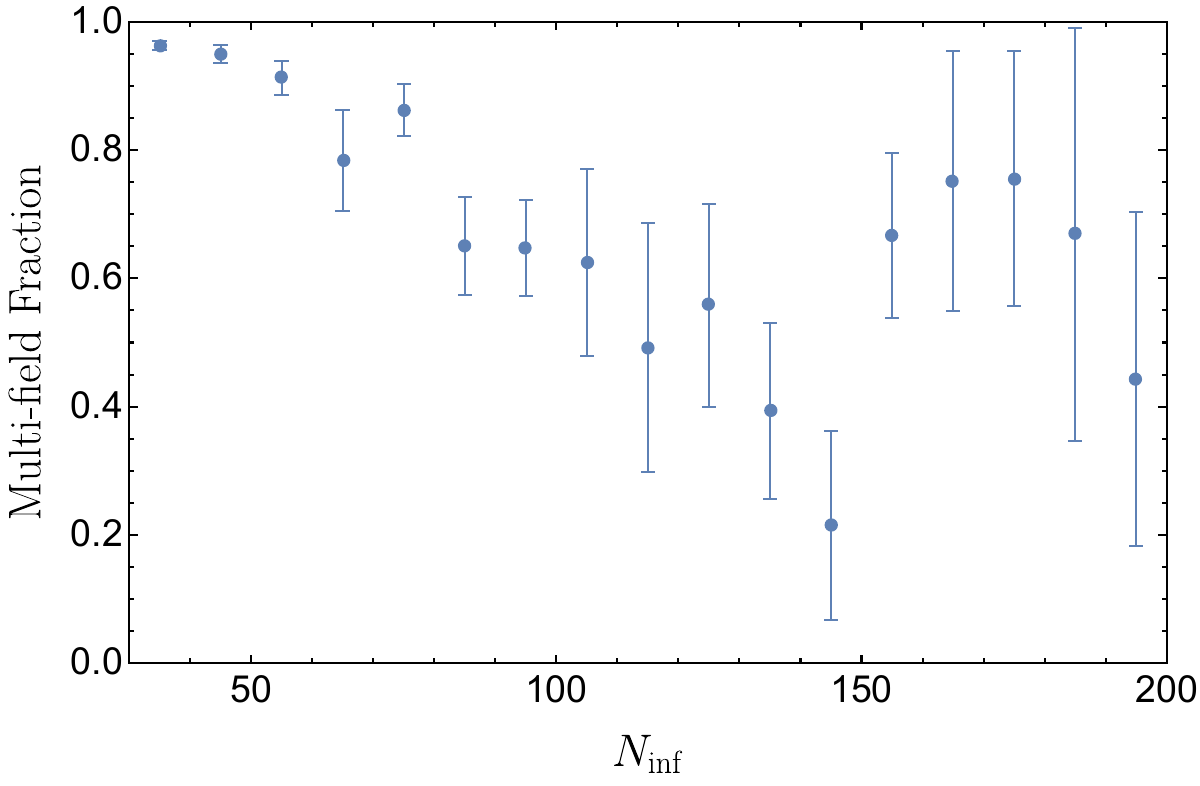}
    \caption{{\em $N_{\rm f}=3$ case $(\xi=0.85\Mp$, $\etamed=0.38)$}. Left panel: The distribution of $\Theta$ of all trajectories, in which the trajectories with $\log_{10}(\Theta/\pi)>-1$ is categorized as multi-field trajectories. Right panel: The binned fraction of multi-field trajectories with bin width $\Delta N=10$, with error bars estimated by the statistical bootstrap method.
    }
  \label{fig:theta_3d}
\end{figure}

\paragraph{Effective single-field and pure multi-field trajectories.}
We can also calculate the values of $\Theta$ of successful trajectories as the previous case, and the distribution of $\log_{10}(\Theta/\pi)$ is shown in the left panel of Figure \ref{fig:theta_3d}. The median of $\Theta$ is $0.51\pi$, and the total number of multi-field trajectories is $24\,440$ ($88.8\%$ of all successful trajectories), both representing substantial increases compared with the two-field case. In the right panel of Figure \ref{fig:theta_3d}, we again find a decrease in the multi-field fraction as $N_{\rm inf}$ increases, which is the same trend as in the two-field case. 

\begin{figure}[ht!]
    \centering  
    \includegraphics[width=0.43\textwidth]{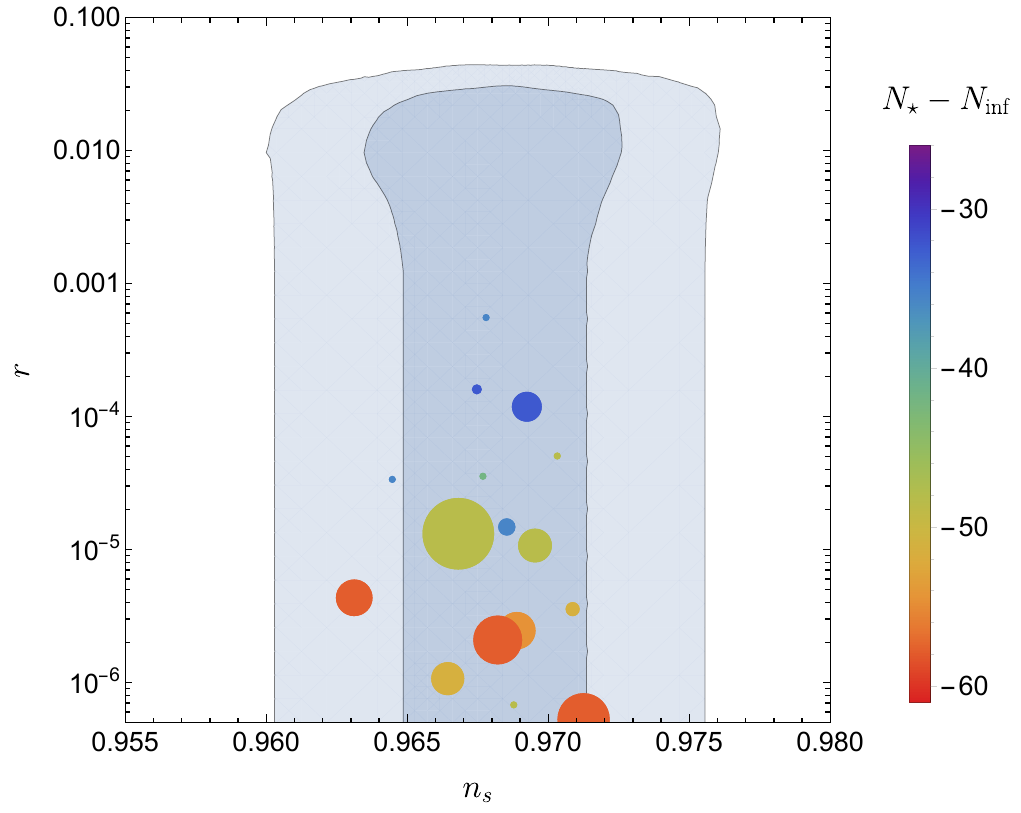}
    \includegraphics[width=0.55\textwidth]{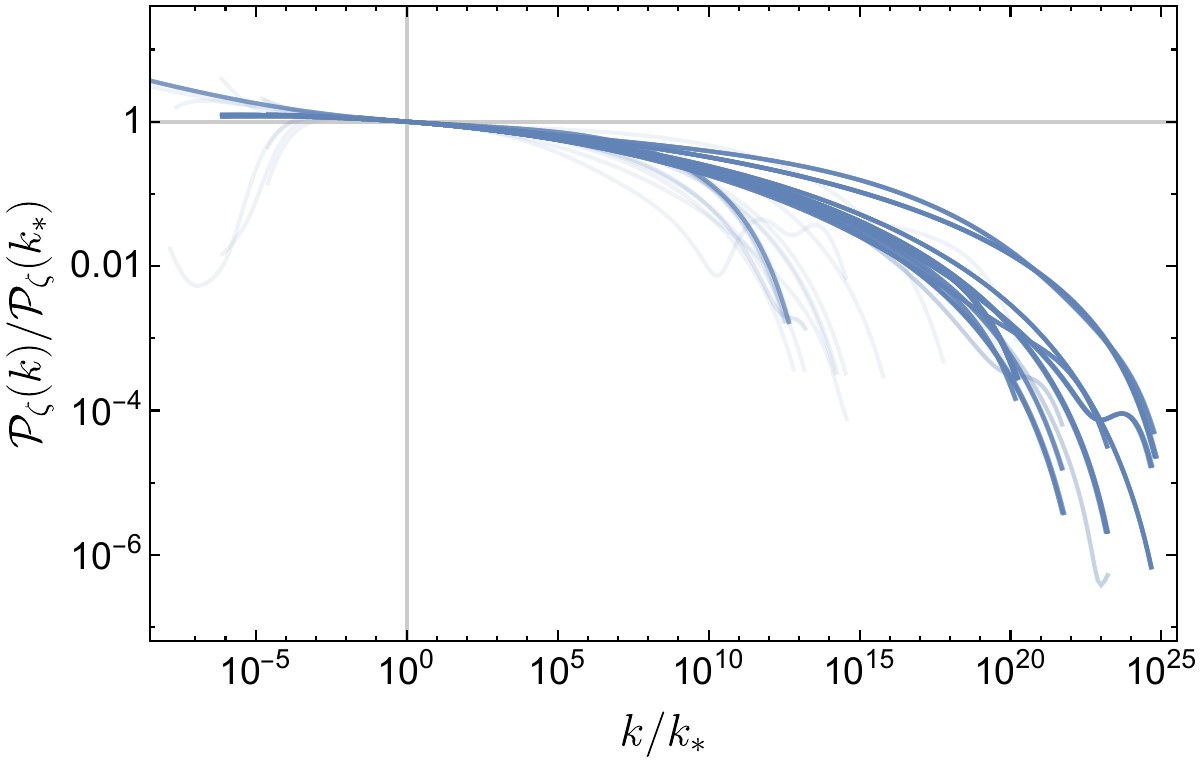}
    \caption{{\em $N_{\rm f}=3$ case} $(\xi=0.85\Mp$, $\etamed=0.38)$. The positions on the $r$-$n_s$ plane of all CMB-compatible trajectories as well as the $68\%$ and $95\%$ CMB constraints. Each point denotes a particular realization, with the area proportional to the number of trajectories and the colour corresponding to the pivot scale. Right panel: The power spectra of all CMB-compatible trajectories. Each trajectory is plotted in the thin colour and the colour is enhanced if there are multiple trajectories in the same inflation attractor.
    }
  \label{fig:cmb_compatible_3d}
\end{figure}

\paragraph{CMB compatibility.} The CMB compatibility search can still be conducted in much the same way as in previous cases, albeit numerically much more demanding. The limited amount of realizations makes the result less statistically informative than previous cases. Indeed, we find only 20 realizations with at least one CMB-compatible trajectory, and the total number of CMB-compatible trajectories is 484 ($1.76\%$ of successful trajectories). The properties of these trajectories are illustrated in Figure \ref{fig:cmb_compatible_3d}, in which the left panel shows the positions of the trajectories in the CMB-compatible contour together with their pivot scales, and the right panel shows their power spectra. Although the amount of distinctive trajectories is quite limited, we can still see a few trajectories having non-trivial features at small scales.

\subsection{Trends with increasing $N_{\rm f}$}
With the one-field, two-field and three-field data at hand, we can now study how the statistical properties of trajectories depend on $N_{\rm f}$. This could provide first clues on these properties on landscapes with much larger $N_{\rm f}$, an interesting regime that may be expected from a UV-complete theory.

\begin{figure}[ht!]
    \centering  
    \includegraphics[width=0.455\textwidth]{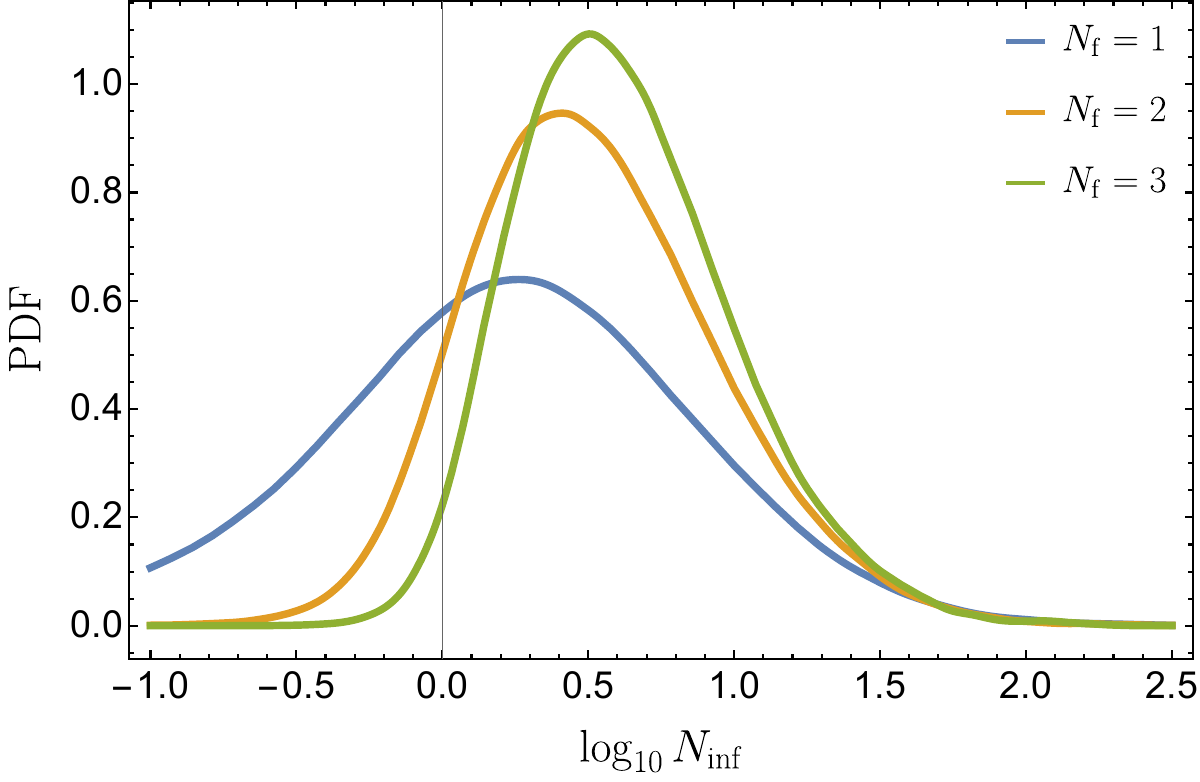}
    \hspace{8pt}
    \includegraphics[width=0.46\textwidth]{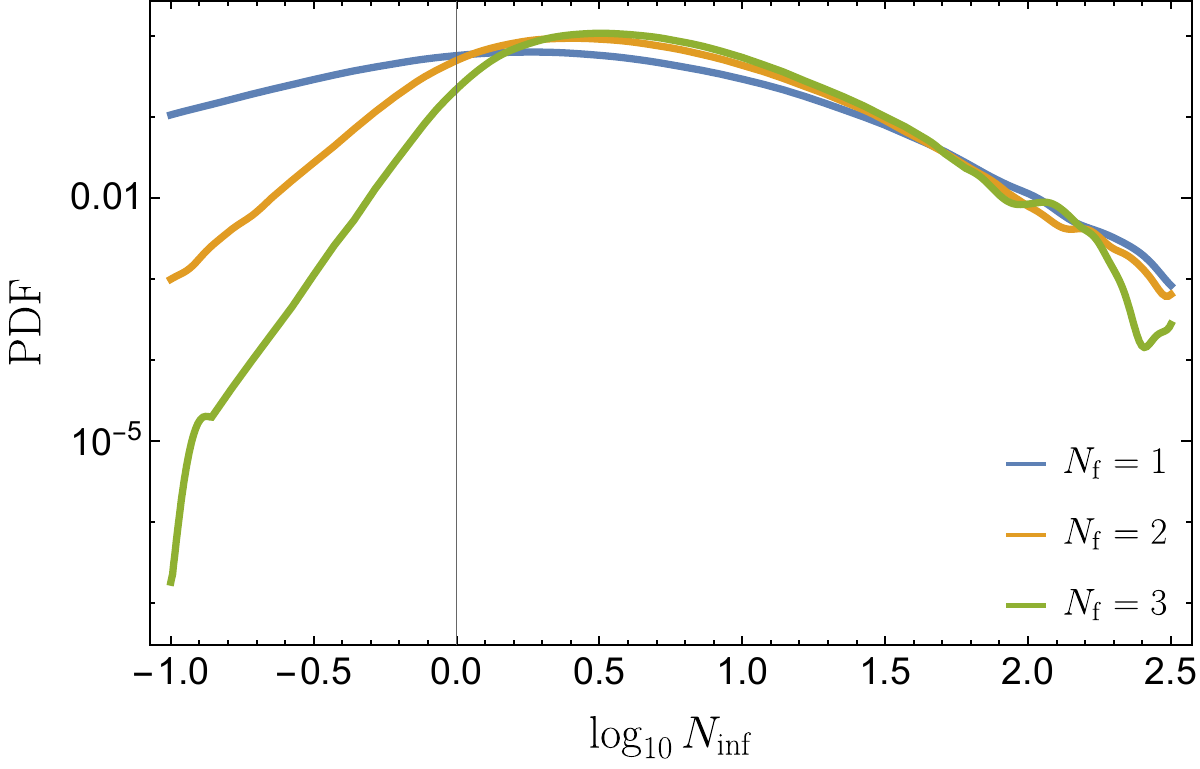}
    \caption{Left panel: The PDFs of $\log_{10}N_{\rm inf}$ of samples with different $N_{\rm f}$ with fixed $\etamed$. Right panel: The same plot on the log scale, in which the coincidence of the three PDFs at large $N_{\rm inf}$ is evident.
    }
  \label{fig:NDist_varying_Nf}
\end{figure}

\paragraph{Distribution of $N_{\rm inf}$.} As we have seen in previous sections, the distribution of $N_{\rm inf}$ of all admissible trajectories appears to have a logarithmic normal profile in all cases. As $N_{\rm f}$ increases, we find that the median value of $N_{\rm inf}$ increases and the standard deviation decreases, \textit{i.e.} the distribution prefers larger $N_{\rm f}$ but has a narrower peak.
However, the large-$N_{\rm inf}$ tail of the distribution deviates from the standard profile as $N_{\rm f}$ increases, making it obscure to see what the trend looks like for long-lasting trajectories. So, in Figure \ref{fig:NDist_varying_Nf}, we plot the numerically obtained PDFs of $\log_{10}N_{\rm f}$ for all three cases on both the linear and log scale. While in the left panel it is obvious that all samples have an approximate normal profile, in the right panel we find that the three PDFs coincide when $N_{\rm inf}\gtrsim30$ ($\log_{10}N_{\rm inf}>1.48$). In other words, it indicates that the $N_{\rm inf}$ distribution of \textit{successful} trajectories is \textit{approximately invariant} with the changes in $N_{\rm f}$, with $\etamed$ fixed. Indeed, the fractions of successful trajectories among admissible trajectories are nearly identical for different $N_{\rm f}$, as can be checked in Figures \ref{fig:NeDist_xi1_log_1d.pdf}, \ref{fig:NeDist_xi1_log_2d.pdf} and \ref{fig:NeDist_log_3d.pdf}.

\begin{figure}[ht!]
    \centering  
    \includegraphics[width=0.7\textwidth]{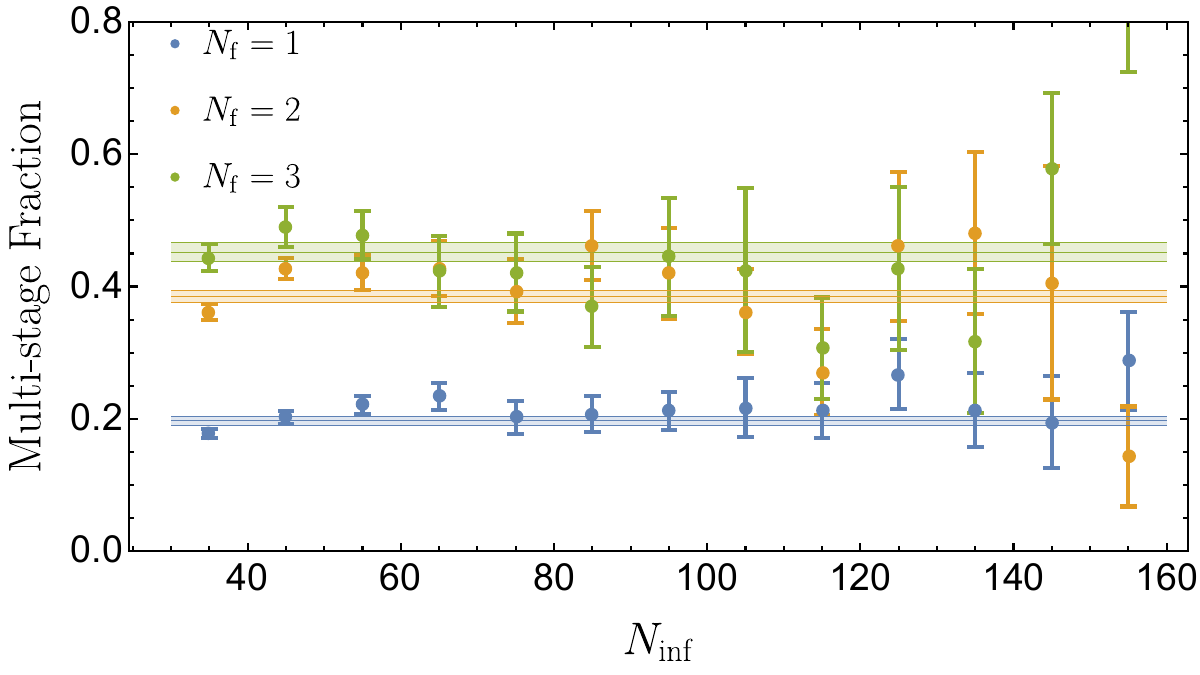}
    \caption{The binned fractions of multi-stage trajectories for samples with different $N_{\rm f}$ but the same $\etamed=0.38$, with bin width $\Delta N_{\rm inf}=10$. The horizontal bands are the total multi-stage fractions and the corresponding errors of these samples.
    }
  \label{fig:Stage_varying_Nf}
\end{figure}

\paragraph{Multi-stage and multi-field properties.}
The binned and total fractions of multi-stage trajectories for different $N_{\rm f}$, holding $\etamed$ fixed, are plotted together in Figure \ref{fig:Stage_varying_Nf}. From this figure, we can see a clear trend that the multi-stage fraction increases with $N_{\rm f}$, with a statistical significance exceeding $3\sigma$. This trend holds across different values of $N_{\rm inf}$, except at larger $N_{\rm inf}$, where the variances increase due to the limited sample sizes and the trend becomes much less clear.
This result is important in the sense that larger fraction of multi-stage trajectories means greater possibility of phenomenologies drastically different from the single-stage slow-roll scenario, especially if this trend holds for even larger $N_{\rm f}$.

This trend may be understood qualitatively in the following way. Generically, each landscape contains only a very small number of, such as one or two, long-lasting slow-roll attractors, which we refer to as the ``main stream(s)" in the following. Most of the multi-stage trajectories appear to arise from a short inflationary stage that connects to this ``main stream" at various locations. As $N_{\rm f}$ increases, trajectories have more directions in which to fall into, and more locations to connect to, the ``main stream'', resulting in a larger fraction of multi-stage inflation models.
However, even if this trend holds, it is hard to tell whether the fraction approaches 1 or converges to a finite value when $N_{\rm f}\gg1$. These questions may only be answered by promoting our practice to higher $N_{\rm f}$. Nevertheless, since we have observed a fraction of $\sim50\%$ of multi-stage trajectories even at $N_{\rm f}=3$, we can at least conclude that multi-stage inflation is \textit{common} in this landscape scenario.

\begin{figure}[t]
\centering
\providecommand{\Mp}{M_{\rm Pl}}
\providecommand{\etamed}{\eta_{\rm med}}
\definecolor{funnel}{RGB}{58,80,112}   

\resizebox{\textwidth}{!}{%
\begin{tikzpicture}[
  x=1cm, y=1cm,
  grid/.style   ={draw=black!10, line width=0.3pt},
  axline/.style ={draw=black!35, line width=0.4pt},
  tick/.style   ={font=\scriptsize, color=black!55, anchor=north, inner sep=2pt},
  axtitle/.style={font=\scriptsize, color=black!55, anchor=north},
  ptitle/.style ={font=\small, anchor=west},
  rname/.style  ={font=\small, anchor=east, inner sep=0pt},
  rsub/.style   ={font=\scriptsize, color=black!55, anchor=east, inner sep=0pt},
  seg/.style    ={font=\scriptsize, anchor=center, inner sep=0pt},
  callout/.style={font=\scriptsize, color=black!65, anchor=south, inner sep=1pt,
                  fill=white},
  lead/.style   ={draw=black!45, line width=0.3pt},
  guide/.style  ={draw=black!45, line width=0.4pt, dotted},
  bnote/.style  ={font=\scriptsize, color=black!50, anchor=north, inner sep=1pt},
  mhead/.style  ={font=\scriptsize, color=black!55, anchor=west, inner sep=0pt},
  mval/.style   ={font=\scriptsize, color=black!75, anchor=west, inner sep=0pt},
  lgroup/.style ={font=\scriptsize, color=black!55, anchor=west, inner sep=0pt},
  lname/.style  ={font=\scriptsize, anchor=west, inner sep=0pt},
]
\node[lgroup] at (0,-0.55) {single-field:};
\fill[draw=funnel!55, line width=0.3pt, fill=funnel!25] (1.72,-0.67) rectangle (2.02,-0.43);
\node[lname] at (2.10,-0.55) {SFSS};
\fill[draw=funnel!55, line width=0.3pt, fill=funnel!48] (3.05,-0.67) rectangle (3.35,-0.43);
\node[lname] at (3.43,-0.55) {SFMS};
\node[lgroup] at (4.70,-0.55) {multi-field:};
\fill[draw=funnel!55, line width=0.3pt, fill=funnel!72] (6.22,-0.67) rectangle (6.52,-0.43);
\node[lname] at (6.60,-0.55) {MFSS};
\fill[draw=funnel!55, line width=0.3pt, fill=funnel!100] (7.55,-0.67) rectangle (7.85,-0.43);
\node[lname] at (7.93,-0.55) {MFMS};
\draw[grid] (0.000,-0.97) -- (0.000,-4.45);
\draw[axline] (0.000,-4.45) -- (0.000,-4.56);
\node[tick] at (0.000,-4.56) {0};
\draw[grid] (2.200,-0.97) -- (2.200,-4.45);
\draw[axline] (2.200,-4.45) -- (2.200,-4.56);
\node[tick] at (2.200,-4.56) {20};
\draw[grid] (4.400,-0.97) -- (4.400,-4.45);
\draw[axline] (4.400,-4.45) -- (4.400,-4.56);
\node[tick] at (4.400,-4.56) {40};
\draw[grid] (6.600,-0.97) -- (6.600,-4.45);
\draw[axline] (6.600,-4.45) -- (6.600,-4.56);
\node[tick] at (6.600,-4.56) {60};
\draw[grid] (8.800,-0.97) -- (8.800,-4.45);
\draw[axline] (8.800,-4.45) -- (8.800,-4.56);
\node[tick] at (8.800,-4.56) {80};
\draw[grid] (11.000,-0.97) -- (11.000,-4.45);
\draw[axline] (11.000,-4.45) -- (11.000,-4.56);
\node[tick] at (11.000,-4.56) {100};
\draw[axline] (0,-4.45) -- (11.00,-4.45);
\node[axtitle] at (5.50,-5.01) {share of the successful trajectories (\%)};
\node[rname] at (-0.35,-1.41) {$N_{\rm f}=1$};
\node[rsub]  at (-0.35,-1.77) {$\xi=1.32\Mp$};
\fill[draw=none, fill=funnel!25] (0.000,-1.85) rectangle (8.819,-1.25);
\node[seg, text=black] at (4.409,-1.55) {80.17\%};
\draw[white, line width=0.7pt] (8.819,-1.85) -- (8.819,-1.25);
\fill[draw=none, fill=funnel!48] (8.819,-1.85) rectangle (11.000,-1.25);
\node[seg, text=black] at (9.909,-1.55) {19.83\%};
\draw[draw=funnel!55, line width=0.3pt] (0,-1.85) rectangle (11.000,-1.25);
\node[rname] at (-0.35,-2.46) {$N_{\rm f}=2$};
\node[rsub]  at (-0.35,-2.82) {$\xi=\Mp$};
\fill[draw=none, fill=funnel!25] (0.000,-2.90) rectangle (1.973,-2.30);
\node[seg, text=black] at (0.987,-2.60) {17.94\%};
\draw[white, line width=0.7pt] (1.973,-2.90) -- (1.973,-2.30);
\fill[draw=none, fill=funnel!48] (1.973,-2.90) rectangle (2.191,-2.30);
\draw[lead] (2.082,-2.30) -- (2.082,-2.14);
\node[callout] at (2.082,-2.13) {SFMS\;1.98\%};
\draw[white, line width=1.1pt] (2.191,-2.90) -- (2.191,-2.30);
\fill[draw=none, fill=funnel!72] (2.191,-2.90) rectangle (6.961,-2.30);
\node[seg, text=black] at (4.576,-2.60) {43.36\%};
\draw[white, line width=0.7pt] (6.961,-2.90) -- (6.961,-2.30);
\fill[draw=none, fill=funnel!100] (6.961,-2.90) rectangle (11.000,-2.30);
\node[seg, text=white] at (8.980,-2.60) {36.72\%};
\draw[draw=funnel!55, line width=0.3pt] (0,-2.90) rectangle (11.000,-2.30);
\node[rname] at (-0.35,-3.51) {$N_{\rm f}=3$};
\node[rsub]  at (-0.35,-3.87) {$\xi=0.85\Mp$};
\fill[draw=none, fill=funnel!25] (0.000,-3.95) rectangle (1.068,-3.35);
\node[seg, text=black] at (0.534,-3.65) {9.71\%};
\draw[white, line width=0.7pt] (1.068,-3.95) -- (1.068,-3.35);
\fill[draw=none, fill=funnel!48] (1.068,-3.95) rectangle (1.234,-3.35);
\draw[lead] (1.151,-3.35) -- (1.151,-3.19);
\node[callout] at (1.151,-3.18) {SFMS\;1.51\%};
\draw[white, line width=1.1pt] (1.234,-3.95) -- (1.234,-3.35);
\fill[draw=none, fill=funnel!72] (1.234,-3.95) rectangle (6.174,-3.35);
\node[seg, text=black] at (3.704,-3.65) {44.91\%};
\draw[white, line width=0.7pt] (6.174,-3.95) -- (6.174,-3.35);
\fill[draw=none, fill=funnel!100] (6.174,-3.95) rectangle (11.000,-3.35);
\node[seg, text=white] at (8.587,-3.65) {43.87\%};
\draw[draw=funnel!55, line width=0.3pt] (0,-3.95) rectangle (11.000,-3.35);
\node[mhead] at (11.34,-0.71) {multi-stage};
\node[mhead] at (11.34,-0.97) {share};
\node[mval] at (11.34,-1.55) {19.83\%};
\node[mval] at (11.34,-2.60) {38.70\%};
\node[mval] at (11.34,-3.65) {45.38\%};
\end{tikzpicture}}
\caption{Composition of the successful trajectories as the number of fields $N_{\rm f}$ is varied (each sample at its own $\xi$, indicated below the label, with $\etamed=0.38$). Bars are stacked in the order SFSS, SFMS, MFSS, MFMS, so the heavier white rule separates the single-field from the multi-field categories. The right-hand column gives the multi-stage share, SFMS$+$MFMS. SFSS: Single-Field-Single-Stage; SFMS: Single-Field-Multi-Stage; MFSS: Multi-Field-Single-Stage; MFMS: Multi-Field-Multi-Stage.}
\label{fig:comp_var_Nf}
\end{figure}
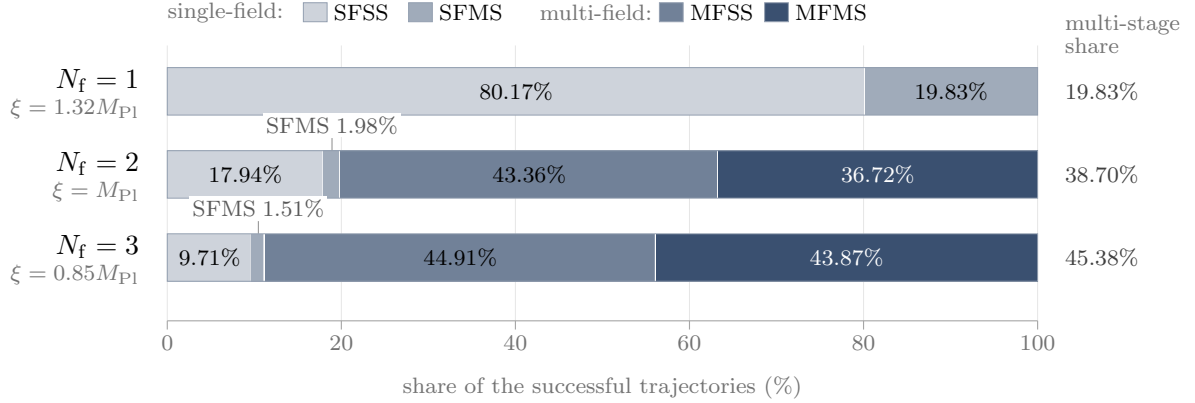

In addition, we can add the information of multi-field fractions to our discussion. We summarize the percentages of different categories in different samples in Fig.~\ref{fig:comp_var_Nf}. Since the multi-field property does not play a role for $N_{\rm f}=1$, it is not as informative as the multi-stage data. Nevertheless, some interesting implications can still be observed. The table shows that the fraction of MFSS trajectories is almost identical for $N_{\rm f}=2$ and $N_{\rm f}=3$ (with $\etamed$ fixed), which means that the increase in multi-field fraction is almost entirely due to the increase in multi-stage trajectories. Furthermore, the increase in the MFSS/SFSS ratio means that single-stage trajectories are more likely to be curved as $N_{\rm f}$ increases. If we further assume that the properties of single-stage trajectories reflects the properties of \textit{attractors},\footnote{That is to say, individual slow-roll attractors in multi-field trajectories follows similar statistics with single-stage trajectories.} it indicates that there is an enhanced probability of finding an attractor being curved than straight with increasing $N_{\rm f}$.


\paragraph{CMB compatibility.}
There are $\sim1\%$ successful trajectories that are compatible with the current CMB observation.
In addition, we find in all three cases that the fraction of power spectra that have non-trivial features is qualitatively smaller than the fraction of multi-stage trajectories.
This is expected since transitions happening before horizon exit for the CMB scales are invisible. However, we are unable to directly test whether the increase in multi-stage fraction with $N_{\rm f}$ still holds for CMB-compatible trajectories and with transitions constrained to $N>N_{\star}$, due to the limited number of trajectories with such properties that we could obtain. We leave this for future study.

\paragraph{Summary and remarks.}
The trends of statistics with varying $N_{\rm f}$ and fixed $\etamed$ are summarized as follows:
\begin{itemize}
    \item The PDF of $\log_{10}N_{\rm inf}$ has a Gaussian profile at $N_{\inf}\lesssim30$, with the mean value increasing and the standard deviation decreasing as $N_{\rm f}$ increases.
    \item In the region relevant to successful trajectories ($N_{\rm inf}>30$), this PDF is nearly invariant with $N_{\rm f}$.
    \item The fraction of multi-stage trajectories increases as $N_{\rm f}$ increases.
    \item Among single-stage trajectories, the fraction of multi-field (curved) trajectories increases as $N_{\rm f}$ increases.
    \item The CMB-compatible trajectories currently represents a generic subset ($\sim1\%$\footnote{The specific value of this fraction depends on how precise the spectral index and tensor mode are measured, and should certainly become smaller if the measurements are made more precise in the future.}) of successful trajectories. This value remains approximately the same for all values of $N_{\rm f}$.
\end{itemize}
As a remark, all the above conclusions are extracted from numerical data with $N_{\rm f}\leq3$, which may not be immediately extrapolated to the $N_{\rm f}\gg1$ regime. Nonetheless, we have tried to understand some of these trends qualitatively using properties of the landscapes and inflationary trajectories. Understanding some of the others, and especially developing analytical understanding of these properties, remains an open challenge. These trends provide the first clues to the behavior of analogous statistics in landscapes with much larger $N_{\rm f}$. Whether these trends can be extrapolated to much larger values of $N_{\rm f}$, and how such an extrapolation should proceed, remain important challenges for future studies.
Most remarkably, if the multi-stage and multi-field trajectories comprise the majority of the possibility, they will have important phenomenological consequences on both the CMB scales and, perhaps even more importantly, on much shorter scales that are increasingly accessible to experiments nowadays.

\section{Conclusions and discussion}
\label{sec-conclusion}

In this paper, we have numerically generated simple models of random inflationary landscapes and conducted a comprehensive statistical study of the properties of inflation trajectories on these landscapes. The properties include the overall statistics of the number of $e$-folds, and detailed properties of trajectories such as the CMB compatibility, multi-field-ness, and, most importantly, multi-stage-ness. In particular, we investigated the dependence of these properties on two main parameters describing the landscape: the field dimension $N_{\rm f}$ and the median of the $\eta$-parameter, $\etamed$.

Firstly, we have employed a comprehensive framework for constructing the landscape from statistical requirements specified through the field-space two-point function. This framework is flexible enough to impose different landscape properties as simple representations of more realistic inflationary landscapes. In this work, we mainly use the Gaussian random field with a Gaussian-like spectrum, but it can be generalized to different forms of spectra. A potential limitation of this framework is that the choice of the vacuum energy is artificial.
We choose to lift the global minimum of each realization to zero so that the entire landscape is non-negative, mostly for practical convenience but also to crudely represent a long wavelength modulation on a local patch.

We solve the background equations of motion of the inflaton with random initial conditions drawn from the many landscape realizations, generating the sample of trajectories that we collect for statistical study. 
The computational demand is acceptable for the low field dimensions explored in this work, but it increases dramatically with $N_{\rm f}$. It requires further optimization and remains a challenge if $N_{\rm f}\gg1$.

We find that only a small fraction ($\sim2\%$) of admissible trajectories are capable of supporting successful inflation in the sense of solving the horizon problem. 
We have analyzed and summarized many interesting properties of these trajectories in the main text.
The most remarkable finding is that a substantial fraction of successful trajectories exhibit multiple stages, and that this fraction increases with the field-space dimension at fixed typical landscape curvature, based on our samples with $N_{\rm f}=1,2,3$. This is a previously unrecognized property of inflationary trajectories in these random landscapes, largely because the landscape was typically constructed only within a local patch around an existing slow-roll attractor, with the end of inflation defined by the violation of the slow-roll conditions. Such a procedure is largely blind to multi-stage models. Indeed, the multi-stage nature of inflation becomes manifest only once we construct a \textit{global} realization of the landscape with a clearly specified true vacuum and search the multi-stage trajectories in the entire collection of inflationary trajectories, which is precisely what we do in this work.

In addition, we frequently find that the inflationary trajectory turns during its evolution. In some cases, the turning persists for several $e$-folds. This opens up interesting phenomenological possibilities, as in models of quasi-single-field inflation or multi-field inflation.

Here we would like to make some more detailed comparisons with previous work on Gaussian random landscapes~\cite{Masoumi:2016eag,Tegmark:2004qd,Frazer:2011tg,Frazer:2011br,Bjorkmo:2017nzd}. Our use of Gaussian random potentials is motivated by the works of Tegmark and of Masoumi, Vilenkin, and Yamada~\cite{Tegmark:2004qd,Masoumi:2016eag}, from which we also adopt some basic definitions and notation. In searching for successful inflationary trajectories, Frazer and Liddle~\cite{Frazer:2011tg,Frazer:2011br} find much lower success rates than we do. This difference is largely due to our choice to uplift each landscape such that the global minimum of the potential is zero and to choose the initial positions of the inflaton not too far from this minimum to reduce unnecessary computational cost. They also find a substantial fraction of multi-field inflation models and emphasize the impact of isocurvature perturbations in these models.
Bjorkmo and Marsh~\cite{Bjorkmo_2019} are able to study landscapes with much higher field-space dimensions by employing a local construction of the landscape in terms of a Taylor expansion, which significantly reduces the number of independent parameters. However, since their construction is valid only over much smaller ranges in field space, it may not be suitable for our purpose of searching for multi-stage inflation models. Overall, as emphasized throughout this work, the main novel aspect of our study of Gaussian random landscapes is the search for multi-stage inflation models and the analysis of their statistics.

There are many important questions that are worth studying in the future.

An important conclusion of this paper is that a substantial fraction of inflationary trajectories in Gaussian random landscapes with field-space dimensions one, two, and three exhibit multiple stages, and that this fraction increases with the field-space dimension. Since a UV-complete theory would typically imply an inflationary landscape with $N_{\rm f}$ much larger than the values considered here, several important questions arise. Does this growth trend persist at much larger $N_{\rm f}$? What is the asymptotic value of this fraction as $N_{\rm f}\to\infty$? As the fraction of multi-stage models increases, does the number of stages in these models also increase? More specifically, how does the distribution of the number of stages depend on $N_{\rm f}$? It would also be interesting to develop a more quantitative analytical understanding of the results.

We have considered a simple toy model of inflationary landscapes. It would also be interesting to investigate these questions, especially the statistics of multi-stage inflation models, across different types of landscapes, field-space geometries, and choices of statistical measure.

These findings and open questions also have important phenomenological consequences for the properties of primordial fluctuations, both on CMB scales and, perhaps even more importantly, on much shorter scales that are becoming increasingly accessible to observations.

\medskip
\section*{Acknowledgments}

We thank Alan Guth and Ling-feng Li for helpful discussions. We thank Haoxiang Guo for checking many of the results presented here and for helpful discussions. We acknowledge FAS Research Computing at Harvard University for the use of its computing cluster for the numerical calculations. ZX is supported by NSFC under Grants No.\ 12275146 and No.\ 12247103, the National Key R\&D Program of China (2021YFC2203100), and the Dushi Program of Tsinghua University. 
YZ was hosted and financially supported by LPENS during this work. 

\appendix

\section{Constraints on the Number of $e$-folds}
\label{app_efolds}

In this appendix, we briefly review the constraints put on the number of $e$-folds $N_{\rm inf}$ by current theory and observations. 

If we assume an instant reheating, the total number of inflationary efolds required to solve the horizon problem, $N_{\rm horizon}$, is related to the reheating energy $E_{\rm reheat}$ by (see e.g.~\cite{Liddle:2000cg})
\begin{equation}
    N_{\rm horizon} = 62 - \ln {\frac{{10^{16} {\rm GeV}}}{E_{\rm reheat}}} ~,
    \label{Eq:Nhorizon-Ereheat}
\end{equation}
and an inflaton trajectory should have $N_{\rm inf}\gtrsim N_{\rm horizon}$ to be a successful one. Therefore, different choices of $E_{\rm reheat}$ place different bounds on $N_{\rm inf}$. 

The minimally required $E_{\rm reheat}$ is set by ensuring a successful Big Bang Nucleosynthesis, requiring $E_{\rm reheat}>1~\rm MeV$, which corresponds to $N_{\rm inf}>18$. Alternatively, if we require $E_{\rm reheat}$ to be larger than the energy scale of baryogensis, estimated as $1~\rm GeV$, it gives $N_{\rm inf}>25$. More conservatively, if we require $E_{\rm reheat}$ to be larger than the electro-weak scale, $100~\rm GeV$, then it gives $N_{\rm inf}>30$, which is set to be the criterion of \textit{successful} trajectories in the main text. 

Meanwhile, the observational upper bound on the Hubble parameter $H$ places an upper bound on $E_{\rm reheat}$, which can be translated to an upper bound on $N_{\rm horizon}$. The current bound on $H$ is $H<4.2\times10^{12}~\rm GeV$, which translates to $E_{\rm reheating}<4.2\times10^{15}~{\rm GeV}$ and $N_{\rm horizon}<61$ upon instant reheating. For a trajectory with $N_{\rm inf}$ greater than 61, the power spectrum corresponding to $N-N_{\rm inf}<-61$ is beyond the observable scale of the CMB, placing an additional bound on the CMB compatibility.

In the case of multi-stage inflation models, the reheating energy is typically determined by the potential energy of the last inflationary stage, because the extra energy from a previous stage tends to be red-shifted away by the subsequent stage. Additionally, allowing non-instant and more complicated reheating processes would typically lower the required number of $e$-folds.

\section{The landscape power spectrum}
\label{app-differentPS}

The choice of the landscape power spectrum $P(\ppi)$ is largely arbitrary. It may potentially be determined by a precise UV-complete theory, but this is unknown to us. In practice, we apply the Gaussian-like power spectrum (\ref{eq:Gaussian-PS}) throughout this work, which is a toy model used for its simplicity and elegance. However, if the statistical properties discovered in this work turns out to be insensitive to the precise form of $P(\ppi)$ to some extent, they would certainly become more interesting and important. Therefore, in this appendix, we initiate some studies on the robustness of our results upon changing the functional form of $P(\ppi)$. The comparison is conducted in the two-field case $N_{\rm f}=2$.

\begin{figure}[ht!]
    \centering  
    \includegraphics[width=0.7\textwidth]{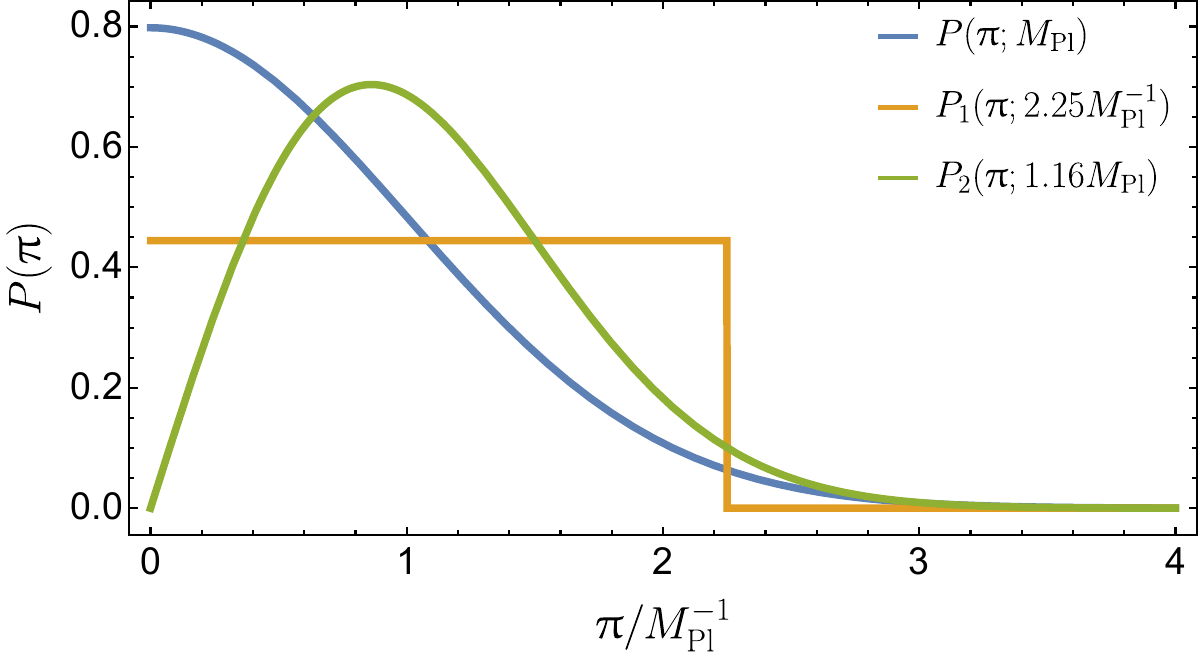}
    \caption{The comparison with the three different forms of $P(\ppi)$ with the same $\etamed=0.38$. Each function is normalized so that $\int_0^\infty P(\ppi)\dd\ppi=1$.
    }
  \label{fig:different_P}
\end{figure}

We apply two alternative forms of $P(\ppi)$, each of which is parameterized by a single parameter:
\begin{equation}
    P_1(\ppi;\ppi_0)=\Theta(\ppi_0-\ppi),
\end{equation}
where $\Theta(x)$ is the Heaviside function, which represents a top-hat form, and
\begin{equation}
    P_2(\ppi;\zeta)=(\zeta\ppi)\exp\left[-\frac{(\zeta\ppi)^2}{2}\right],
\end{equation}
which is a Rayleigh-like function. To make the comparison on equal footing, we pick the parameters of the alternative forms so that they have the same values of $\etamed$ as that of $P(\ppi,\Mp)$. As a result, the parameters are found to be $\ppi_0=2.25\Mp^{-1}$ and $\zeta=1.16\Mp$. A direct comparison of the three landscape power spectra is shown in Figure \ref{fig:different_P}. 

\begin{figure}[ht!]
    \centering  
    \includegraphics[width=0.46\textwidth]{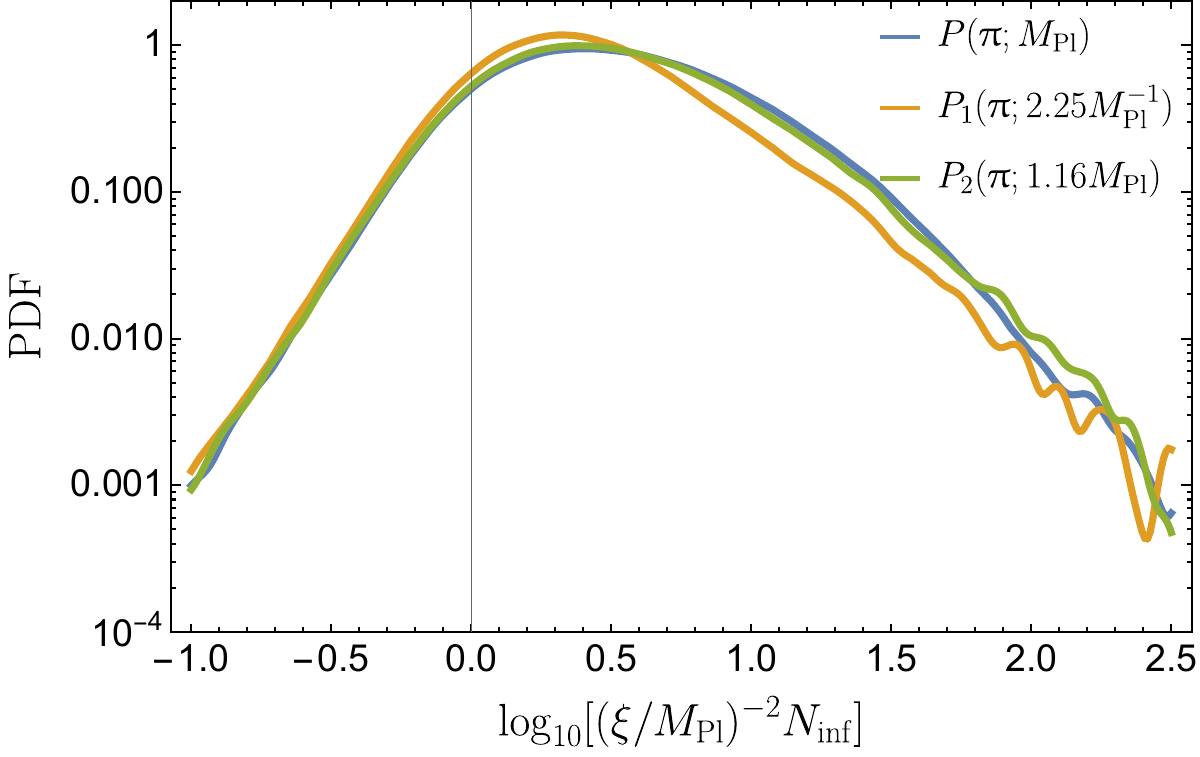}
    \hspace{8pt}
    \includegraphics[width=0.45\textwidth]{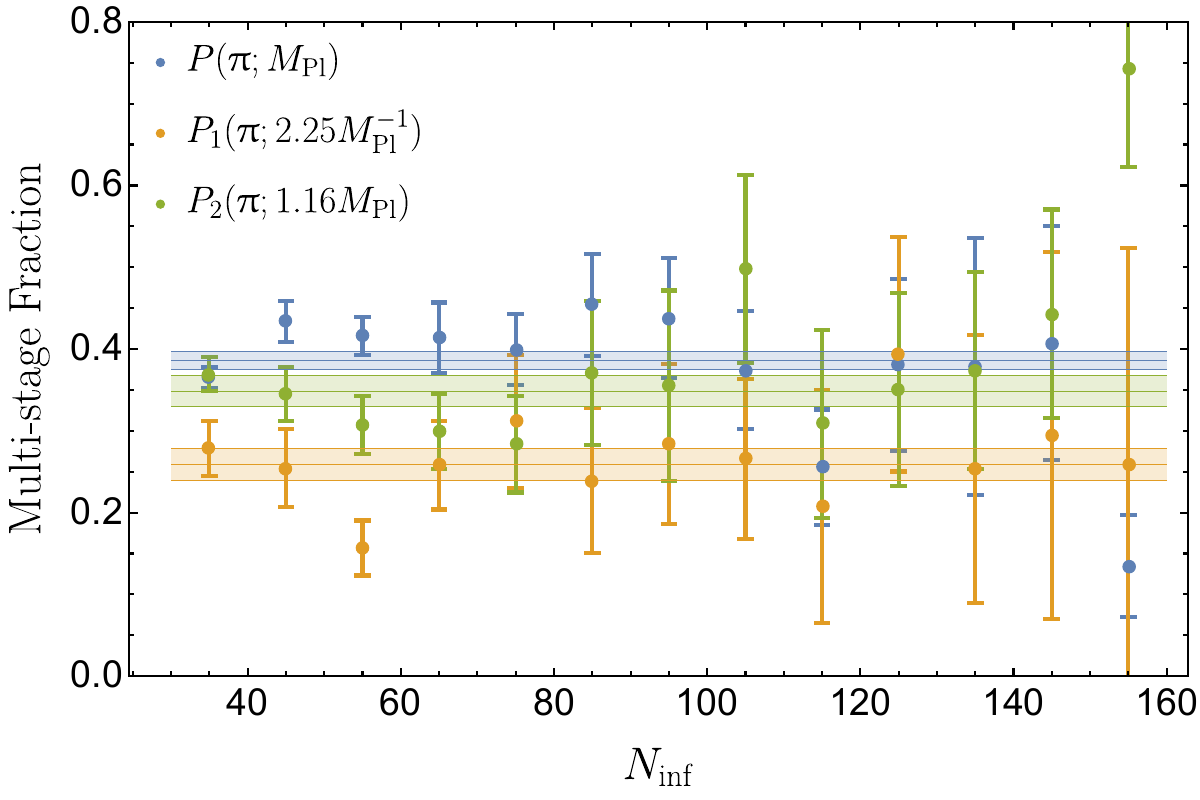}
    \caption{Left panel: The PDFs of $\log_{10}N_{\rm inf}$ of samples with different functional forms of $P(\ppi)$ with fixed $\etamed=0.38$. Right panel: The binned fractions of multi-stage trajectories of these samples, with horizontal bands the total multi-stage fractions and corresponding errors.
    }
  \label{fig:varP_comparison}
\end{figure}

For each of $P_1$ and $P_2$, we calculate a sample with the size of $4000$ realizations. In this appendix, we choose to focus on a few essential properties as proxies of robustness, instead of making a comprehensive study of these samples. We choose to compute the PDF of the distribution of $N_{\rm inf}$ and the multi-stage fractions, and the results can be found in Figure \ref{fig:varP_comparison}. In the left panel, we find that the PDFs obtained from the three samples coincide fairly well, albeit with a slight deviation in the top-hat case. In the right panel, the fractions of multi-stage trajectories appear to be all substantial. In particular, the Rayleigh form power spectrum gives very similar results as the Gaussian form, and the top-hat form deviates a bit more but still gives a substantial fraction of multi-stage models.

We leave comprehensive comparisons of other properties for future works.

\section{Intermediate steps in characterizing $\eta_{V}$}
\label{appB}
In this appendix, we fill the missing gaps of some derivations related to the discussion of characterizing $\eta_{V}$ in Section \ref{sec-landscapes}.

\subsection{Mean of $v_{\rm min}$}

We have quoted the expression of $\langle v_{\rm min}\rangle$ in (\ref{eq:vmin}), of which we provide here a derivation based on the extreme value theory. 

Our task is to derive the expectation value of the \textit{global} minimum of a Gaussian random field $v(\boldsymbol{\phi})$ in a finite-sized domain of size $\Lambda$. Heuristically, the Gaussian random function can be divided into boxes with the size being the correlation length $\xi$, and each of the boxes can be viewed as an independent Gaussian variable whose variance is given by $\sigma^2_v=\braket{v^2}$. Therefore, the problem reduces to determining the expectation of the maximum of a number of $\sim(\Lambda/\xi)^2$ random Gaussian variables with variance $\sigma^2_v$.

Mathematically, if $(X_1,X_2,\dots,X_n)$ is a sample of size $n$ of independent and identically distributed random variables, each with cumulative distribution function (CDF) $F$, the maximum of the sample is described by the Fisher-Tippett-Gnedenko theorem \cite{FisherTippett1928,Gnedenko1943}. The theorem says that if there exist two sequences of real numbers $a_n>0$ and $b_n\in\mathbb{R}$ and a non-degenerate CDF $G$ such that, for 
every continuity point $x\in\mathbb{R}$ of $G$,
\begin{equation}\label{eq:max_limit}
    \lim_{n\to\infty} P\left(\frac{\max\left\{X_1,X_2,\dots,X_n\right\}-b_n}{a_n}\leq x\right)=G(x),
\end{equation}
then $G$ is the CDF of one of the three families of distributions: the Fr\'echet, the Gumbel, or the Weibull distribution. This condition can be equivalently translated into the following form:
\begin{equation}
    \lim_{n\to\infty}\left[F(a_n x+b_n)\right]^n=G(x).
\end{equation}
In particular, if $F$ is the CDF of the Gaussian distribution, the resulting distribution of $G$ belongs to the Gumbel distribution, which takes the form of 
\begin{equation}\label{eq:Gumbel}
    G(x)=\exp\left[-\exp(-x)\right].
\end{equation}
If we can find the forms of the sequences $a_n$ and $b_n$, the expectation value of the maximum of an $n$-sized sample can be read from $b_n$ according to (\ref{eq:max_limit}). Indeed, for a Gaussian distribution with PDF being
\begin{equation}
    f(y)=\frac{1}{\sqrt{2\pi\sigma^2}}\exp\left[-\frac{(y-\mu)^2}{2\sigma^2}\right],
\end{equation}
the CDF takes the standard form of
\begin{equation}
    F(y)=\Phi\left(\frac{y-\mu}{\sigma}\right)=\frac{1}{2}\left[1+\mathrm{erf}\left(\frac{y-\mu}{\sqrt{2}\sigma}\right)\right].
\end{equation}
This function leads to an asymptotic form of $\ln\left[F(y)\right]^n$ as $y\to\infty$ that
\begin{equation}
    \ln\left[F(y)\right]^n=n\ln F(y)\to-\frac{n\sigma}{\sqrt{2\pi}(y-\mu)}\exp\left[-\frac{(y-\mu)^2}{2\sigma^2}\right].
\end{equation}
Therefore, the task is to find sequences of $a_n$ and $b_n$ such that if $y=a_n x+b_n$, the expression above coincides with $\ln G(x)=-\exp(-x)$. In fact, if we choose $b_n$ such that
\begin{equation}\label{eq:b_n_for_Gumbel}
    \frac{n\sigma}{\sqrt{2\pi}(b_n-\mu)}\exp\left[-\frac{(b_n-\mu)^2}{2\sigma^2}\right]=1,
\end{equation}
we will find when $a_n x/(b_n-\mu)\ll1$ that
\begin{equation}
    \ln\left[F(a_n x+b_n)\right]^n\to-\left(1+O\left(\frac{a_n x}{b_n-\mu}\right)\right)\exp\left[-\frac{(b_n-\mu)a_n x}{\sigma^2}\right].
\end{equation}
Therefore, if we choose $a_n$ to be
\begin{equation}
    a_n=\frac{\sigma^2}{b_n-\mu},
\end{equation}
we will obtain the Gumbel distribution as expected. In the limit of $n\to\infty$, the equation (\ref{eq:b_n_for_Gumbel}) can be solved asymptotically to get
\begin{equation}
    b_n-\mu=\sigma\sqrt{2\ln n},\quad a_n=\frac{\sigma}{\sqrt{2\ln n}}.
\end{equation}
In conclusion, for a Gaussian variable with an expectation value $\mu$ and a standard deviation $\sigma$, the expectation value of the \textit{maximum} of an $n$-sized sample is given by $\mu+\sigma\sqrt{2\ln n}$ in the limit $n\to\infty$. Due to the symmetry of the Gaussian distribution, the expectation value of the minimum is $\mu-\sigma\sqrt{2\ln n}$.

In the case of the Gaussian random field $v(\boldsymbol{\phi})$, we have $\mu=0$ and $\sigma=\sigma_v$. The size of the sample $n$ is taken as the effective number of independent Gaussian variables in a realization, which can be estimated as $N_{\rm eff}=(\Lambda/\xi)^{N_{\rm f}}$. As a result, the value of $\braket{v_{\rm min}}$ is given by
\begin{equation}
    \braket{v_{\rm min}}=-\sigma_v\sqrt{2\ln N_{\rm eff}},
\end{equation}
which gives Eq.~\eqref{eq:vmin} of the main text.
For our fiducial set of parameters, this formula gives $-\braket{v_{\rm min}}/\sigma_v \in\{3.03,\, 3.72,\, \ldots\}$ for $N_{\rm f} \in \{2,\,3,\, \ldots \}$ with expected relative errors of order 10\%.
In the ensemble of landscapes that we have generated for our fiducial set of parameters, we instead find $-\braket{v_{\rm min}}/\sigma_v \in\{3.11,\, 4.14,\, \ldots\}$ for $N_{\rm f} \in \{2,\,3,\, \ldots \}$ which indeed agrees well with the theoretical formula up to the expected relative error.

\subsection{Mean of $\eta_V$}

Now, we want to compute $\braket{\eta_V}_U$ on the $U({\bm \phi})$-landscape. The joint probability density function for ${\bm X}_U=(U,\nabla^2 U)$ is the same as the one for $p_{\bm X}({\bm X})$ in Eq.~\eqref{eq:pxX} under $v\rightarrow U+\braket{v_{\rm min}}$.
In particular, we have $p_U(U) = (2 \pi \sigma_v^2)^{-1/2}\exp\left[-(U+\braket{v_{\rm min}})^2/(2 \sigma_v^2)\right]$ and therefore $\braket{U}=-\braket{v_{\rm min}}$ as expected.
We find
\begin{align}
    \braket{\eta_V}_U &= \frac{\Mp^2}{N_{\mathrm{f}}} \int \dd U \, \frac{p_U(U)}{U} \underbrace{\int \dd (\nabla^2 U ) \sqrt{\frac{1}{2\pi \, {\rm det} \Sigma/\sigma_v^2}} \exp\left[-\frac{\left(\nabla^2 U + N_{\rm f} \,(U+\braket{v_{\rm min}})/\xi^2  \right)^2}{2 \, {\rm det} \Sigma /\sigma_v^2}\right] \, \nabla^2 U}_{-N_{\rm f}( U+\braket{v_{\rm min}})/\xi^2} \nonumber\\ 
    & =- \frac{\Mp^2}{\xi^2} \left[1 +\braket{v_{\rm min}} \, \int \dd U \,  \frac{p_U(U)}{U} \right] \,.
\end{align}
Strictly speaking, the correction proportional to $\braket{v_{\rm min}}$ is a divergent integral, but it can (again, see footnote~\ref{footnote:eta_integral}) be regulated by either: taking its principal value; evaluating it on a grid with finite lattice spacing and then let the number of points go to infinity.
Either way, we find the same result, which is 
\begin{equation}
    \braket{\eta_V}_U = - \frac{\Mp^2}{\xi^2} \left[1 - \sqrt{2}\, \frac{\braket{v_{\rm min}}}{\sigma_v} F\left(\frac{\braket{v_{\rm min}}}{\sqrt{2} \sigma_v}\right) \right]\,,
\end{equation}
where we remind that $F(x)=e^{-x^2} \int_0^x \dd t \, e^{t^2} $ is Dawson's $F$ integral.
On a grid, using Eq.~\eqref{eq:vmin}, we find Eq.~\eqref{eq:mean-etaV-final} of the main text.
As already said in the main body of this article, for our fiducial set of parameters, this corresponds to $ \braket{\eta_V}_U \in\{0.17,\, 0.10,\, \ldots\} \Mp^2/\xi^2$ for $N_{\rm f} \in \{2,\,3,\, \ldots \}$, to be compared with the values that we find in our numerical simulations $ \braket{\eta_V}_U \in\{0.22,\,0.08,\, \ldots\} \Mp^2/\xi^2$ for $N_{\rm f} \in \{2, 3, \ldots\}$.

\section{Statistical Bootstrap Technique}
\label{app_bootstrap}
Statistical bootstrap \cite{Efron:1979Bootstrap} is a powerful numerical technique of estimating the statistical variance of a quantity from a given sample of data, which we have used intensively throughout this work to derive the variances of quantities such as the fractions of multi-stage or multi-field trajectories. In this appendix, we provide an introduction of how this technique works.

\subsection{Theoretical framework}
Suppose an independent and identically-distributed random variable $X$ follows a certain distribution $P$, and a quantity $\theta$ is dictated by the distribution, \textit{i.e.} $\theta=t(P)$. The quantity $\theta$ can be various quantities of interest, such as the mean, the median, or other derived quantities.\footnote{There are some exceptions that are not applicable, for example the maximum.} We draw a size-$n$ sample $\boldsymbol{X}=(X_1,X_2,\dots,X_n)$ from the distribution $P$, and an estimator $\hat{\theta}$ of the quantity of interest can be constructed from the sample by $\hat{\theta}=s(\boldsymbol{X})$. The question is how we can obtain an estimate of the variance of $\hat{\theta}$ from the existing sample $\boldsymbol{X}$, when drawing multiple samples of the same size from $P$ is technically difficult or even impossible.

The fundamental idea is that we use the empirical distribution $\hat{P}$ derived from the sample $\boldsymbol{X}$ as a direct estimate of the true distribution $P$. From the empirical distribution $\hat{P}$, we can draw a bootstrap sample $\boldsymbol{X}^*=(X_1^*,X_2^*,\dots,X_n^*)$, which is equivalent to drawing a sample of size $n$ \textit{with replacement} from the original data set $\boldsymbol{X}=(X_1,X_2,\dots,X_n)$.\footnote{That is, in some observations $X_i$ can be drawn multiple times, while some may not appear in $\boldsymbol{X}^*$ at all.} With each bootstrap sample, an estimator of the quantity $\theta$ can be obtained by $\hat{\theta}=s(\boldsymbol{X}^*)$.

An estimate of the variance of $\hat{\theta}$ can be obtained in the following steps:
\begin{itemize}
    \item Draw a number of bootstrap samples $\boldsymbol{X}^*_b=(X_{b,1}^*,X_{b,1}^*,\dots,X_{b,n}^*)$ for $b=1,2,\dots,B$.
    \item Calculate the estimator $\hat{\theta}_b^*$ of the $b$-th bootstrap sample.
    \item The bootstrap estimate of the standard deviation of $\hat{\theta}$, denoted by $\hat{\sigma}(\hat{\theta})$, is given by
    \begin{equation}\label{app_std_var}
        \hat{\sigma}(\hat{\theta})=\sqrt{\frac{1}{B-1}\sum_{b=1}^B\left(\hat{\theta}_b^*-\bar{\theta}^*\right)^2},
    \end{equation}
    in which $\bar\theta^*=\frac{1}{B}\sum_{b=1}^B\hat{\theta}_b^*$,
\end{itemize}
The value of $B$ should be taken sufficiently large so that the estimated value of $\hat{\sigma}$ is convergent.

\subsection{Application to the multi-stage (multi-field) fraction}
We demonstrate here how the bootstrap technique applies to the estimation of the statistical variance of the fraction of multi-stage (multi-field) trajectories in this work. In this case, each observation $X$ is a realization of the landscape. Therefore, the underlying distribution $P$ of $X$ is implicitly encoded in the way we generate the realizations. The original sample $\boldsymbol{X}$ contains $M$ realizations, with $N_i$ successful trajectories in the $i$-th realization. Therefore, the total number of trajectories in the sample is $N_{\rm tot}=\sum_{i=1}^{M}N_i$. The quantity $\theta$ in question is the fraction of multi-stage (multi-field) trajectories, which is fundamentally dictated by the distribution $P$. On the other hand, an estimator $\hat{\theta}$ can be derived from the sample by counting the total number of multi-stage (multi-field) trajectories $N_{\rm m}$ among all successful trajectories, and the estimator is given by $\hat{\theta}=N_{\rm m}/N_{\rm tot}$.

The empirical distribution $\hat{P}$ consists of the original sample $\boldsymbol{X}$ of $M$ realizations, from which a bootstrap sample $\boldsymbol{X}^*$ can be constructed by drawing a sample of realizations of size $M$ \textit{with replacement} from $\boldsymbol{X}$. Since the $i$-th realization in the bootstrap sample carries $N_i^*$ successful trajectories,\footnote{If the $i$-th realization in the bootstrap sample is given by the $j$-th realization in the original sample, we will have $N_i^*=N_j$.} the total number of successful trajectories is given by $N_{\rm tot}^*=\sum_{i=1}^M N_i^*$. After counting the number of multi-stage (multi-field) trajectories $N_{\rm m}^*$, the estimator is given by $\hat{\theta}^*=N_{\rm m}^*/N_{\rm tot}^*$. After drawing $B$ bootstrap samples, the estimated variance $\hat{\sigma}(\hat{\theta})$ is directly obtained by (\ref{app_std_var}). In our case, the result is sufficiently convergent with $B=100$, in the sense that the deviation of $\hat{\sigma}$ among different trials is far less than $\hat{\sigma}$ itself.

This procedure can be easily generalized to the binned case, in which $N_{\rm tot}$ and $N_{\rm m}$ are calculated in a constraint range of $N_{\rm inf}$. Indeed, in each bootstrap sample, we can obtain $\hat{\theta}^{(j)*}=N_{\rm m}^{(j)*}/N_{\rm tot}^{(j)*}$ in the $j$-th bin, and the variance of each $\hat{\theta}^{(j)*}$ is readily obtained after drawing $B$ times.

\section{A Gallery of Trajectories}
\label{app_gallery}
\begin{figure}[ht!]
    \centering  
    \includegraphics[width=\textwidth]{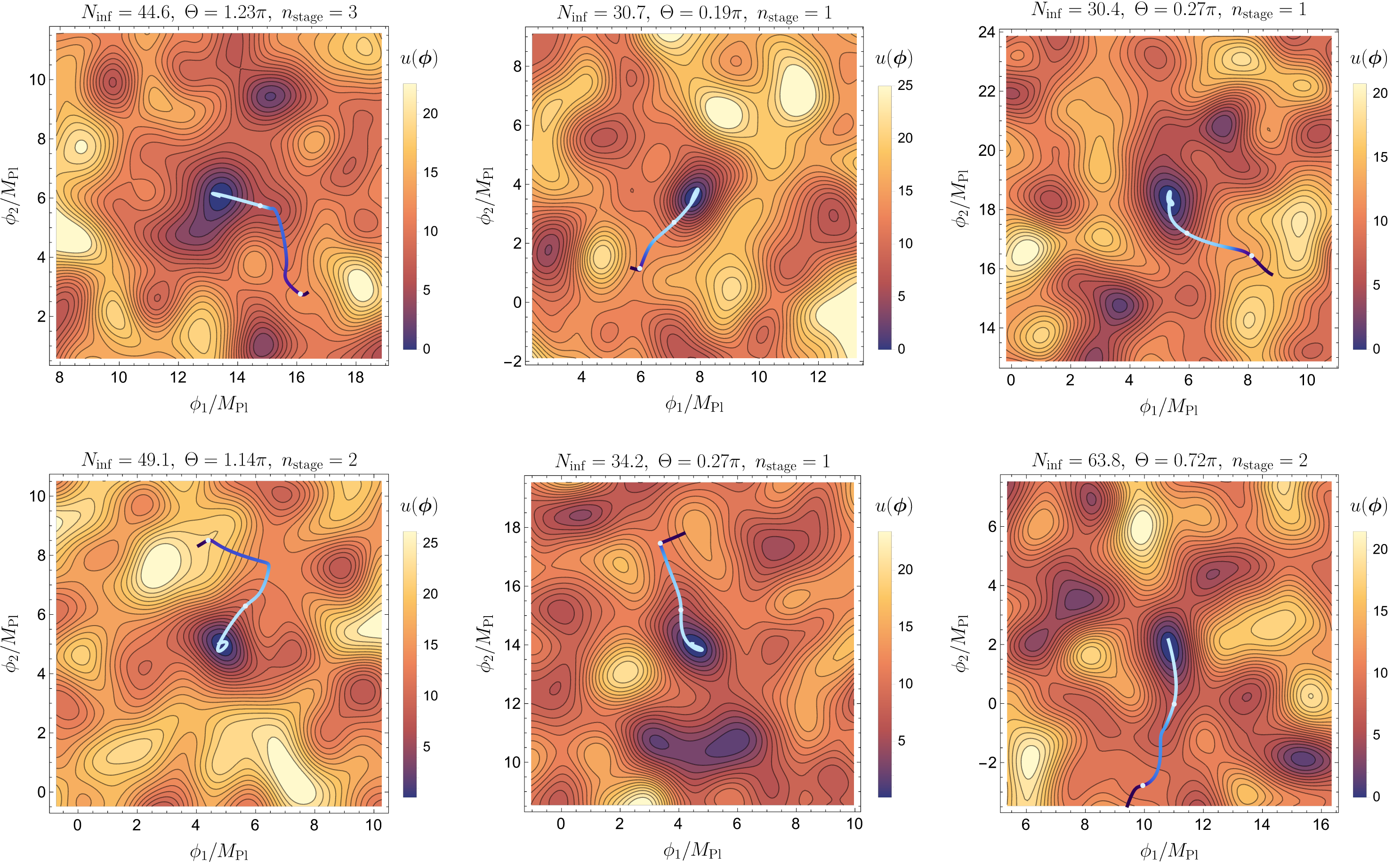}
    \caption{The appearance of randomly selected successful trajectories in the two-field case. Each panel is labelled with the values of $N_{\rm inf}$, $\Theta$ and $n_{\rm stage}$ of the trajectory, and the highlighted points on the trajectory are the instants of $N_{\rm ini}$ and $N_{\rm fin}$ within which the value of $\Theta$ is calculated.
    }
  \label{fig:trajectory_gallery}
\end{figure}

We have uncovered a number of statistical properties of the ensemble of trajectories in the main text. To provide a more intuitive physical picture of what these trajectories look like beyond their statistical properties, in this appendix we show several successful trajectories for the $N_{\rm f}=2$ case in Figure \ref{fig:trajectory_gallery}. These trajectories are randomly selected from the ensemble and are broadly representative of typical trajectories. Some of them contain multiple stages and clearly exhibit multiple slow-roll attractors. Along each trajectory, we highlight the instants $N_{\rm ini}$ and $N_{\rm fin}$, and make the geometrical meaning of $\Theta$ readily apparent. From these examples, we can also identify two sources of $\Theta$: non-slow-roll transitions between different slow-roll stages and curving of a slow-roll attractor.

\begin{figure}[ht!]
    \centering  
    \includegraphics[width=\textwidth]{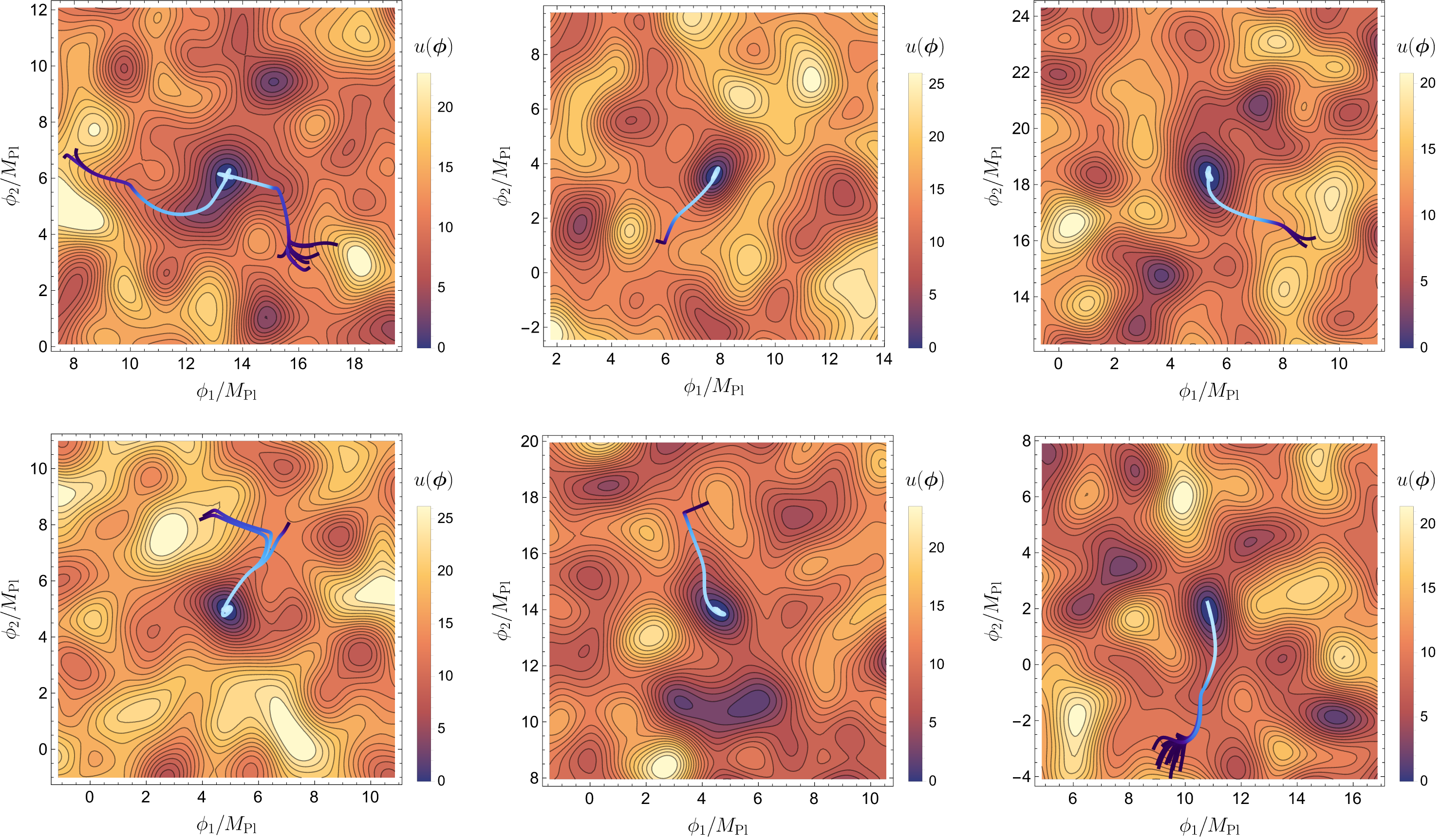}
    \caption{A visualization of all successful trajectories of the same landscape realizations in Figure \ref{fig:trajectory_gallery}. See the main text for descriptions.
    }
  \label{fig:trajectory_gallery_successful}
\end{figure}

There are cases in which multiple successful trajectories share a common section of their routes, as well as cases in which multiple distinct routes exist within the same realization. To build intuition for these possibilities, we showcase all successful trajectories in each landscape realization in Figure \ref{fig:trajectory_gallery_successful} (while in Figure \ref{fig:trajectory_gallery}, only one successful trajectory is shown for each landscape realization). From these examples, we can clearly see that, in most realizations, there is only one mainstream route, with multiple trajectories joining it partway through and giving rise to several successful trajectories. Each mainstream route is associated with certain patches of the landscape such that, if the inflaton starts within one of these patches, it flows into the corresponding mainstream route. The sizes of these patches vary among different mainstream routes, contributing different weights to the statistical properties. A particularly interesting situation is found in the upper-left panel of Figure \ref{fig:trajectory_gallery_successful}, where two distinct routes are available, each containing a group of successful trajectories.

\bibliographystyle{JHEP}
\bibliography{Ref}

\providecommand{\href}[2]{#2}\begingroup\raggedright\begin{thebibliography}{10}

\bibitem{Starobinsky:1979ty}
A.~A. Starobinsky, \emph{{Spectrum of relict gravitational radiation and the early state of the universe}}, {\emph{JETP Lett.} {\bfseries 30} (1979) 682}.

\bibitem{Mukhanov:1981xt}
V.~F. Mukhanov and G.~V. Chibisov, \emph{{Quantum Fluctuations and a Nonsingular Universe}}, {\emph{JETP Lett.} {\bfseries 33} (1981) 532}.

\bibitem{Guth1981inflation}
A.~H. Guth, \emph{Inflationary universe: A possible solution to the horizon and flatness problems}, \href{https://doi.org/10.1103/PhysRevD.23.347}{\emph{Phys. Rev. D} {\bfseries 23} (1981) 347}.

\bibitem{Linde:1982uu}
A.~D. Linde, \emph{{Scalar Field Fluctuations in Expanding Universe and the New Inflationary Universe Scenario}}, \href{https://doi.org/10.1016/0370-2693(82)90293-3}{\emph{Phys. Lett. B} {\bfseries 116} (1982) 335}.

\bibitem{Albrecht:1982mp}
A.~Albrecht, P.~J. Steinhardt, M.~S. Turner and F.~Wilczek, \emph{{Reheating an Inflationary Universe}}, \href{https://doi.org/10.1103/PhysRevLett.48.1437}{\emph{Phys. Rev. Lett.} {\bfseries 48} (1982) 1437}.

\bibitem{Hawking:1982cz}
S.~W. Hawking, \emph{{The Development of Irregularities in a Single Bubble Inflationary Universe}}, \href{https://doi.org/10.1016/0370-2693(82)90373-2}{\emph{Phys. Lett. B} {\bfseries 115} (1982) 295}.

\bibitem{Starobinsky:1982ee}
A.~A. Starobinsky, \emph{{Dynamics of Phase Transition in the New Inflationary Universe Scenario and Generation of Perturbations}}, \href{https://doi.org/10.1016/0370-2693(82)90541-X}{\emph{Phys. Lett. B} {\bfseries 117} (1982) 175}.

\bibitem{Guth1982inflation-perturbations}
A.~H. Guth and S.-Y. Pi, \emph{Fluctuations in the new inflationary universe}, \href{https://doi.org/10.1103/PhysRevLett.49.1110}{\emph{Phys. Rev. Lett.} {\bfseries 49} (1982) 1110}.

\bibitem{Vilenkin:1982wt}
A.~Vilenkin and L.~H. Ford, \emph{{Gravitational Effects upon Cosmological Phase Transitions}}, \href{https://doi.org/10.1103/PhysRevD.26.1231}{\emph{Phys. Rev. D} {\bfseries 26} (1982) 1231}.

\bibitem{Linde:1983gd}
A.~D. Linde, \emph{{Chaotic Inflation}}, \href{https://doi.org/10.1016/0370-2693(83)90837-7}{\emph{Phys. Lett. B} {\bfseries 129} (1983) 177}.

\bibitem{Mukhanov:1985rz}
V.~F. Mukhanov, \emph{{Gravitational Instability of the Universe Filled with a Scalar Field}}, {\emph{JETP Lett.} {\bfseries 41} (1985) 493}.

\bibitem{Sasaki:1986hm}
M.~Sasaki, \emph{{Large Scale Quantum Fluctuations in the Inflationary Universe}}, \href{https://doi.org/10.1143/PTP.76.1036}{\emph{Prog. Theor. Phys.} {\bfseries 76} (1986) 1036}.

\bibitem{Planck-legacy-2020}
{Planck Collaboration}, N.~{Aghanim}, Y.~{Akrami}, F.~{Arroja}, M.~{Ashdown}, J.~{Aumont} et~al., \emph{{Planck 2018 results. I. Overview and the cosmological legacy of Planck}}, \href{https://doi.org/10.1051/0004-6361/201833880}{\emph{\aap} {\bfseries 641} (2020) A1} [\href{https://arxiv.org/abs/1807.06205}{{\ttfamily 1807.06205}}].

\bibitem{Copeland:1994vg}
E.~J. Copeland, A.~R. Liddle, D.~H. Lyth, E.~D. Stewart and D.~Wands, \emph{{False vacuum inflation with Einstein gravity}}, \href{https://doi.org/10.1103/PhysRevD.49.6410}{\emph{Phys. Rev.} {\bfseries D49} (1994) 6410} [\href{https://arxiv.org/abs/astro-ph/9401011}{{\ttfamily astro-ph/9401011}}].

\bibitem{Chen:2008hz}
X.~Chen, \emph{{Fine-Tuning in DBI Inflationary Mechanism}}, \href{https://doi.org/10.1088/1475-7516/2008/12/009}{\emph{JCAP} {\bfseries 12} (2008) 009} [\href{https://arxiv.org/abs/0807.3191}{{\ttfamily 0807.3191}}].

\bibitem{Masoumi:2016eag}
A.~Masoumi, A.~Vilenkin and M.~Yamada, \emph{{Inflation in random Gaussian landscapes}}, \href{https://doi.org/10.1088/1475-7516/2017/05/053}{\emph{JCAP} {\bfseries 05} (2017) 053} [\href{https://arxiv.org/abs/1612.03960}{{\ttfamily 1612.03960}}].

\bibitem{Tegmark:2004qd}
M.~Tegmark, \emph{{What does inflation really predict?}}, \href{https://doi.org/10.1088/1475-7516/2005/04/001}{\emph{JCAP} {\bfseries 04} (2005) 001} [\href{https://arxiv.org/abs/astro-ph/0410281}{{\ttfamily astro-ph/0410281}}].

\bibitem{Frazer:2011tg}
J.~Frazer and A.~R. Liddle, \emph{{Exploring a string-like landscape}}, \href{https://doi.org/10.1088/1475-7516/2011/02/026}{\emph{JCAP} {\bfseries 02} (2011) 026} [\href{https://arxiv.org/abs/1101.1619}{{\ttfamily 1101.1619}}].

\bibitem{Frazer:2011br}
J.~Frazer and A.~R. Liddle, \emph{{Multi-field inflation with random potentials: field dimension, feature scale and non-Gaussianity}}, \href{https://doi.org/10.1088/1475-7516/2012/02/039}{\emph{JCAP} {\bfseries 02} (2012) 039} [\href{https://arxiv.org/abs/1111.6646}{{\ttfamily 1111.6646}}].

\bibitem{McAllister:2012am}
L.~McAllister, S.~Renaux-Petel and G.~Xu, \emph{{A Statistical Approach to Multifield Inflation: Many-field Perturbations Beyond Slow Roll}}, \href{https://doi.org/10.1088/1475-7516/2012/10/046}{\emph{JCAP} {\bfseries 1210} (2012) 046} [\href{https://arxiv.org/abs/1207.0317}{{\ttfamily 1207.0317}}].

\bibitem{Bjorkmo:2017nzd}
T.~Bjorkmo and M.~C.~D. Marsh, \emph{{Manyfield Inflation in Random Potentials}}, \href{https://doi.org/10.1088/1475-7516/2018/02/037}{\emph{JCAP} {\bfseries 02} (2018) 037} [\href{https://arxiv.org/abs/1709.10076}{{\ttfamily 1709.10076}}].

\bibitem{Planck:2018jri}
{\scshape Planck} collaboration, Y.~Akrami et~al., \emph{{Planck 2018 results. X. Constraints on inflation}}, \href{https://doi.org/10.1051/0004-6361/201833887}{\emph{Astron. Astrophys.} {\bfseries 641} (2020) A10} [\href{https://arxiv.org/abs/1807.06211}{{\ttfamily 1807.06211}}].

\bibitem{Quevedo:2002xw}
F.~Quevedo, \emph{{Lectures on string/brane cosmology}}, \href{https://doi.org/10.1088/0264-9381/19/22/304}{\emph{Class. Quant. Grav.} {\bfseries 19} (2002) 5721} [\href{https://arxiv.org/abs/hep-th/0210292}{{\ttfamily hep-th/0210292}}].

\bibitem{Linde:2005dd}
A.~D. Linde, \emph{{Inflation and string cosmology}}, \href{https://doi.org/10.1143/PTPS.163.295}{\emph{Prog. Theor. Phys. Suppl.} {\bfseries 163} (2006) 295} [\href{https://arxiv.org/abs/hep-th/0503195}{{\ttfamily hep-th/0503195}}].

\bibitem{McAllister:2007bg}
L.~McAllister and E.~Silverstein, \emph{{String Cosmology: A Review}}, \href{https://doi.org/10.1007/s10714-007-0556-6}{\emph{Gen. Rel. Grav.} {\bfseries 40} (2008) 565} [\href{https://arxiv.org/abs/0710.2951}{{\ttfamily 0710.2951}}].

\bibitem{Baumann:2014nda}
D.~Baumann and L.~McAllister, \emph{{Inflation and String Theory}}. Cambridge University Press, 2015, [\href{https://arxiv.org/abs/1404.2601}{{\ttfamily 1404.2601}}].

\bibitem{Cicoli:2023opf}
M.~Cicoli, J.~P. Conlon, A.~Maharana, S.~Parameswaran, F.~Quevedo and I.~Zavala, \emph{{String cosmology: From the early universe to today}}, \href{https://doi.org/10.1016/j.physrep.2024.01.002}{\emph{Phys. Rept.} {\bfseries 1059} (2024) 1} [\href{https://arxiv.org/abs/2303.04819}{{\ttfamily 2303.04819}}].

\bibitem{Achucarro:2018vey}
A.~Ach{\'u}carro and G.~A. Palma, \emph{{The string swampland constraints require multi-field inflation}}, \href{https://doi.org/10.1088/1475-7516/2019/02/041}{\emph{JCAP} {\bfseries 1902} (2019) 041} [\href{https://arxiv.org/abs/1807.04390}{{\ttfamily 1807.04390}}].

\bibitem{Bravo:2020wdr}
R.~Bravo, G.~A. Palma and S.~Riquelme, \emph{{A Tip for Landscape Riders: Multi-Field Inflation Can Fulfill the Swampland Distance Conjecture}}, \href{https://doi.org/10.1088/1475-7516/2020/02/004}{\emph{JCAP} {\bfseries 02} (2020) 004} [\href{https://arxiv.org/abs/1906.05772}{{\ttfamily 1906.05772}}].

\bibitem{Obied:2018sgi}
G.~Obied, H.~Ooguri, L.~Spodyneiko and C.~Vafa, \emph{{De Sitter Space and the Swampland}},  \href{https://arxiv.org/abs/1806.08362}{{\ttfamily 1806.08362}}.

\bibitem{Klaewer:2016kiy}
D.~Klaewer and E.~Palti, \emph{{Super-Planckian Spatial Field Variations and Quantum Gravity}}, \href{https://doi.org/10.1007/JHEP01(2017)088}{\emph{JHEP} {\bfseries 01} (2017) 088} [\href{https://arxiv.org/abs/1610.00010}{{\ttfamily 1610.00010}}].

\bibitem{Grimm:2018ohb}
T.~W. Grimm, E.~Palti and I.~Valenzuela, \emph{{Infinite Distances in Field Space and Massless Towers of States}}, \href{https://doi.org/10.1007/JHEP08(2018)143}{\emph{JHEP} {\bfseries 08} (2018) 143} [\href{https://arxiv.org/abs/1802.08264}{{\ttfamily 1802.08264}}].

\bibitem{Weinberg:2008hq}
S.~Weinberg, \emph{{Effective Field Theory for Inflation}}, \href{https://doi.org/10.1103/PhysRevD.77.123541}{\emph{Phys. Rev. D} {\bfseries 77} (2008) 123541} [\href{https://arxiv.org/abs/0804.4291}{{\ttfamily 0804.4291}}].

\bibitem{Cheung:2007st}
C.~Cheung, P.~Creminelli, A.~L. Fitzpatrick, J.~Kaplan and L.~Senatore, \emph{{The Effective Field Theory of Inflation}}, \href{https://doi.org/10.1088/1126-6708/2008/03/014}{\emph{JHEP} {\bfseries 03} (2008) 014} [\href{https://arxiv.org/abs/0709.0293}{{\ttfamily 0709.0293}}].

\bibitem{Senatore:2010wk}
L.~Senatore and M.~Zaldarriaga, \emph{{The Effective Field Theory of Multifield Inflation}}, \href{https://doi.org/10.1007/JHEP04(2012)024}{\emph{JHEP} {\bfseries 04} (2012) 024} [\href{https://arxiv.org/abs/1009.2093}{{\ttfamily 1009.2093}}].

\bibitem{Noumi:2012vr}
T.~Noumi, M.~Yamaguchi and D.~Yokoyama, \emph{{Effective field theory approach to quasi-single field inflation and effects of heavy fields}}, \href{https://doi.org/10.1007/JHEP06(2013)051}{\emph{JHEP} {\bfseries 06} (2013) 051} [\href{https://arxiv.org/abs/1211.1624}{{\ttfamily 1211.1624}}].

\bibitem{Pinol:2024arz}
L.~Pinol, \emph{{Effective field theory of multifield inflationary fluctuations}}, \href{https://doi.org/10.1103/PhysRevD.110.L041302}{\emph{Phys. Rev. D} {\bfseries 110} (2024) L041302} [\href{https://arxiv.org/abs/2405.02190}{{\ttfamily 2405.02190}}].

\bibitem{Kaiser:2010ps}
D.~I. Kaiser, \emph{{Conformal Transformations with Multiple Scalar Fields}}, \href{https://doi.org/10.1103/PhysRevD.81.084044}{\emph{Phys. Rev. D} {\bfseries 81} (2010) 084044} [\href{https://arxiv.org/abs/1003.1159}{{\ttfamily 1003.1159}}].

\bibitem{Mulryne:2016mzv}
D.~J. Mulryne and J.~W. Ronayne, \emph{{PyTransport: A Python package for the calculation of inflationary correlation functions}}, \href{https://doi.org/10.21105/joss.00494}{\emph{J. Open Source Softw.} {\bfseries 3} (2018) 494} [\href{https://arxiv.org/abs/1609.00381}{{\ttfamily 1609.00381}}].

\bibitem{Seery:2012vj}
D.~Seery, D.~J. Mulryne, J.~Frazer and R.~H. Ribeiro, \emph{{Inflationary perturbation theory is geometrical optics in phase space}}, \href{https://doi.org/10.1088/1475-7516/2012/09/010}{\emph{JCAP} {\bfseries 09} (2012) 010} [\href{https://arxiv.org/abs/1203.2635}{{\ttfamily 1203.2635}}].

\bibitem{Mulryne:2013uka}
D.~J. Mulryne, \emph{{Transporting non-Gaussianity from sub to super-horizon scales}}, \href{https://doi.org/10.1088/1475-7516/2013/09/010}{\emph{JCAP} {\bfseries 09} (2013) 010} [\href{https://arxiv.org/abs/1302.3842}{{\ttfamily 1302.3842}}].

\bibitem{Dias:2016rjq}
M.~Dias, J.~Frazer, D.~J. Mulryne and D.~Seery, \emph{{Numerical evaluation of the bispectrum in multiple field inflation\textemdash{}the transport approach with code}}, \href{https://doi.org/10.1088/1475-7516/2016/12/033}{\emph{JCAP} {\bfseries 12} (2016) 033} [\href{https://arxiv.org/abs/1609.00379}{{\ttfamily 1609.00379}}].

\bibitem{Butchers:2018hds}
S.~Butchers and D.~Seery, \emph{{Numerical evaluation of inflationary 3-point functions on curved field space---with the transport method \& CppTransport}}, \href{https://doi.org/10.1088/1475-7516/2018/07/031}{\emph{JCAP} {\bfseries 1807} (2018) 031} [\href{https://arxiv.org/abs/1803.10563}{{\ttfamily 1803.10563}}].

\bibitem{Chen:2009we}
X.~Chen and Y.~Wang, \emph{{Large non-Gaussianities with Intermediate Shapes from Quasi-Single Field Inflation}}, \href{https://doi.org/10.1103/PhysRevD.81.063511}{\emph{Phys. Rev. D} {\bfseries 81} (2010) 063511} [\href{https://arxiv.org/abs/0909.0496}{{\ttfamily 0909.0496}}].

\bibitem{Chen:2009zp}
X.~Chen and Y.~Wang, \emph{{Quasi-Single Field Inflation and Non-Gaussianities}}, \href{https://doi.org/10.1088/1475-7516/2010/04/027}{\emph{JCAP} {\bfseries 04} (2010) 027} [\href{https://arxiv.org/abs/0911.3380}{{\ttfamily 0911.3380}}].

\bibitem{GrootNibbelink:2000vx}
S.~Groot~Nibbelink and B.~J.~W. van Tent, \emph{{Density perturbations arising from multiple field slow- roll inflation}},  \href{https://arxiv.org/abs/hep-ph/0011325}{{\ttfamily hep-ph/0011325}}.

\bibitem{GrootNibbelink:2001qt}
S.~Groot~Nibbelink and B.~van Tent, \emph{{Scalar perturbations during multiple field slow-roll inflation}}, \href{https://doi.org/10.1088/0264-9381/19/4/302}{\emph{Class. Quant. Grav.} {\bfseries 19} (2002) 613} [\href{https://arxiv.org/abs/hep-ph/0107272}{{\ttfamily hep-ph/0107272}}].

\bibitem{Kaiser:2012ak}
D.~I. Kaiser, E.~A. Mazenc and E.~I. Sfakianakis, \emph{{Primordial Bispectrum from Multifield Inflation with Nonminimal Couplings}}, \href{https://doi.org/10.1103/PhysRevD.87.064004}{\emph{Phys. Rev. D} {\bfseries 87} (2013) 064004} [\href{https://arxiv.org/abs/1210.7487}{{\ttfamily 1210.7487}}].

\bibitem{Achucarro:2018ngj}
A.~Ach{\'u}carro, S.~C{\'e}spedes, A.-C. Davis and G.~A. Palma, \emph{{Constraints on Holographic Multifield Inflation and Models Based on the Hamilton-Jacobi Formalism}}, \href{https://doi.org/10.1103/PhysRevLett.122.191301}{\emph{Phys. Rev. Lett.} {\bfseries 122} (2019) 191301} [\href{https://arxiv.org/abs/1809.05341}{{\ttfamily 1809.05341}}].

\bibitem{Pinol:2020kvw}
L.~Pinol, \emph{{Multifield inflation beyond $N_\mathrm{field}=2$: non-Gaussianities and single-field effective theory}}, \href{https://doi.org/10.1088/1475-7516/2021/04/002}{\emph{JCAP} {\bfseries 04} (2021) 002} [\href{https://arxiv.org/abs/2011.05930}{{\ttfamily 2011.05930}}].

\bibitem{Balkenhol:2025wms}
L.~Balkenhol et~al., \emph{{Inflation at the End of 2025: Constraints on $r$ and $n_s$ Using the Latest CMB and BAO Data}},  \href{https://arxiv.org/abs/2512.10613}{{\ttfamily 2512.10613}}.

\bibitem{Chen:2011zf}
X.~Chen, \emph{{Primordial Features as Evidence for Inflation}}, \href{https://doi.org/10.1088/1475-7516/2012/01/038}{\emph{JCAP} {\bfseries 01} (2012) 038} [\href{https://arxiv.org/abs/1104.1323}{{\ttfamily 1104.1323}}].

\bibitem{Chen:2014cwa}
X.~Chen, M.~H. Namjoo and Y.~Wang, \emph{{Models of the Primordial Standard Clock}}, \href{https://doi.org/10.1088/1475-7516/2015/02/027}{\emph{JCAP} {\bfseries 02} (2015) 027} [\href{https://arxiv.org/abs/1411.2349}{{\ttfamily 1411.2349}}].

\bibitem{Bjorkmo_2019}
T.~Bjorkmo, \emph{Rapid-turn inflationary attractors}, \href{https://doi.org/10.1103/physrevlett.122.251301}{\emph{Physical Review Letters} {\bfseries 122} (2019) }.

\bibitem{Liddle:2000cg}
A.~R. Liddle and D.~H. Lyth, \emph{{Cosmological inflation and large scale structure}}. 2000, \href{https://doi.org/10.1017/CBO9781139175180}{10.1017/CBO9781139175180}.

\bibitem{FisherTippett1928}
R.~A. Fisher and L.~H.~C. Tippett, \emph{Limiting forms of the frequency distribution of the largest or smallest member of a sample}, \href{https://doi.org/10.1017/S0305004100015681}{\emph{Proceedings of the Cambridge Philosophical Society} {\bfseries 24} (1928) 180}.

\bibitem{Gnedenko1943}
B.~Gnedenko, \emph{Sur la distribution limite du terme maximum d'une s\'erie al\'eatoire}, \href{https://doi.org/10.2307/1968974}{\emph{Annals of Mathematics} {\bfseries 44} (1943) 423}.

\bibitem{Efron:1979Bootstrap}
B.~Efron, \emph{Bootstrap methods: Another look at the jackknife}, \href{https://doi.org/10.1214/aos/1176344552}{\emph{The Annals of Statistics} {\bfseries 7} (1979) 1}.

\end{thebibliography}\endgroup

\end{document}